\documentclass[%
reprint,
superscriptaddress,
 amsmath,amssymb,
 aps,
noeprints,
floats,
10pt]{revtex4-2}

\usepackage[utf8]{inputenc}
\usepackage{physics}
\usepackage{amsmath}
\usepackage{caption}
\usepackage{bm}
\usepackage{upgreek}
\usepackage{xr}
\usepackage[dvipsnames]{xcolor}
\usepackage{subcaption}
\usepackage{float}
\usepackage{tablefootnote}
\usepackage{footnote}
\usepackage{graphicx}
\usepackage{braket}
\usepackage[symbol]{footmisc}
\usepackage[export]{adjustbox}
\usepackage{tabularx}
\usepackage{lipsum}  
\newcommand{\comment}[1]{}
\usepackage{multirow}
\usepackage{makecell}

\begin{document}

\title{Optical absorption spectra of metal oxides from time-dependent density functional theory and many-body perturbation theory based on optimally-tuned hybrid functionals}

\author{Guy Ohad}
    \affiliation{Department of Molecular Chemistry and Materials Science, Weizmann Institute of Science, Rehovoth 76100, Israel}
\author{Stephen E. Gant}
    \affiliation{Department of Physics, University of California, Berkeley, CA 94720}%
    \affiliation{Materials Sciences Division, Lawrence Berkeley National Laboratory, Berkeley, CA 94720}
\author{Dahvyd Wing}
    \affiliation{Department of Molecular Chemistry and Materials Science, Weizmann Institute of Science, Rehovoth 76100, Israel}
\author{Jonah B. Haber}
    \affiliation{Department of Physics, University of California, Berkeley, CA 94720}%
    \affiliation{Materials Sciences Division, Lawrence Berkeley National Laboratory, Berkeley, CA 94720}
\author{Mar\'ia Camarasa-G\'omez}
    \affiliation{Department of Molecular Chemistry and Materials Science, Weizmann Institute of Science, Rehovoth 76100, Israel}
\author{Francisca Sagredo}
    \affiliation{Department of Physics, University of California, Berkeley, CA 94720}%
    \affiliation{Materials Sciences Division, Lawrence Berkeley National Laboratory, Berkeley, CA 94720}
\author{Marina R. Filip}
    \affiliation{Department of Physics, University of Oxford, Oxford OX1 3PJ, United Kingdom}
\author{Jeffrey B. Neaton}
    \affiliation{Department of Physics, University of California, Berkeley, CA 94720}
    \affiliation{Materials Sciences Division, Lawrence Berkeley National Laboratory, Berkeley, CA 94720}
    \affiliation{Kavli Energy NanoSciences Institute at Berkeley, University of California, Berkeley, CA 94720}
\author{Leeor Kronik}
   \affiliation{Department of Molecular Chemistry and Materials Science, Weizmann Institute of Science, Rehovoth 76100, Israel}

\begin{abstract}
Using both time-dependent density functional theory (TDDFT) and the ``single-shot" $GW$ plus Bethe-Salpeter equation ($GW$-BSE) approach, we compute optical band gaps and optical absorption spectra from first principles for eight common binary and ternary closed-shell metal oxides (MgO, Al$_2$O$_3$, CaO, TiO$_2$, Cu$_2$O, ZnO, BaSnO$_3$, and BiVO$_4$), based on the non-empirical Wannier-localized optimally-tuned screened range-separated hybrid functional. Overall, we find excellent agreement between our TDDFT and $GW$-BSE results and experiment, with a mean absolute error less than 0.4 eV, including for Cu$_2$O and ZnO, traditionally considered to be challenging for both methods.
\end{abstract}

\maketitle

\section{Introduction}

The optical absorption spectrum is a solid-state property of critical importance in optoelectronic materials. A state-of-the-art \textit{ab-initio} methodology for predicting accurate optical spectra of solids is the $GW$ plus Bethe-Salpeter equation (BSE) approach, where $G$ is the single particle Green's function and $W$ is the dynamically screened Coulomb interaction \cite{albrecht1998excitonic, rohlfing2000electron, onida2002electronic, hedin1965new, hybertsenElectronCorrelationSemiconductors1986, onida2002electronic}. The accuracy of $GW$-BSE calculations comes at a large computational cost that scales roughly as $N^4$, where $N$ is the number of atoms in the system. Time-dependent density functional theory (TDDFT) \cite{runge1984density, ullrich2011time, burke2012perspective, maitra2016perspective, Byun_Ullrich_2020} can be a viable alternative due to its substantially reduced computational cost \cite{onida2002electronic}. However, it suffers from serious inaccuracies when applied to the solid state using standard exchange-correlation functionals \cite{gavrilenko1997optical,maitra2016perspective}.

Excited-state properties of solids from linear-response TDDFT are typically obtained by solving the Casida equation based on Kohn-Sham (KS) orbitals \cite{casida1995}. The adiabatic approximation is typically employed, by using the static KS approximation for the exchange-correlation potential, $V_{xc}$, to obtain the exchange-correlation kernel, $f_{xc}$, defined as the functional derivative of $V_{xc}$ with respect to the electron density. This kernel is a key quantity in the Casida equation and highly affects the accuracy of the resulting optical spectra. This is manifested in two major challenges in predicting optical spectra that are in good agreement with experiment and with $GW$-BSE calculations. First, TDDFT based on KS (semi-)local functionals inherits the underlying KS band gap, which is known to be severely underestimated \cite{onida2002electronic, kuemmel_kronik_2008}. The resulting optical spectra are then typically red-shifted with respect to experiment \cite{onida2002electronic, botti2004long, botti2005energy, botti2007time}. Second, the exchange-correlation kernel derived from (semi-)local functionals lacks the correct long wavelength limit, namely $f_{xc}(q\rightarrow0)\propto 1/q^2$ (where $q$ is a reciprocal space vector in the Brillouin zone), which is an essential property for an accurate description of excitonic effects \cite{ghosez1997long, reining2002excitonic, ullrich2011time, Byun_Ullrich_2020}. Using (semi-)local approximations for optical spectra calculations then results in incorrect line shapes \cite{ghosez1997long, onida2002electronic, botti2004long, botti2005energy, botti2007time, maitra2016perspective}.

Within KS TDDFT, several approaches for overcoming these two challenges have been proposed in recent years. In many cases, the two aforementioned challenges are treated separately. The band gap problem is often solved based on a fit to a target value, e.g. by using a scissors operator to correct the eigenvalues \cite{levineLinearOpticalResponse1989}. Subsequently, several ideas have been put forth for constructing a kernel that recovers the correct long wavelength limit (see Refs.~\cite{maitra2016perspective,Byun_Ullrich_2020} and references therein). While good results can be obtained using such methods, they can be computationally complex and usually at least one of the aforementioned challenges is solved empirically, limiting the predictive power of these methods. Therefore, a broader, non-empirical and simple formalism that can solve both challenges at the same time is desirable. We note a recent non-empirical approach, proposed by Cavo \textit{et al.}~\cite{cavo2020accurate}, based on the link between the exchange-correlation kernel and the derivative discontinuity. While their approach treats the band gap problem explicitly, excitonic effects are captured by using the polarization functional within the framework of time-dependent current DFT. 

An alternative approach, still entirely within TDDFT, is based on the use of hybrid functionals within generalized KS (GKS) theory \cite{seidl_levy_1996, tretiak2003resonant, baer_kronik_2018}. The inclusion of non-local effects in GKS, or more specifically the incorporation of exact exchange in hybrid functionals, has the potential to solve the two fundamental problems described above simultaneously. This is because the free parameters that control the amount of exact (Fock) exchange in a hybrid functional can be chosen such that the band gap description is improved and the correct long wavelength limit is accounted for. The latter is achieved by preserving a non-zero fraction of exact exchange in the long range such that the functional possesses the correct asymptotic behavior \cite{shimazaki_asai_2008, kronik2016excited, refaely-abramson_kronik_2013,refaely-abramson_kronik_2015, zheng_cororpceanu_2017, kronik_kuemmel_2018, kronik_kummel_2020} and the kernel behaves as $1/q^2$ in the long wavelength limit \cite{refaely-abramson_kronik_2015, kronik2016excited}. Clearly, a key issue is then how to determine the parameters of a hybrid functional.

Several non-empirical, hybrid-functional based methods for optical spectra calculations have been proposed in recent years. Yang \textit{et al.}~\cite{yang2015simple} proposed a screened exact-exchange (SXX) approach to replace the full dielectric function in the BSE kernel with a single screening parameter that can be calculated within the random phase approximation (RPA) \cite{hybertsenElectronCorrelationSemiconductors1986}. Sun \textit{et al.}~\cite{sun_ullrich_2020_prr, sun_ullrich_2020} then proposed constructing a hybrid kernel by combining SXX and (semi-)local exchange and correlation kernels. Tal \textit{et al.}~\cite{tal_pasquarello_2020} used dielectric-dependent hybrid functionals \cite{chen_pasquarello_2018} where the parameters are determined self-consistently based on fitting to a dielectric function calculated via the RPA. 

A promising hybrid functional in the context of optical spectra calculations is the screened range-separated hybrid (SRSH) functional \cite{refaely-abramson_kronik_2013, kronik2016excited}, as it has a potential that by construction behaves as $\frac{1}{\varepsilon_\infty r}$ for a large interelectronic distance $r$, where $\varepsilon_\infty$ is the high-frequency dielectric constant of the material. It has been demonstrated repeatedly that when the SRSH parameters are empirically fitted to reproduce the $GW$ or the experimental band gap, one can obtain highly accurate optical absorption spectra of solids \cite{refaely-abramson_kronik_2015, wing_kronik_2019, ramasubramaniam_2019, wing2020narrowgap, lewis2020tuned, camarasa2023transferable}. 

Recently, we removed the empiricism in SRSH fundamental band gap calculations in the solid state by choosing the parameters of SRSH based on a Wannier-localized, optimally-tuned SRSH (WOT-SRSH) functional \cite{wing_2021}. In this method, the range-separation parameter is selected to satisfy an \textit{ansatz} that generalizes the ionization potential theorem to the removal of an electron from a localized Wannier function \cite{ma_wang_2016}. This method has been shown to yield highly accurate quasiparticle (QP) band gaps for prototypical semiconductors and insulators \cite{wing_2021} and for halide perovskites \cite{ohad_2022_wotsrsh_haps}, that are in excellent agreement with experimental and $GW$ results. Furthermore, the merit of using an optimally-tuned eigensystem as a starting point to single-shot $G_0W_0$ calculations has been recently demonstrated by Gant \textit{et al.}~\cite{gant2022optimally}, who obtained highly accurate band gaps, band widths and $d$-band locations for a variety of semiconductors. In light of this success, and based on the accuracy of the prior empirical SRSH calculations discussed above, it is evident that WOT-SRSH holds a significant potential for accurate, non-empirical optical spectra predictions for solids.

An interesting application is the case of metal oxides (MOs), which are of much importance in various applications, including solar cells, catalysts, batteries, and sensors \cite{fierro2005metal, yu2016metal}. From a computational perspective, the accurate prediction of the electronic structure and optical properties of MOs is challenging, and has been widely studied (see, e.g., Refs.~\cite{das2019band, gerosa_2017, chevrier2010hybrid, li_2013, liu_2019, mandal2019systematic, massidda1997quasiparticle, ostrom2022designing, samsonidze2014insights, weng_wang_2020, coulter2013limitations, shih2010quasiparticle,brunevalExchangeCorrelationEffects2006,kangQuasiparticleOpticalProperties2010,shishkinSelfconsistentGWCalculations2007,vanschilfgaardeQuasiparticleSelfConsistentGW2006b,rangelReproducibilityCalculationsSolids2020,schleifeBandstructureOpticaltransitionParameters2009,golzeGWCompendiumPractical2019, wu2018first, wu2020theoretical, park2022applicability}). The major challenges with MOs are attributed to the localized nature of the electrons in the $d$-orbitals. The well known self-interaction error \cite{perdew1981self} and delocalization error \cite{mori-sanchez_yang_2008} associated with (semi-)local functionals are more significant for MOs, leading to DFT calculations that predict unphysical metallic behavior for some systems \cite{li_2013, liu_2019, ostrom2022designing}. Promisingly, the fraction of exact exchange employed in hybrid functionals directly reduces these errors, and has been shown to offer a better description of their electronic structure \cite{das2019band, gerosa_2017, chevrier2010hybrid, li_2013, liu_2019, ostrom2022designing}.

In this article, we assess the accuracy of the WOT-SRSH method in predicting the optical absorption spectra of a set of MO crystals. We perform both TDDFT and $GW$-BSE study of eight common binary and ternary closed-shell MOs, using the WOT-SRSH formalism as a non-empirical foundation for both sets of calculations. We find that both methods agree well with one another and predict optical absorption spectra in good agreement with experiment. Our calculations demonstrate the applicability of WOT-SRSH to complex systems, either in itself, using TDDFT, or as a starting point for $GW$-BSE calculations. 

\section{Methods}

\subsection{Materials}
We focus on eight abundant closed-shell metal oxides for which both computational and experimental data is available in the literature: MgO, Al$_2$O$_3$, CaO, TiO$_2$, Cu$_2$O, ZnO, BaSnO$_3$ \cite{aggoune2022consistent}, and BiVO$_4$ \cite{wiktor2017BiVO4}. We use experimental crystal structures at room temperature, the details of which are given in Table \ref{tab:structural_details_TMOs}.

\begin{table*}[hbtp!]
\centering
\caption{\label{tab:structural_details_TMOs} Structural details of the crystals used in the calculations.}
\setlength{\tabcolsep}{0.06 in}

\begin{tabular}{cccc}
  & Crystal structure & Space group & Unit cell parameters (\AA) \\
\hline
MgO$^\text{a}$            & Rock salt  & Fm-3m   & a=b=c=4.22    \\
Al$_2$O$_3$$^\text{b}$   & Corundum   & R-3cH   & a=b=4.76, c=13.00 \\
CaO$^\text{a}$            & Rock salt  & Fm-3m   & a=b=c=4.81   \\
TiO$_2$$^\text{c}$        & Rutile     & P42/mnm & a=b=4.59, c=2.96  \\
Cu$_2$O$^\text{d}$        & Cubic      & Pn-3mZ  & a=b=c=4.27  \\
ZnO$^\text{e}$            & Wurtzite   & P63mc   & a=b=3.25, c=5.21  \\
BaSnO$_3$$^\text{f}$      & Perovskite & Pm-3m   & a=b=c=4.11   \\
BiVO$_4$$^\text{g}$       & Monoclinic & C2/c    & \makecell{a=b=6.88, c=5.09 \\ $\alpha=68.45^{\circ}$, $\beta=111.55^{\circ}$,$\gamma=63.56^{\circ}$} \\
\hline\hline
\end{tabular}
\caption*{
\textsuperscript{a}Ref.~\cite{Madelung_2004}.
\textsuperscript{b}Ref.~\cite{kondo2008structural}.
\textsuperscript{c}Ref.~\cite{sugiyama1991crystal}.
\textsuperscript{d}Ref.~\cite{foo2006synthesis}.
\textsuperscript{e}Ref.~\cite{garcia1993microstructural}.
\textsuperscript{f}Ref.~\cite{mizoguchi2004strong}.
\textsuperscript{g}Ref.~\cite{sleight1979crystal}.
}
\end{table*}

\subsection{DFT}

\subsubsection{WOT-SRSH}

The SRSH functional \cite{refaely-abramson_kronik_2013} splits the Coulomb operator via the identity
\begin{equation}
\label{eq:srsh}
\begin{split}
     \frac{1}{\abs{\bm{r}-\bm{r}'}} =\underbrace{ \alpha\frac{\textrm{erfc}(\gamma \abs{\bm{r}-\bm{r}'})}{\abs{\bm{r}-\bm{r}'}} }_{\textrm{xx, SR}}+ \underbrace{(1-\alpha )\frac{\textrm{erfc}(\gamma \abs{\bm{r}-\bm{r}'})}{\abs{\bm{r}-\bm{r}'}}}_{\textrm{KSx, SR}}\\ +\underbrace{\varepsilon_\infty^{-1}\frac{ \erf(\gamma \abs{\bm{r}-\bm{r}'})}{ \abs{\bm{r}-\bm{r}'}}}_{\textrm{xx, LR}}  +\underbrace{\left(1-\varepsilon_\infty^{-1}\right)\frac{ \erf(\gamma \abs{\bm{r}-\bm{r}'})}{\abs{\bm{r}-\bm{r}'}}}_{\textrm{KSx, LR}}\,,
\end{split}
\end{equation}
where the exchange expressions that result from the four terms are evaluated with exact exchange (xx) integrals for the first and third terms and with semi-local Kohn-Sham exchange (KSx) integrals (in this work, the Perdew–Burke–Ernzerhof, PBE, functional \cite{perdew_burke_ernzerhof_1996}) for the second and fourth terms. In this construct, the fraction of exact exchange in the short-range (SR) is $\alpha$ and the fraction of exact exchange in the long-range (LR) is the inverse of the dielectric constant, $\varepsilon_\infty^{-1}$. In this manner, a different balance between exchange and correlation is obtained in the SR and LR, the transition between which is controlled by the range-separation parameter, $\gamma$. The default choice for $\alpha$ is 0.25, adopted from the hybrid Perdew-Burke-Ernzerhof (PBE0) \cite{perdew_burke_hybrid_1996, adamo_barone_1999} and the Heyd–Scuseria–Ernzerhof (HSE06) \cite{heyd_ernzerhof_2006} functionals, although it may vary based on considerations discussed below. The choice of $\varepsilon_\infty^{-1}$ as the fraction of exact exchange in the LR attains the asymptotically correct potential of the SRSH functional \cite{shimazaki_asai_2008, kronik2016excited, refaely-abramson_kronik_2013,refaely-abramson_kronik_2015, zheng_cororpceanu_2017, kronik_kuemmel_2018, kronik_kummel_2020}.

The procedure of selecting $\gamma$ is often carried out in a non-empirical fashion by enforcing an exact physical condition, the ionization potential theorem (IPT) \cite{perdew_balduz_1982, Almbladh_von_Barth_1985, perdew_levy_1997, levy_sahni_1984}. This procedure, known as optimal tuning, has shown great success in the prediction of fundamental gaps of molecules \cite{stein_baer_2010, kronik_stein_refaely-abramson_baer_2012, refaely_kronik_2011, autschbach_srebro_2014, phillips_dunietz_2014, foster_allendorf_2014, korzdorfer_bredas_2014, faber2014excited}. In the bulk limit, however, optimal tuning fails because the IPT is trivially satisfied for every parametrization of SRSH (or indeed any functional) \cite{mori-sanchez_yang_2008, kraisler_kronik_2014, vlcek_eisenberg_steinle-neumann_baer_2015, gorling_2015}, such that the uniqueness of the optimally tuned $\gamma$ that is achieved in molecules is lost.

The reason for the failure of optimal tuning in the bulk limit is the natural delocalization of the electronic orbitals. Recently, a number of studies have exploited different localization schemes for electronic structure predictions \cite{anisimov_kozhevnikov_2005,ma_wang_2016, weng_wang_2017, li_yang_2017, miceli_pasquarello_2018, nguyen_marzari_2018, bischoff_pasquarello_2019, bischoff_pasquarello_2019_perovskites, elliott2019koopmans, weng_wang_2020, su2020preserving, bischoff2021band, colonna2022koopmans, mahler2022localized, yang2022one, degennaro2022bloch, linscott2023koopmans}. Similarly, the WOT-SRSH approach adopts a criterion that generalizes the IPT to the removal of charge from a maximally localized Wannier function \cite{wing_2021}. This \textit{ansatz}, inspired by Ma and Wang \cite{ma_wang_2016}, is given by
\begin{equation}
\label{eq:deltaI}
  \Delta I^\gamma=  E_{\textrm{constr}}^\gamma[\phi](N-1)- E^\gamma(N)+ \bra{\phi} \hat{H}_{\textrm{SRSH}}^\gamma \ket{\phi} = 0,
\end{equation}
where $E^\gamma(N)$ is the total energy of the system with $N$ electrons and $E_{\textrm{constr}}^\gamma[\phi](N-1)$ is the total energy of a system with one electron removed from a Wannier function $\phi$, including an image charge correction (see Supplementary Material, SM \cite{SM}, for further details). $\bra{\phi} \hat{H}_{\textrm{SRSH}}^\gamma \ket{\phi}$ is the expectation value for the energy of the Wannier function with respect to the SRSH Hamiltonian of an $N$ electron system. The energy of the charged system is calculated under a constraint that allows one to control the occupation of the Wannier function via the Lagrange multiplier $\lambda$ \cite{wing_2021}. The constraint is imposed using the equation
\begin{equation}
\label{eq:hamiltonian}
\hat{H}_\textrm{SRSH} \ket{\psi_i}+\lambda \ket{\phi} \braket{\phi|\psi_i} = \epsilon_i \ket{\psi_i},
\end{equation}
where $\{\psi_i\}$ and $\{\epsilon_i\}$ are the GKS eigenfunctions and eigenvalues, respectively, of the constrained $(N-1)$-electron system.

Here, the WOT-SRSH procedure is carried out in an iterative manner, based on the four-step scheme suggested by Wing \textit{et al.}~\cite{wing_2021}. In step 1, the orientationally-averaged ion-clamped dielectric constant, $\varepsilon_\infty$, is calculated in the primitive unit cell. In step 2, we compose maximally localized Wannier functions from the topmost valence bands in a supercell. We then select the Wannier function with highest energy in the manifold and use it in step 3, where we enforce the \textit{ansatz} given in Eq.~\eqref{eq:deltaI} by selecting the range-separation parameter $\gamma$ so that $\Delta I^\gamma=0$ for the supercell. Finally, in step 4 we calculate properties of interest with the selected $\gamma$. This scheme is repeated iteratively: $\varepsilon_\infty$ in step 1 is initially calculated using HSE06, and after performing steps 2-4, $\varepsilon_\infty$ is calculated again using the optimally tuned parameters found in step 3.

In the scheme described above, $\alpha$ is kept fixed. As can be seen in Table \ref{tab:parameters}, we do not always use the default choice of 0.25. There are two scenarios where $\alpha$ has to be changed, already encountered in previous WOT-SRSH studies \cite{wing_2021, ohad_2022_wotsrsh_haps}. The first scenario is that the fraction of LR exact exchange, $\varepsilon_\infty^{-1}$, is close to 0.25, resulting in the insensitivity of $\Delta I$ to variations in $\gamma$. The second scenario is that there is no $\gamma$ for which the generalized IPT is satisfied. In this work these two issues are solved by slightly increasing $\alpha$ from the default value in three of the materials. For further discussion see the SM \cite{SM}.

We emphasize that while the parameters $\alpha$, $\varepsilon_\infty$, and $\gamma$ are system-dependent, they are non-empirical throughout. The self-consistent WOT-SRSH parameters used in this work are reported in Table \ref{tab:parameters}. They have been obtained for QP band gap convergence to within 50 meV, a condition achieved with up to three iterations. See SM \cite{SM} for additional computational details.

\begin{table}[hbtp!]
\centering
\caption{\label{tab:parameters} Self-consistent WOT-SRSH parameters obtained in this work. $\varepsilon_\infty$ is orientationally-averaged.}

\begin{tabular}{c c c c}
& $\alpha$ & $\varepsilon_\infty$ & $\gamma$ (\AA$^{-1}$)  \\
\hline
MgO & 0.25 & 2.85 & 2.40 \\
Al$_2$O$_3$ & 0.40 & 2.94 & 1.40\\
CaO & 0.25 & 3.25 & 1.70 \\
TiO$_2$ & 0.25 & 6.25 & 0.85\\
Cu$_2$O & 0.25 & 6.51 & 0.95 \\
ZnO & 0.30 & 3.57 & 1.30\\
BaSnO$_3$ & 0.30 & 3.92 & 1.40 \\
BiVO$_4$ & 0.25 & 5.92 & 2.00 \\
\hline\hline
\end{tabular}

\end{table}

\subsubsection{TDDFT}

Optical spectra are computed using linear-response TDDFT by solving the Casida equation within the Tamm-Dancoff approximation \cite{hirata1999time, ullrich2011time}. The Casida equation then has the following form \cite{casida1995, tretiak2003resonant,refaely-abramson_kronik_2015,sun2021pros}
\begin{equation}
\label{eq:casida_main}
\begin{aligned}
&\Omega^S A_{vc\bm{k}}^S=\left(\epsilon_{c\bm{k}}^{\textrm{GKS}}-\epsilon_{v\bm{k}}^{\textrm{GKS}}\right)A_{vc\bm{k}}^S\\
&+\sum_{v'c'\bm{k}'}\Bigr[\braket{v\bm{k},c\bm{k}|K_{Hxc}\left(\alpha,\varepsilon_\infty,\gamma\right)|v'\bm{k}',c'\bm{k}'} \\ & -\braket{v\bm{k},v'\bm{k}'|K_{\textrm{sxx}}\left(\alpha,\varepsilon_\infty,\gamma\right)|c\bm{k},c'\bm{k}'} \Bigr]A_{v'c'\bm{k}'}^S,
\end{aligned}
\end{equation}
where $v$ and $c$ denote valence and conduction band states, respectively, $\epsilon^{\textrm{GKS}}$ are the GKS eigenvalues, $\Omega^S$ are the excitation energies, and $A_{vc\bm{k}}^S$ are the expansion coefficients of the exciton wavefunction $\Psi_{S}$ in terms of valence and conduction band state pairs at the same $\bm{k}$-point, namely: 
\begin{equation}
\label{eq:exciton_wavefunction}
    \Psi_{S}(\bm{r}_e,\bm{r}_h)=\sum_{vc\bm{k}}A_{vc\bm{k}}^S\psi_{c\bm{k}}(\bm{r}_e)\psi_{v\bm{k}}^{*}(\bm{r}_h).
\end{equation}

As expressed in Eq.~\eqref{eq:casida_main}, the TDDFT kernel is composed of two parts: the Hartree-exchange-correlation kernel, $K_{Hxc}$, and the screened exact exchange kernel $K_{\textrm{sxx}}$, defined as
\begin{equation}
\label{eq:hxc_kernel}
K_{Hxc}\left(\alpha,\varepsilon_\infty,\gamma\right)=\frac{1}{\abs{\bm{r}-\bm{r}'}}+(1-\alpha)f_{xc}^{\textrm{SR},\gamma}+(1-\varepsilon_\infty^{-1})f_{xc}^{\textrm{LR},\gamma}
\end{equation}
and
\begin{equation}
\label{xx_kernel}
K_{\textrm{sxx}}\left(\alpha,\varepsilon_\infty,\gamma\right)=\alpha\frac{\textrm{erfc}(\gamma \abs{\bm{r}-\bm{r}'})}{\abs{\bm{r}-\bm{r}'}}+\varepsilon_\infty^{-1}\frac{\erf(\gamma \abs{\bm{r}-\bm{r}'})}{\abs{\bm{r}-\bm{r}'}},
\end{equation}
where $f_{xc}^{\textrm{SR},\gamma}$ and $f_{xc}^{\textrm{LR},\gamma}$ are the short- and long-range contributions, respectively, of the exchange-correlation kernel of the (semi-)local Kohn-Sham approximation.

We note that the bracket notation in Eq.~\eqref{eq:casida_main} represents real space integrals of the form
\begin{flalign}
\label{eq:braket_notation}
&\braket{b_1\bm{k}_1,b_2\bm{k}_2|K|b_3\bm{k}_3,b_4\bm{k}_4}= && \nonumber \\
&\quad\quad\int d^3r d^3r' \psi_{b_1\bm{k}_1}^{*}(\bm{r}) \psi_{b_2\bm{k}_2}(\bm{r}) K(\bm{r},\bm{r}') \psi_{b_3\bm{k}_3}(\bm{r}') \psi_{b_4\bm{k}_4}^{*}(\bm{r}'), &&
\end{flalign}
where $b_i$ can be a valence or conduction band index and it is understood that the wavefunctions on the LHS always have position $\bm{r}$ and the wavefunctions on the RHS position $\bm{r}'$. 

Once the linear-response equation is solved, optical absorption spectra (i.e. the imaginary part of the dielectric function, $\varepsilon_2$) can be obtained by
\begin{equation}
    \label{eq:bse_eps_2}
    \varepsilon_2(\omega)=\frac{16\pi^2}{\omega^2} \sum_S\left|\hat{\bm{p}}\cdot\braket{0|\mathbf{v}|S}\right|^2 \delta(\omega-\Omega^S),
\end{equation}
where
\begin{equation}
    \label{eq:bse_inner_prod}    \braket{0|\mathbf{v}|S}=\sum_{vc\bm{k}}A_{vc\bm{k}}^S\braket{v\bm{k}|\mathbf{v}|c\bm{k}},
\end{equation}
$S$ is a neutral excitation, $\mathbf{v}$ is the single-particle velocity operator and $\hat{\bm{p}}$ is the direction of the polarization of light.

TDDFT calculations in this work are both performed at the PBE level (denoted TDPBE), the equation for which is obtained by using the PBE eigenvalues and setting $\alpha=\varepsilon_\infty^{-1}=0$ in Eq.~\eqref{eq:casida_main}, and at the WOT-SRSH level (denoted TDWOT-SRSH), the equation for which is obtained by using the WOT-SRSH eigenvalues and the optimally tuned $\alpha$, $\varepsilon_\infty$, and $\gamma$ parameters in Eq.~\eqref{eq:casida_main}. See SM \cite{SM} for additional computational details.

\subsection{Many-Body Perturbation Theory}
\subsubsection{\textit{GW} Approximation}
Within the framework of many-body perturbation theory (MBPT), the electron self-energy $\Sigma$ can be approximated to first order as the convolution of $G$ and $W$, written symbolically as $\Sigma=iGW$ \cite{hedin1965new}. $\Sigma$ is usually constructed from an underlying DFT eigensystem, $\left\{\psi_{n\bm{k}},\epsilon^{\text{DFT}}_{n\bm{k}}\right\}$, at varying levels of self-consistency, with the choice of self-consistency usually having significant implications for the accuracy and variability of results \cite{hybertsenElectronCorrelationSemiconductors1986,luoQuasiparticleBandStructure2002a,faleevAllElectronSelfConsistentGW2004,vanschilfgaardeQuasiparticleSelfConsistentGW2006b,kotaniQuasiparticleSelfconsistentGW2007,shishkinSelfconsistentGWCalculations2007,rangelReproducibilityCalculationsSolids2020, golzeGWCompendiumPractical2019}. The simplest approach, and the one employed in this work, is the ``single-shot" method (denoted $G_0W_0$), where the QP energies are calculated as a first-order perturbative correction to a DFT eigensystem \cite{hybertsenFirstPrinciplesTheoryQuasiparticles1985,hybertsenElectronCorrelationSemiconductors1986, onida2002electronic, golzeGWCompendiumPractical2019}.

Specifically, the single-particle Green's function, $G_0$, is constructed directly from the DFT eigensystem, and the dynamically screened Coulomb interaction $W_0$ is given by
\begin{equation}
    W_0(\bm{r},\bm{r}';\omega)=\int d\bm{r}''\varepsilon^{-1}(\bm{r},\bm{r}'';\omega)\frac{1}{\left|\bm{r}'-\bm{r}''\right|},
\end{equation}
where the dielectric function is computed within the RPA based on the polarizability, $\chi_0(\bm{r},\bm{r}',\omega)$, given by the Adler-Wiser expression \cite{adlerQuantumTheoryDielectric1962,wiserDielectricConstantLocal1963}.

In practice, $\chi_0(\bm{r},\bm{r}',\omega)$ can be evaluated explicitly, via a full-frequency (FF) calculation, or approximately modeled using a plasmon-pole model (PPM). In the FF approach, the convolution of $G_0$ with $W_0$ is handled via contour deformation \cite{godbySelfenergyOperatorsExchangecorrelation1988,govoniLargeScaleGW2015a} using explicitly sampled frequencies along the imaginary axis. To mitigate the substantial cost of computing the FF dielectric function, we employ the static subspace approximation \cite{nguyenImprovingAccuracyEfficiency2012a,phamGWCalculationsUsing2013,wilsonEfficientIterativeMethod2008,wilsonIterativeCalculationsDielectric2009,delbenStaticSubspaceApproximation2019b}, where $\chi_0(\bm{r},\bm{r}',\omega)$ is efficiently but approximately represented using the leading eigenvectors of a low-rank decomposition of the static polarizability $\chi_0(\bm{r},\bm{r}',0)$. In the PPM approach, $\chi_0$ is evaluated statically ($\omega=0$), and extended to finite frequencies via a simplified model \cite{hybertsenElectronCorrelationSemiconductors1986,godbyMetalinsulatorTransitionKohnSham1989a,oschliesGWSelfenergyCalculations1995}. Here we employ FF calculations for all materials except Cu$_2$O, where use the PPM. See SM \cite{SM} for further details.

With the above quantities, the $G_0W_0$ self-energy can be used to correct the DFT eigenvalues perturbatively via
\begin{equation}\label{eq:gw_corection}
\epsilon_{n\bm{k}}^\text{QP} = \epsilon^\text{DFT}_{n\bm{k}} + \braket{n\bm{k}|\Sigma(\epsilon_{n\bm{k}}^\text{QP}) - V_{xc}|n\bm{k}}. 
\end{equation}
Due to the fact that $\epsilon_{n\bm{k}}^\text{QP}$ in Eq.~\eqref{eq:gw_corection} depends on itself, evaluating this expression can be non-trivial. For FF calculations, $\braket{n\bm{k}|\Sigma(\omega)|n\bm{k}}$ is accurately known for a range of frequencies, allowing for a solution of Eq.~\eqref{eq:gw_corection}. However, if a plasmon-pole model for the frequency dependence of the screening is used, we employ the common practice of expanding Eq.~\eqref{eq:gw_corection} to first order about $\epsilon^\text{DFT}_{n\bm{k}}$ to evaluate it \cite{giantomassiElectronicPropertiesInterfaces2011,liuCubicScalingGW2016,wilhelmGWGaussianPlane2016}.

The single-shot approach has the advantage of being the least computationally demanding $GW$ approach, and, typically, the QP band structures computed within $G_0W_0$ are in substantially better agreement with experiment than those computed from their underlying DFT functionals \cite{aulburQuasiparticleCalculationsSolids2000a,louie2005quasiparticle, fuchsQuasiparticleBandStructure2007b,chenAccurateBandGaps2015a, jiangGWLinearizedAugmented2016,grumetQuasiparticleApproximationFully2018,golzeGWCompendiumPractical2019}.
However, the single-shot approach also suffers from a  sensitivity to the starting point, i.e. the (G)KS eigensystem used to construct $\Sigma$. Hence, the question of how to choose an appropriate DFT starting point for $G_0W_0$ calculations has been actively debated \cite{fuchsQuasiparticleBandStructure2007b,rinkeCombiningGWcalculationsExactexchangeDensityfunctional2005b,brunevalBenchmarkingStartingPoints2013a,vansettenAutomationMethodologiesLargescale2017,golzeGWCompendiumPractical2019,leppertPredictiveBandGaps2019a,maromBenchmarkGWMethods2012a,sharifzadehManybodyPerturbationTheory2018}. In this work, we focus on the WOT-SRSH eigensystem as a starting point for $G_0W_0$ (denoted $G_0W_0$@WOT-SRSH), as done in Ref.~\cite{gant2022optimally}, where it was demonstrated to be highly accurate over a broad range of systems. For the sake of comparison, we also examine results obtained from using the PBE functional \cite{perdew_burke_ernzerhof_1996} as a starting point (denoted $G_0W_0$@PBE). Additional computational details, including convergence details, can be found in the SM \cite{SM}.

\subsubsection{\textit{Ab} initio BSE Method}
The ab initio Bethe-Salpeter equation, within the the Tamm-Dancoff approximation \cite{onida2002electronic,rohlfing2000electron,rohlfingElectronHoleExcitationsSemiconductors1998b}, has a standard form that is very similar to the Casida equation. It can be constructed from Eq.~\eqref{eq:casida_main} by substituting $\epsilon^{\textrm{GKS}}$ with $\epsilon^{\textrm{QP}}$, $K_{Hxc}$ with the bare exchange interaction kernel $K_x=\frac{1}{\abs{\bm{r}-\bm{r}'}}$, and $K_{\textrm{sxx}}$ with the static screened direct interaction kernel $K_d=W_0(\bm{r},\bm{r}';\omega=0)$ \cite{deslippeBerkeleyGWMassivelyParallel2012b,rohlfing2000electron}. In practice, when constructing $K_x$ and $K_d$ we interpolate from coarse $\Gamma$-centered $\bm{k}$-grids to fine shifted $\bm{k}$-grids, as specified in the SM \cite{SM}. After solving the BSE, the exciton wavefunction and the imaginary part of the dielectric function are obtained from Eqs.~\eqref{eq:exciton_wavefunction} and \eqref{eq:bse_eps_2}, respectively. Additional computational details can be found in the SM \cite{SM}.

\begin{figure*}[htp!]
\begin{centering}
\includegraphics[width=\textwidth]{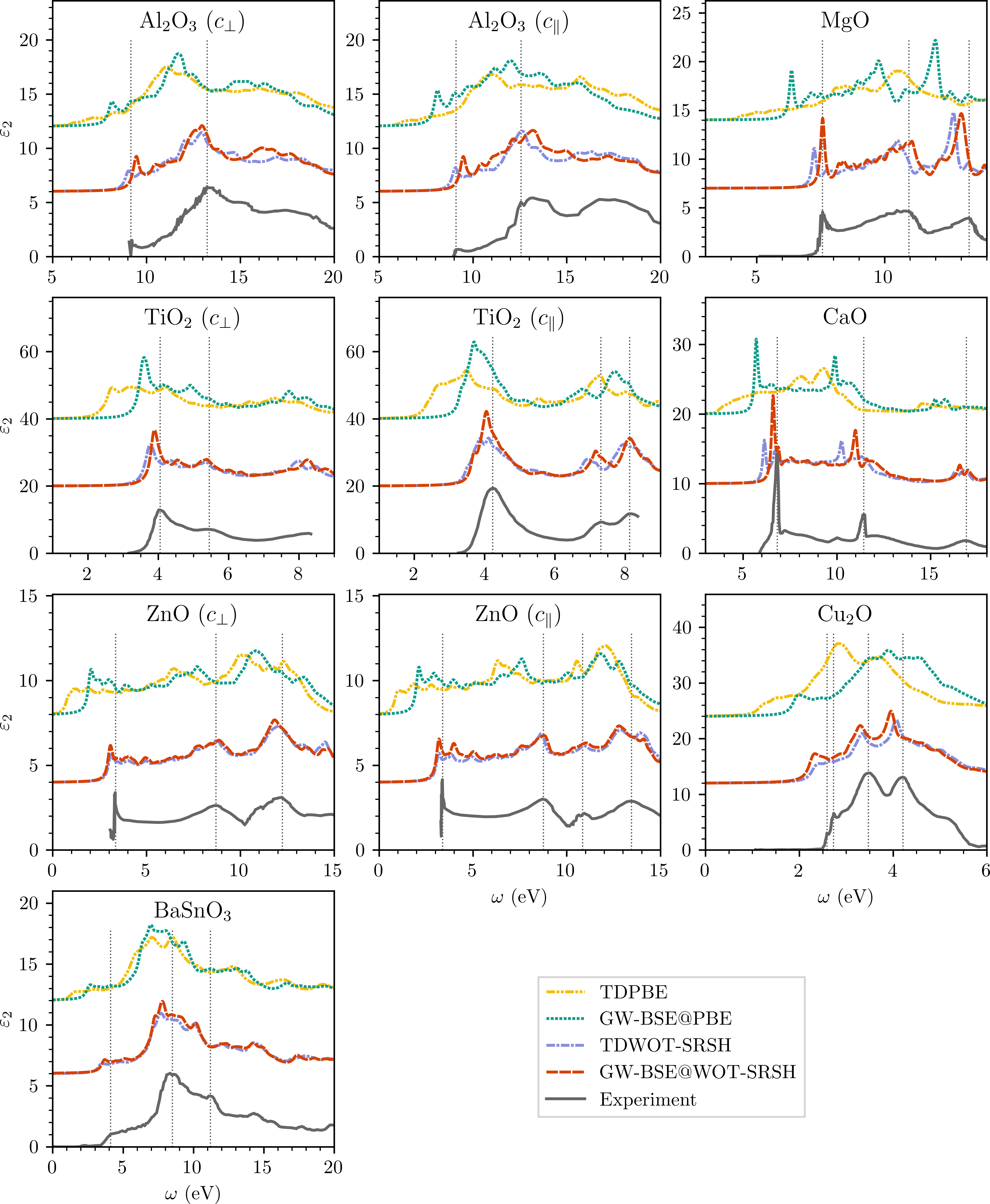}
\caption{\label{fig:spectra} Imaginary part of the dielectric function, computed with TDPBE (yellow dot-dashed line), $G_0W_0$-BSE@PBE (green dotted line), TDWOT-SRSH (purple dot-dashed line), and $G_0W_0$-BSE@WOT-SRSH (red dashed line), compared to experiment (gray solid line). Vertical dotted lines indicate the main spectral features in experiment. The anisotropy in Al$_2$O$_3$, TiO$_2$, and ZnO is accounted for by considering polarization perpendicular to the optic axis (ordinary component, $c_\perp$) and parallel to the optic axis (extraordinary component, $c_\parallel$) explicitly. Computed spectra are rigidly shifted in the energy axis by the vibrational renormalization reported in Table \ref{tab:thermal_renorm}. They are also shifted in the vertical axis such that the zero absorption tail exactly begins where indicated by an axis tick. Experimental data are taken from the following sources: Al$_2$O$_3$: Ref.~\cite{tomiki1993anisotropic}; MgO: Ref.~\cite{bortz1990temperature}; TiO$_2$: Ref.~\cite{tiwald2000measurement}; CaO: Ref.~\cite{whited_walker_1969_cao}; ZnO: Ref.~\cite{gori2010optical}; Cu$_2$O: Ref.~\cite{haidu2011dielectric}; BaSnO$_3$: Ref.~\cite{aggoune2022consistent}.}
\end{centering}
\end{figure*}

\subsection{Vibrational Renormalization of Band Gaps and Optical Spectra}
To make a meaningful comparison with experimental band gaps and optical spectra, two effects should be taken into account: zero-point renormalization (ZPR) energy and finite temperature fluctuations (FTF). Both are inherently excluded in calculations that use the fixed ion approximation, but can have a significant effect on electronic properties \cite{wiktor2017perovskites, wiktor2017BiVO4, wing_2021, aggoune2022consistent, wang2022accurate, giustino2017electron, Karsai_kresse_2018, miglio2020predominance, antonius2022theory, engel2022zero, cardona_thewalt_2005, zacharias&giustino2016, zacharias&giustino2020, chen_pasquarello_2018, wu2018first, wu2020theoretical, park2022applicability}. These effects can be understood from methods that go beyond static DFT, such as molecular dynamics \cite{wiktor2017perovskites, wiktor2017BiVO4} and electron-phonon self-energy approaches \cite{giustino2017electron, Karsai_kresse_2018, miglio2020predominance, antonius2022theory, engel2022zero, wu2018first, wu2020theoretical, park2022applicability}.

Accurate state-dependent calculations of ZPR and FTF effects are beyond the scope of this work. To account for them, we exploit values from the literature obtained based on different methods, the details of which are given in Table \ref{tab:thermal_renorm}. These renormalization values are used as rigid shifts for the computed optical band gaps and optical absorption spectra.

\begin{table}[hbtp!]
\centering
\caption{\label{tab:thermal_renorm} vibrational renormalization values, taken from prior literature, used in this work as a rigid shift for the computed band gaps and optical absorption spectra. All values include both the ZPR and FTF effects, except for MgO and CaO, where the values include the ZPR effect alone.}

\begin{tabular}{c c}
& \begin{tabular}[c]{@{}c@{}}Thermal renorm.\\ $\textrm{[meV]}$ \end{tabular} \\
\hline
MgO & -533$^\text{a}$ \\
Al$_2$O$_3$ & -310$^\text{b}$ \\
CaO & -357$^\text{a}$ \\
TiO$_2$ & -290$^\text{c}$ \\
Cu$_2$O & -210$^\text{b}$ \\
ZnO & -190$^\text{b}$ \\
BaSnO$_3$ & -367$^\text{d}$ \\
BiVO$_4$ & -920$^\text{e}$ \\
\hline\hline
\end{tabular}
\caption*{
\textsuperscript{a}Ref.~\cite{engel2022zero}, from non-adiabatic Allen-Heine-Cardona theory. 
\textsuperscript{b}Ref.~\cite{park2022applicability}, from non-adiabatic Allen-Heine-Cardona theory. The FTF correction is extracted graphically at 300K.
\textsuperscript{c}Ref.~\cite{wu2018first}, from non-adiabatic Allen-Heine-Cardona theory. The FTF correction is extracted graphically at 300K.
\textsuperscript{d}Ref.~\cite{aggoune2022consistent}, from temperature-dependent optical absorption onset measurements.
\textsuperscript{e}Ref.~\cite{wiktor2017BiVO4}, from path-integral molecular dynamics at the PBE0 level, including nuclear quantum effects.
}
\end{table}

All values in Table \ref{tab:thermal_renorm} represent the renormalization of the QP band gap due to electron-phonon interactions, except for the case of BaSnO$_3$ where the value corresponds to renormalization of the optical band gap due to exciton-phonon interactions. By applying the same rigid shift to all features in the optical spectra (including the optical band gap itself), we implicitly assume the size of the renormalization \cite{filip2021phonon} of exciton binding energies are negligible relative to the energy scales of interest in this work. To demonstrate the validity of this assumption, we calculate phonon screening corrections to the binding energy of the lowest-lying exciton according to the expression derived in Ref.~\cite{filip2021phonon} and found that they are smaller than 0.1 eV. We note, however, that these corrections serve as an approximate lower bound to the exciton binding energy renormalization, because they are based on a model expression, applicable to $1s$ excitons at $0$ K. Thus, the validity of our estimates may be more questionable for materials that exhibit significant thermal fluctuations. For more details see the SM \cite{SM}.

\section{Results and Discussion}

Fig.~\ref{fig:spectra} shows the optical absorption spectra obtained from TDWOT-SRSH, $G_0W_0$-BSE@WOT-SRSH, and experiment for all materials studied in this work, except BiVO$_4$ which is discussed separately below. For reference, Fig.~\ref{fig:spectra} also shows spectra from TDPBE and $G_0W_0$-BSE@PBE. As expected, the PBE-based results are unsatisfactory. TDPBE significantly underestimates the reported measured absorption onset and the line shapes also deviate significantly from experiment. The $G_0W_0$-BSE@PBE line shapes are more accurate, owing to the correct description of electron-hole interactions in BSE, but suffer from a red-shifted absorption onset relative to experiment, due to the PBE starting point. Most notably, TDWOT-SRSH considerably outperforms $G_0W_0$-BSE@PBE.

It is readily apparent that both TDWOT-SRSH and $G_0W_0$-BSE@WOT-SRSH predict peak positions and line shapes in close agreement with each other and with the experimental data. The agreement is consistently good both for the absorption onsets and for higher energy spectral features. Notably, excitonic peak positions are well captured in both methods. In most cases the BSE excitonic peak position is slightly blue-shifted compared to the TDDFT one, most notably for Al$_2$O$_3$, MgO, and CaO, where the shift is $\sim$0.3-0.4 eV. This shift can be explained primarily at the electronic level, where $G_0W_0$ corrections tend to blue-shift the lowest direct gaps, as seen in the SM \cite{SM} and in prior work \cite{gant2022optimally}. This blue-shift is largely caused by the under-screening of the Coulomb interaction in $W_0$, brought about by the use of the RPA in conjunction with an accurate hybrid functional \cite{leppertPredictiveBandGaps2019a, blase2011first}. This can be seen when comparing the values of $\varepsilon_\infty$ used in WOT-SRSH and the high-frequency RPA dielectric constant (obtained from the same eigensystem) reported in the SM \cite{SM}. Relatedly, the under-screening present in $W_0$ also manifests in an about $10\%$ increase, on average, of the computed $G_0W_0$-BSE@WOT-SRSH exciton binding energy. This competing effect red-shifts the resulting spectra, but by much less than the blue-shift at the electronic level.

It can be seen that, in the three cases where there are larger deviations between the WOT-SRSH-based methods, namely Al$_2$O$_3$, MgO and CaO, the BSE spectra predict peak positions that are in better overall agreement with experiment, suggesting possible improved predictive accuracy associated with $G_0W_0$-BSE@WOT-SRSH. However, this improved accuracy can in part be attributed to a cancellation of errors resulting from underscreening, as discussed above.

A notable success of both methods is their accuracy for ZnO, a system known to have significant convergence issues in MBPT that resulted in a range of different reported band gap values \cite{usuda2002all,Rinke_2005,vanschilfgaardeAdequacyApproximationsMathbitGW2006,shishkinSelfconsistentGWCalculations2007,fuchsQuasiparticleBandStructure2007b,shih2010quasiparticle,stankovski0WBandGap2011,friedrich2012hybrid, samsonidze2014insights, rangelReproducibilityCalculationsSolids2020}. Here, using both WOT-SRSH and $G_0W_0$@WOT-SRSH, we obtain the optical absorption spectra for ZnO in excellent agreement with experiment (after approximately accounting for vibrational effects) and between the two methods via a straightforward application of the WOT-SRSH functional.

Another general trend we observe is that the oscillator strength of the first excitonic peak is reduced in TDDFT, compared to BSE, while other features at higher energies are in better agreement. This reflects an underestimation of electron-hole interaction and more delocalized exciton in TDDFT, in line with previous comparisons between the two methods \cite{wing_kronik_2019}.

\begin{figure}[h!]
\begin{centering}
\includegraphics[width=0.95\linewidth]{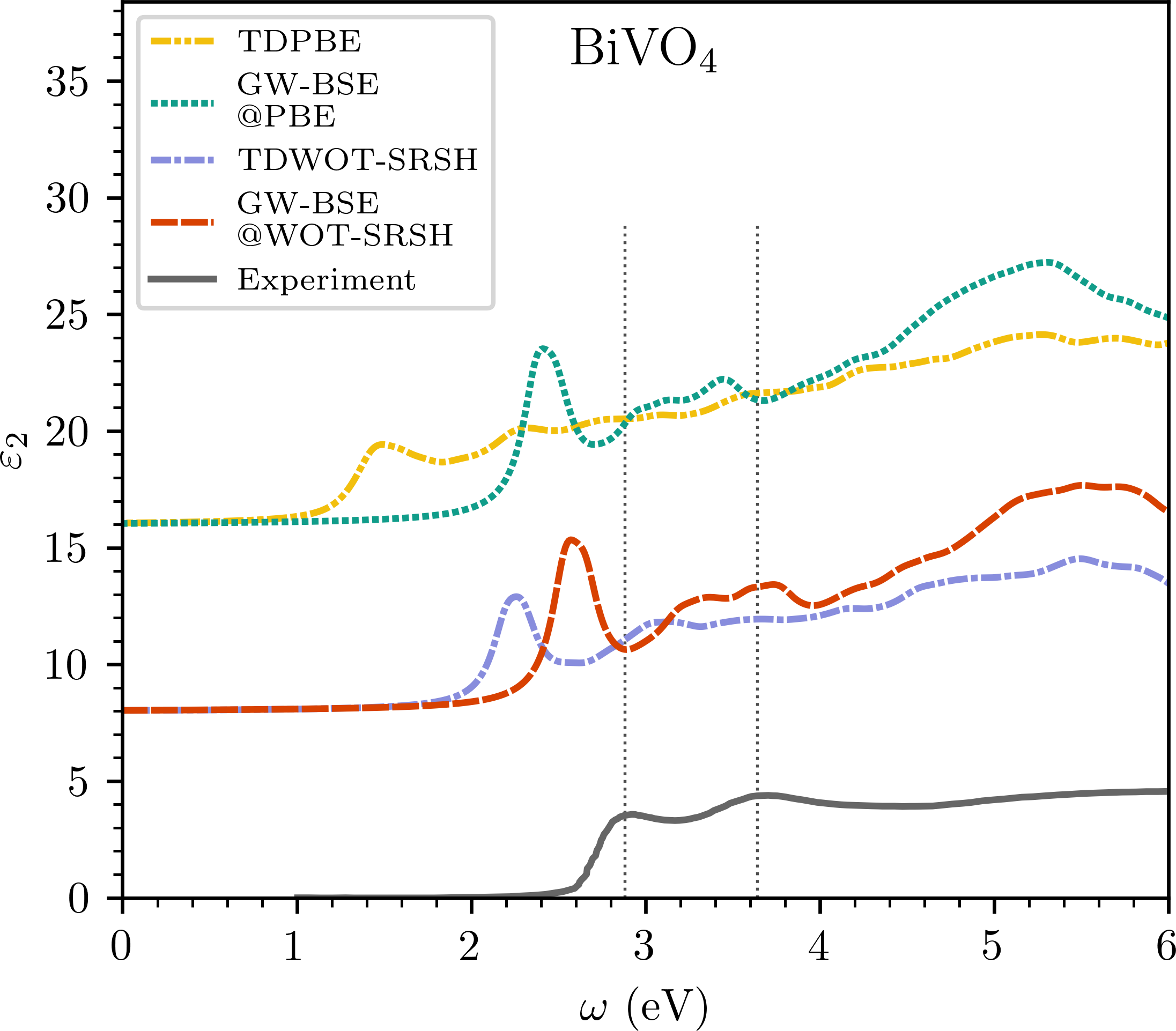}
\caption{\label{fig:bivo4_spectra} Same as Fig.~\ref{fig:spectra}, but for BiVO$_4$. The anisotropy in the optical response is directionally averaged. Experimental data are taken from Ref.~\cite{cooper2015indirect}.}
\end{centering}
\end{figure}

For BiVO$_4$, we observe a larger deviation between the WOT-SRSH-based spectra and experiment, as can be seen in Fig.~\ref{fig:bivo4_spectra}. This system was comprehensively studied by Wiktor \textit{et al.}~\cite{wiktor2017BiVO4}, with a special emphasis on the effect of thermal fluctuations on the electronic structure. Excluding these effects and the effect of spin-orbit coupling, they obtained a QP band gap of 3.64 eV using quasiparticle self-consistent $GW$, in good agreement with our results (3.5 eV and 3.8 eV from WOT-SRSH and $G_0W_0$@WOT-SRSH, respectively). Using path-integral molecular dynamics (including nuclear quantum effects) at the PBE0 level, they found a large QP band gap renormalization of -0.92 eV at 300 K, a value which we adopted in this work. Shifting the QP band gap by this amount brings it very close to the experimental optical indirect band gap of 2.5 eV \cite{cooper2015indirect}. While the effect of thermal fluctuations on the QP band gap in BiVO$_4$ has been explored, their effects on the optical absorption spectra, beyond causing a scissor-shift in the electronic bands, has not been studied to the best of our knowledge. Using the aforementioned QP thermal shift in the absorption spectrum may be insufficient for such a complex system with significant thermal fluctuations, because exciton-phonon interactions may also renormalize the exciton binding energy significantly. We therefore leave the question of thermal effects on the optical properties of BiVO$_4$ for the future, noting the agreement between the WOT-SRSH-based QP band gaps computed in this work and the one obtained by Wiktor \textit{et al.}~\cite{wiktor2017BiVO4}. 

Comparing the absorption onset of TDDFT and BSE with experiment in the case of BaSnO$_3$ and BiVO$_4$, we observe sharp excitonic peaks at the onset in both TDDFT and BSE, as opposed to shallow ``shoulders" in experiment. This can be directly attributed to significant finite temperature effects in those systems \cite{aggoune2022consistent, wiktor2017BiVO4} that can substantially alter the exciton, reduce the exciton binding energy and oscillator strength of excitonic peaks, and are not taken into account in our calculations. We note that peak shapes in agreement with our results have been obtained in Ref.~\cite{aggoune2022consistent} for BaSnO$_3$ and in Ref.~\cite{wiktor2017BiVO4} for BiVO$_4$, from $GW$-BSE.

In the context of comparing computed band gaps with experiment, we point out that a comparison of fundamental band gaps with optical experiments is inconsistent for MOs, because the exciton binding energy cannot be neglected. One can in principle compare fundamental band gaps with values obtained from, e.g., combined photoemission and inverse photoemission spectroscopy, but such experiments often suffer from significant experimental resolution uncertainties that amount to $\sim$0.4-0.5 eV \cite{tezuka1994photoemission,zimmermann1999electronic} and from sensitivity to surface effects and crystal dynamics \cite{sharifzadeh2012quasiparticle}. For these reasons, in this work we focus on optical band gaps for the comparison with experiment. Still, as fundamental band gaps are of general interest, we list them in the SM \cite{SM}.

The optical band gap is defined in most cases in this work as the onset of absorption, where a bright (dipole allowed) excitonic transition can be observed. As our optical spectra calculations do not account for momentum transfer, we choose as a benchmark experimental values that represent minimal direct transitions obtained in optical measurements. Table \ref{tab:optical_gaps} summarizes the optical band gaps predicted from TDWOT-SRSH and $G_0W_0$-BSE@WOT-SRSH, compared to experimental values. The optical band gap predictions are in overall good agreement between the two methods and experimental values, indicated by mean absolute errors of $\sim$0.3-0.4 eV with respect to experiment. We note that some discrepancies with respect to experimental gaps are to be anticipated, because there can be ambiguity associated with the choice of the model and fitting method used to analyze the absorption edge or the spectral features in experimental data. We also highlight that this work primarily focuses on the optical absorption spectra as a whole, where extrapolation is not needed to make a direct comparison. Additionally, we emphasize that while the shifted fine $\bm{k}$-grids used to compute the optical absorption spectra are relatively converged with respect to the overall peak positions and line-shape in the scale of the plot, the absorption onset obtained from our calculations are likely somewhat under-converged \cite{rohlfing2000electron,sun_ullrich_2020_prr} (see SM \cite{SM} for more details).

\begin{table*}[hbtp!]
\renewcommand{\arraystretch}{1.1}
\setlength{\tabcolsep}{0.07 in}
\centering
\caption{\label{tab:optical_gaps} Computed optical band gaps, compared with experimental optical measurements of direct transitions. Computed values refer to bright excited state energies at the onset of absorption, unless mentioned otherwise. Corrected values are obtained by adding the vibrational renormalization values taken from Table \ref{tab:thermal_renorm}. Spin-orbit coupling effects are not included. The mean absolute error (MAE) with respect to experiment is also given. All values are given in eV.}

\begin{tabular}{cccccc}
             &
  TDWOT-SRSH   &
  $G_0W_0$@WOT-SRSH &
  \begin{tabular}[c]{@{}c@{}}Corrected\\ TDWOT-SRSH\end{tabular} &
  \begin{tabular}[c]{@{}c@{}}Corrected\\ $G_0W_0$@WOT-SRSH\end{tabular} &
  Experiment \\ \hline
MgO                          & 7.8 & 8.1  &  7.2 & 7.6 & 7.7$^\text{a}$         \\
Al$_2$O$_3$                  & 9.3 & 9.8  &  9.0 & 9.4 & 8.8$^\text{b}$         \\
CaO                          & 6.5 & 6.9  &  6.1 & 6.6 & 6.9$^\text{a}$         \\
TiO$_2$$^\text{*}$           & 3.4 & 3.6  &  3.1 & 3.3 & 3.0$^\text{c}$         \\
Cu$_2$O$^\text{\#}$          & 2.5 & 2.4  &  2.3 & 2.2 & 2.6$^\text{d}$         \\
ZnO                          & 3.2 & 3.3  &  3.1 & 3.1 & 3.5$^\text{e}$         \\
BaSnO$_3$                    & 3.8 & 4.0  &  3.4 & 3.6 & 3.6$^\text{f}$         \\
BiVO$_4$                     & 3.1 & 3.5  &  2.2 & 2.5 & 2.7$^\text{g}$         \\
MAE                          &      &                          & 0.37 & 0.31 &                        \\
\hline
\end{tabular}

\caption*{
\textsuperscript{*}Values are dark excitons. See text for additional information.
\textsuperscript{\#}Values are first bright excited state. See text for additional information.
\textsuperscript{a}Ref.~\cite{whited_walker_1973}, from thermoreflectance spectra at 85K.
\textsuperscript{b}Ref.~\cite{french1990electronic}, from VUV reflectance at 300K. 
\textsuperscript{c}Ref.~\cite{pascual1977resolved}, from absorption spectra at 1.6K.
\textsuperscript{d}Ref.~\cite{takahata2018photoluminescence}, from photoluminescence spectra at 6K.
\textsuperscript{e}Ref.~\cite{tsoi2006isotopic}, from wavelength-modulated reflectivity measurements at low temperature.
\textsuperscript{f}Ref.~\cite{aggoune2022consistent}, from electron energy loss spectroscopy at 300K.
\textsuperscript{g}Ref.~\cite{cooper2015indirect}, from UV-vis absorption spectroscopy.
}
\end{table*}

There are two exceptional cases to the above definition for the optical band gap. These are rutile TiO$_2$ and Cu$_2$O, where the onset of absorption is a dark (dipole forbidden) transition. In TiO$_2$, the dark bound $1s$ exciton has been resolved by Pascual \textit{et al.}~\cite{pascual1977resolved}, allowing for direct comparison with TDDFT and BSE results. Both methods predict other in-gap brighter transitions, but those are less directly comparable with existing experimental data. Nonetheless, the shape and position of the first absorption peak (near 4 eV) is in good agreement with experiment for both TDWOT-SRSH and $GW$-BSE@WOT-SRSH.

The second exception to the above definition is Cu$_2$O, where the in-gap transitions from the topmost valence bands to the lowest conduction band (the so-called yellow/green exciton series) are dipole-forbidden transitions between states of orbital character of $3d$ and $4s$ respectively. These bound excitons, which have a $p$-like orbital character, occur just below the fundamental band gap \cite{kazimierczuk2014giant, takahata2018photoluminescence}. Experimentally, these low energy transitions are found to occur at 2.03 eV (1s exciton) and 2.15 eV (2p exciton) \cite{uihlein1981investigation, kazimierczuk2014giant}, whereas we observe the onset at 1.7 eV and 1.8 eV via TDWOT-SRSH and $GW$-BSE@WOT-SRSH respectively. However, the so-called blue/violet excitonic series in Cu$_2$O, associated with transitions from the topmost valence bands to the second lowest conduction bands, are dipole-allowed and manifest as the lowest energy resonant bright transitions that are clearly apparent in the optical spectra. Thus, we choose to define the optical band gap as the first of these bright transitions, which is experimentally observed at 2.6 eV \cite{takahata2018photoluminescence}. This value is in good agreement with the corresponding first bright transitions obtained in theory (see Table \ref{tab:optical_gaps}).

\section{Conclusions}

We have demonstrated the accuracy of the non-empirical WOT-SRSH functional for the prediction of the optical absorption spectra of MOs, a group of materials known for their computational complexity. By applying a simple, computationally efficient scheme for choosing the parameters of the SRSH functional, we find excellent agreement between TDWOT-SRSH and $G_0W_0$-BSE@WOT-SRSH, with slightly increased accuracy of the latter relative to experiment. These results suggest that the range of applicability of WOT-SRSH extends beyond computing band gaps of relatively simple semiconductors and insulators. It can be used with predictive accuracy to compute both electronic and optical properties of more challenging metal oxide systems.

\section*{Acknowledgments}
This work was supported via U.S.-Israel NSF–Binational Science Foundation Grant No. DMR-2015991 and by the Israel Science Foundation. Computational resources were provided by the Extreme Science and Engineering Discovery Environment (XSEDE) \cite{towns2014xsede} supercomputer Stampede2 at the Texas Advanced Computing Center (TACC) through the allocation TG-DMR190070. M.C.-G. is grateful to the Azrieli Foundation for the award of an Azrieli International Postdoctoral Fellowship. M.R.F acknowledges support from the UK Engineering and Physical Sciences Research Council (EPSRC), Grant EP/V010840/1. L.K. thanks the Aryeh and Mintzi Katzman Professorial Chair and the Helen and Martin Kimmel Award for Innovative Investigation.

\bibliographystyle{apsrev4-2_new.bst}
\bibliography{references.bib}

\begin{thebibliography}{183}%
\makeatletter
\providecommand \@ifxundefined [1]{%
 \@ifx{#1\undefined}
}%
\providecommand \@ifnum [1]{%
 \ifnum #1\expandafter \@firstoftwo
 \else \expandafter \@secondoftwo
 \fi
}%
\providecommand \@ifx [1]{%
 \ifx #1\expandafter \@firstoftwo
 \else \expandafter \@secondoftwo
 \fi
}%
\providecommand \natexlab [1]{#1}%
\providecommand \enquote  [1]{``#1''}%
\providecommand \bibnamefont  [1]{#1}%
\providecommand \bibfnamefont [1]{#1}%
\providecommand \citenamefont [1]{#1}%
\providecommand \href@noop [0]{\@secondoftwo}%
\providecommand \href [0]{\begingroup \@sanitize@url \@href}%
\providecommand \@href[1]{\@@startlink{#1}\@@href}%
\providecommand \@@href[1]{\endgroup#1\@@endlink}%
\providecommand \@sanitize@url [0]{\catcode `\\12\catcode `\$12\catcode
  `\&12\catcode `\#12\catcode `\^12\catcode `\_12\catcode `\%12\relax}%
\providecommand \@@startlink[1]{}%
\providecommand \@@endlink[0]{}%
\providecommand \url  [0]{\begingroup\@sanitize@url \@url }%
\providecommand \@url [1]{\endgroup\@href {#1}{\urlprefix }}%
\providecommand \urlprefix  [0]{URL }%
\providecommand \Eprint [0]{\href }%
\providecommand \doibase [0]{https://doi.org/}%
\providecommand \selectlanguage [0]{\@gobble}%
\providecommand \bibinfo  [0]{\@secondoftwo}%
\providecommand \bibfield  [0]{\@secondoftwo}%
\providecommand \translation [1]{[#1]}%
\providecommand \BibitemOpen [0]{}%
\providecommand \bibitemStop [0]{}%
\providecommand \bibitemNoStop [0]{.\EOS\space}%
\providecommand \EOS [0]{\spacefactor3000\relax}%
\providecommand \BibitemShut  [1]{\csname bibitem#1\endcsname}%
\let\auto@bib@innerbib\@empty
\bibitem [{\citenamefont {Albrecht}\ \emph {et~al.}(1998)\citenamefont
  {Albrecht}, \citenamefont {Reining}, \citenamefont {Del~Sole},\ and\
  \citenamefont {Onida}}]{albrecht1998excitonic}%
  \BibitemOpen
  \bibfield  {author} {\bibinfo {author} {\bibfnamefont {S.}~\bibnamefont
  {Albrecht}}, \bibinfo {author} {\bibfnamefont {L.}~\bibnamefont {Reining}},
  \bibinfo {author} {\bibfnamefont {R.}~\bibnamefont {Del~Sole}},\ and\
  \bibinfo {author} {\bibfnamefont {G.}~\bibnamefont {Onida}},\ }\href@noop {}
  {\bibfield  {journal} {\bibinfo  {journal} {Phys. Status Solidi A}\ }\textbf
  {\bibinfo {volume} {170}},\ \bibinfo {pages} {189} (\bibinfo {year}
  {1998})}\BibitemShut {NoStop}%
\bibitem [{\citenamefont {Rohlfing}\ and\ \citenamefont
  {Louie}(2000)}]{rohlfing2000electron}%
  \BibitemOpen
  \bibfield  {author} {\bibinfo {author} {\bibfnamefont {M.}~\bibnamefont
  {Rohlfing}}\ and\ \bibinfo {author} {\bibfnamefont {S.~G.}\ \bibnamefont
  {Louie}},\ }\href@noop {} {\bibfield  {journal} {\bibinfo  {journal} {Phys.
  Rev. B}\ }\textbf {\bibinfo {volume} {62}},\ \bibinfo {pages} {4927}
  (\bibinfo {year} {2000})}\BibitemShut {NoStop}%
\bibitem [{\citenamefont {Onida}\ \emph {et~al.}(2002)\citenamefont {Onida},
  \citenamefont {Reining},\ and\ \citenamefont {Rubio}}]{onida2002electronic}%
  \BibitemOpen
  \bibfield  {author} {\bibinfo {author} {\bibfnamefont {G.}~\bibnamefont
  {Onida}}, \bibinfo {author} {\bibfnamefont {L.}~\bibnamefont {Reining}},\
  and\ \bibinfo {author} {\bibfnamefont {A.}~\bibnamefont {Rubio}},\
  }\href@noop {} {\bibfield  {journal} {\bibinfo  {journal} {Rev. Mod. Phys.}\
  }\textbf {\bibinfo {volume} {74}},\ \bibinfo {pages} {601} (\bibinfo {year}
  {2002})}\BibitemShut {NoStop}%
\bibitem [{\citenamefont {Hedin}(1965)}]{hedin1965new}%
  \BibitemOpen
  \bibfield  {author} {\bibinfo {author} {\bibfnamefont {L.}~\bibnamefont
  {Hedin}},\ }\href@noop {} {\bibfield  {journal} {\bibinfo  {journal} {Phys.
  Rev.}\ }\textbf {\bibinfo {volume} {139}},\ \bibinfo {pages} {A796} (\bibinfo
  {year} {1965})}\BibitemShut {NoStop}%
\bibitem [{\citenamefont {Hybertsen}\ and\ \citenamefont
  {Louie}(1986)}]{hybertsenElectronCorrelationSemiconductors1986}%
  \BibitemOpen
  \bibfield  {author} {\bibinfo {author} {\bibfnamefont {M.~S.}\ \bibnamefont
  {Hybertsen}}\ and\ \bibinfo {author} {\bibfnamefont {S.~G.}\ \bibnamefont
  {Louie}},\ }\href {https://doi.org/10.1103/PhysRevB.34.5390} {\bibfield
  {journal} {\bibinfo  {journal} {Phys. Rev. B}\ }\textbf {\bibinfo {volume}
  {34}},\ \bibinfo {pages} {5390} (\bibinfo {year} {1986})}\BibitemShut
  {NoStop}%
\bibitem [{\citenamefont {Runge}\ and\ \citenamefont
  {Gross}(1984)}]{runge1984density}%
  \BibitemOpen
  \bibfield  {author} {\bibinfo {author} {\bibfnamefont {E.}~\bibnamefont
  {Runge}}\ and\ \bibinfo {author} {\bibfnamefont {E.~K.}\ \bibnamefont
  {Gross}},\ }\href@noop {} {\bibfield  {journal} {\bibinfo  {journal} {Phys.
  Rev. Lett.}\ }\textbf {\bibinfo {volume} {52}},\ \bibinfo {pages} {997}
  (\bibinfo {year} {1984})}\BibitemShut {NoStop}%
\bibitem [{\citenamefont {Ullrich}(2011)}]{ullrich2011time}%
  \BibitemOpen
  \bibfield  {author} {\bibinfo {author} {\bibfnamefont {C.~A.}\ \bibnamefont
  {Ullrich}},\ }\href@noop {} {\emph {\bibinfo {title} {Time-dependent
  density-functional theory: concepts and applications}}}\ (\bibinfo
  {publisher} {OUP Oxford},\ \bibinfo {year} {2011})\BibitemShut {NoStop}%
\bibitem [{\citenamefont {Burke}(2012)}]{burke2012perspective}%
  \BibitemOpen
  \bibfield  {author} {\bibinfo {author} {\bibfnamefont {K.}~\bibnamefont
  {Burke}},\ }\href@noop {} {\bibfield  {journal} {\bibinfo  {journal} {J.
  Chem. Phys.}\ }\textbf {\bibinfo {volume} {136}},\ \bibinfo {pages} {150901}
  (\bibinfo {year} {2012})}\BibitemShut {NoStop}%
\bibitem [{\citenamefont {Maitra}(2016)}]{maitra2016perspective}%
  \BibitemOpen
  \bibfield  {author} {\bibinfo {author} {\bibfnamefont {N.~T.}\ \bibnamefont
  {Maitra}},\ }\href@noop {} {\bibfield  {journal} {\bibinfo  {journal} {J.
  Chem. Phys.}\ }\textbf {\bibinfo {volume} {144}},\ \bibinfo {pages} {220901}
  (\bibinfo {year} {2016})}\BibitemShut {NoStop}%
\bibitem [{\citenamefont {Byun}\ \emph {et~al.}(2020)\citenamefont {Byun},
  \citenamefont {Sun},\ and\ \citenamefont {Ullrich}}]{Byun_Ullrich_2020}%
  \BibitemOpen
  \bibfield  {author} {\bibinfo {author} {\bibfnamefont {Y.-M.}\ \bibnamefont
  {Byun}}, \bibinfo {author} {\bibfnamefont {J.}~\bibnamefont {Sun}},\ and\
  \bibinfo {author} {\bibfnamefont {C.~A.}\ \bibnamefont {Ullrich}},\
  }\href@noop {} {\bibfield  {journal} {\bibinfo  {journal} {Electron.
  Struct.}\ }\textbf {\bibinfo {volume} {2}},\ \bibinfo {pages} {023002}
  (\bibinfo {year} {2020})}\BibitemShut {NoStop}%
\bibitem [{\citenamefont {Gavrilenko}\ and\ \citenamefont
  {Bechstedt}(1997)}]{gavrilenko1997optical}%
  \BibitemOpen
  \bibfield  {author} {\bibinfo {author} {\bibfnamefont {V.}~\bibnamefont
  {Gavrilenko}}\ and\ \bibinfo {author} {\bibfnamefont {F.}~\bibnamefont
  {Bechstedt}},\ }\href@noop {} {\bibfield  {journal} {\bibinfo  {journal}
  {Phys. Rev. B}\ }\textbf {\bibinfo {volume} {55}},\ \bibinfo {pages} {4343}
  (\bibinfo {year} {1997})}\BibitemShut {NoStop}%
\bibitem [{\citenamefont {Casida}(1995)}]{casida1995}%
  \BibitemOpen
  \bibfield  {author} {\bibinfo {author} {\bibfnamefont {M.~E.}\ \bibnamefont
  {Casida}},\ }in\ \href@noop {} {\emph {\bibinfo {booktitle} {Recent Advances
  in Density Functional Methods Part I}}},\ \bibinfo {editor} {edited by\
  \bibinfo {editor} {\bibfnamefont {D.~P.}\ \bibnamefont {Chong}}}\ (\bibinfo
  {publisher} {World Scientific, Singapore},\ \bibinfo {year} {1995})\
  Chap.~\bibinfo {chapter} {5}, pp.\ \bibinfo {pages} {155--192}\BibitemShut
  {NoStop}%
\bibitem [{\citenamefont {K\"ummel}\ and\ \citenamefont
  {Kronik}(2008)}]{kuemmel_kronik_2008}%
  \BibitemOpen
  \bibfield  {author} {\bibinfo {author} {\bibfnamefont {S.}~\bibnamefont
  {K\"ummel}}\ and\ \bibinfo {author} {\bibfnamefont {L.}~\bibnamefont
  {Kronik}},\ }\href@noop {} {\bibfield  {journal} {\bibinfo  {journal} {Rev.
  Mod. Phys.}\ }\textbf {\bibinfo {volume} {80}},\ \bibinfo {pages} {3}
  (\bibinfo {year} {2008})}\BibitemShut {NoStop}%
\bibitem [{\citenamefont {Botti}\ \emph {et~al.}(2004)\citenamefont {Botti},
  \citenamefont {Sottile}, \citenamefont {Vast}, \citenamefont {Olevano},
  \citenamefont {Reining}, \citenamefont {Weissker}, \citenamefont {Rubio},
  \citenamefont {Onida}, \citenamefont {Del~Sole},\ and\ \citenamefont
  {Godby}}]{botti2004long}%
  \BibitemOpen
  \bibfield  {author} {\bibinfo {author} {\bibfnamefont {S.}~\bibnamefont
  {Botti}}, \bibinfo {author} {\bibfnamefont {F.}~\bibnamefont {Sottile}},
  \bibinfo {author} {\bibfnamefont {N.}~\bibnamefont {Vast}}, \bibinfo {author}
  {\bibfnamefont {V.}~\bibnamefont {Olevano}}, \bibinfo {author} {\bibfnamefont
  {L.}~\bibnamefont {Reining}}, \bibinfo {author} {\bibfnamefont {H.-C.}\
  \bibnamefont {Weissker}}, \bibinfo {author} {\bibfnamefont {A.}~\bibnamefont
  {Rubio}}, \bibinfo {author} {\bibfnamefont {G.}~\bibnamefont {Onida}},
  \bibinfo {author} {\bibfnamefont {R.}~\bibnamefont {Del~Sole}},\ and\
  \bibinfo {author} {\bibfnamefont {R.}~\bibnamefont {Godby}},\ }\href@noop {}
  {\bibfield  {journal} {\bibinfo  {journal} {Phys. Rev. B}\ }\textbf {\bibinfo
  {volume} {69}},\ \bibinfo {pages} {155112} (\bibinfo {year}
  {2004})}\BibitemShut {NoStop}%
\bibitem [{\citenamefont {Botti}\ \emph {et~al.}(2005)\citenamefont {Botti},
  \citenamefont {Fourreau}, \citenamefont {Nguyen}, \citenamefont {Renault},
  \citenamefont {Sottile},\ and\ \citenamefont {Reining}}]{botti2005energy}%
  \BibitemOpen
  \bibfield  {author} {\bibinfo {author} {\bibfnamefont {S.}~\bibnamefont
  {Botti}}, \bibinfo {author} {\bibfnamefont {A.}~\bibnamefont {Fourreau}},
  \bibinfo {author} {\bibfnamefont {F.}~\bibnamefont {Nguyen}}, \bibinfo
  {author} {\bibfnamefont {Y.-O.}\ \bibnamefont {Renault}}, \bibinfo {author}
  {\bibfnamefont {F.}~\bibnamefont {Sottile}},\ and\ \bibinfo {author}
  {\bibfnamefont {L.}~\bibnamefont {Reining}},\ }\href@noop {} {\bibfield
  {journal} {\bibinfo  {journal} {Phys. Rev. B}\ }\textbf {\bibinfo {volume}
  {72}},\ \bibinfo {pages} {125203} (\bibinfo {year} {2005})}\BibitemShut
  {NoStop}%
\bibitem [{\citenamefont {Botti}\ \emph {et~al.}(2007)\citenamefont {Botti},
  \citenamefont {Schindlmayr}, \citenamefont {Del~Sole},\ and\ \citenamefont
  {Reining}}]{botti2007time}%
  \BibitemOpen
  \bibfield  {author} {\bibinfo {author} {\bibfnamefont {S.}~\bibnamefont
  {Botti}}, \bibinfo {author} {\bibfnamefont {A.}~\bibnamefont {Schindlmayr}},
  \bibinfo {author} {\bibfnamefont {R.}~\bibnamefont {Del~Sole}},\ and\
  \bibinfo {author} {\bibfnamefont {L.}~\bibnamefont {Reining}},\ }\href@noop
  {} {\bibfield  {journal} {\bibinfo  {journal} {Rep. Prog. Phys.}\ }\textbf
  {\bibinfo {volume} {70}},\ \bibinfo {pages} {357} (\bibinfo {year}
  {2007})}\BibitemShut {NoStop}%
\bibitem [{\citenamefont {Ghosez}\ \emph {et~al.}(1997)\citenamefont {Ghosez},
  \citenamefont {Gonze},\ and\ \citenamefont {Godby}}]{ghosez1997long}%
  \BibitemOpen
  \bibfield  {author} {\bibinfo {author} {\bibfnamefont {P.}~\bibnamefont
  {Ghosez}}, \bibinfo {author} {\bibfnamefont {X.}~\bibnamefont {Gonze}},\ and\
  \bibinfo {author} {\bibfnamefont {R.}~\bibnamefont {Godby}},\ }\href@noop {}
  {\bibfield  {journal} {\bibinfo  {journal} {Phys. Rev. B}\ }\textbf {\bibinfo
  {volume} {56}},\ \bibinfo {pages} {12811} (\bibinfo {year}
  {1997})}\BibitemShut {NoStop}%
\bibitem [{\citenamefont {Reining}\ \emph {et~al.}(2002)\citenamefont
  {Reining}, \citenamefont {Olevano}, \citenamefont {Rubio},\ and\
  \citenamefont {Onida}}]{reining2002excitonic}%
  \BibitemOpen
  \bibfield  {author} {\bibinfo {author} {\bibfnamefont {L.}~\bibnamefont
  {Reining}}, \bibinfo {author} {\bibfnamefont {V.}~\bibnamefont {Olevano}},
  \bibinfo {author} {\bibfnamefont {A.}~\bibnamefont {Rubio}},\ and\ \bibinfo
  {author} {\bibfnamefont {G.}~\bibnamefont {Onida}},\ }\href@noop {}
  {\bibfield  {journal} {\bibinfo  {journal} {Phys. Rev. Lett.}\ }\textbf
  {\bibinfo {volume} {88}},\ \bibinfo {pages} {066404} (\bibinfo {year}
  {2002})}\BibitemShut {NoStop}%
\bibitem [{\citenamefont {Levine}\ and\ \citenamefont
  {Allan}(1989)}]{levineLinearOpticalResponse1989}%
  \BibitemOpen
  \bibfield  {author} {\bibinfo {author} {\bibfnamefont {Z.~H.}\ \bibnamefont
  {Levine}}\ and\ \bibinfo {author} {\bibfnamefont {D.~C.}\ \bibnamefont
  {Allan}},\ }\href {https://doi.org/10.1103/PhysRevLett.63.1719} {\bibfield
  {journal} {\bibinfo  {journal} {Phys. Rev. Lett.}\ }\textbf {\bibinfo
  {volume} {63}},\ \bibinfo {pages} {1719} (\bibinfo {year}
  {1989})}\BibitemShut {NoStop}%
\bibitem [{\citenamefont {Cavo}\ \emph {et~al.}(2020)\citenamefont {Cavo},
  \citenamefont {Berger},\ and\ \citenamefont {Romaniello}}]{cavo2020accurate}%
  \BibitemOpen
  \bibfield  {author} {\bibinfo {author} {\bibfnamefont {S.}~\bibnamefont
  {Cavo}}, \bibinfo {author} {\bibfnamefont {J.}~\bibnamefont {Berger}},\ and\
  \bibinfo {author} {\bibfnamefont {P.}~\bibnamefont {Romaniello}},\
  }\href@noop {} {\bibfield  {journal} {\bibinfo  {journal} {Phys. Rev. B}\
  }\textbf {\bibinfo {volume} {101}},\ \bibinfo {pages} {115109} (\bibinfo
  {year} {2020})}\BibitemShut {NoStop}%
\bibitem [{\citenamefont {Seidl}\ \emph {et~al.}(1996)\citenamefont {Seidl},
  \citenamefont {G{\"o}rling}, \citenamefont {Vogl}, \citenamefont {Majewski},\
  and\ \citenamefont {Levy}}]{seidl_levy_1996}%
  \BibitemOpen
  \bibfield  {author} {\bibinfo {author} {\bibfnamefont {A.}~\bibnamefont
  {Seidl}}, \bibinfo {author} {\bibfnamefont {A.}~\bibnamefont {G{\"o}rling}},
  \bibinfo {author} {\bibfnamefont {P.}~\bibnamefont {Vogl}}, \bibinfo {author}
  {\bibfnamefont {J.}~\bibnamefont {Majewski}},\ and\ \bibinfo {author}
  {\bibfnamefont {M.}~\bibnamefont {Levy}},\ }\href@noop {} {\bibfield
  {journal} {\bibinfo  {journal} {Phys. Rev. B}\ }\textbf {\bibinfo {volume}
  {53}},\ \bibinfo {pages} {3764} (\bibinfo {year} {1996})}\BibitemShut
  {NoStop}%
\bibitem [{\citenamefont {Tretiak}\ and\ \citenamefont
  {Chernyak}(2003)}]{tretiak2003resonant}%
  \BibitemOpen
  \bibfield  {author} {\bibinfo {author} {\bibfnamefont {S.}~\bibnamefont
  {Tretiak}}\ and\ \bibinfo {author} {\bibfnamefont {V.}~\bibnamefont
  {Chernyak}},\ }\href@noop {} {\bibfield  {journal} {\bibinfo  {journal} {J.
  Chem. Phys.}\ }\textbf {\bibinfo {volume} {119}},\ \bibinfo {pages} {8809}
  (\bibinfo {year} {2003})}\BibitemShut {NoStop}%
\bibitem [{\citenamefont {Baer}\ and\ \citenamefont
  {Kronik}(2018)}]{baer_kronik_2018}%
  \BibitemOpen
  \bibfield  {author} {\bibinfo {author} {\bibfnamefont {R.}~\bibnamefont
  {Baer}}\ and\ \bibinfo {author} {\bibfnamefont {L.}~\bibnamefont {Kronik}},\
  }\href@noop {} {\bibfield  {journal} {\bibinfo  {journal} {Eur. Phys. J. B}\
  }\textbf {\bibinfo {volume} {91}},\ \bibinfo {pages} {1} (\bibinfo {year}
  {2018})}\BibitemShut {NoStop}%
\bibitem [{\citenamefont {Shimazaki}\ and\ \citenamefont
  {Asai}(2008)}]{shimazaki_asai_2008}%
  \BibitemOpen
  \bibfield  {author} {\bibinfo {author} {\bibfnamefont {T.}~\bibnamefont
  {Shimazaki}}\ and\ \bibinfo {author} {\bibfnamefont {Y.}~\bibnamefont
  {Asai}},\ }\href@noop {} {\bibfield  {journal} {\bibinfo  {journal} {Chem.
  Phys. Lett.}\ }\textbf {\bibinfo {volume} {466}},\ \bibinfo {pages} {91 }
  (\bibinfo {year} {2008})}\BibitemShut {NoStop}%
\bibitem [{\citenamefont {Kronik}\ and\ \citenamefont
  {Neaton}(2016)}]{kronik2016excited}%
  \BibitemOpen
  \bibfield  {author} {\bibinfo {author} {\bibfnamefont {L.}~\bibnamefont
  {Kronik}}\ and\ \bibinfo {author} {\bibfnamefont {J.~B.}\ \bibnamefont
  {Neaton}},\ }\href@noop {} {\bibfield  {journal} {\bibinfo  {journal} {Annu.
  Rev. Phys. Chem.}\ }\textbf {\bibinfo {volume} {67}},\ \bibinfo {pages} {587}
  (\bibinfo {year} {2016})}\BibitemShut {NoStop}%
\bibitem [{\citenamefont {Refaely-Abramson}\ \emph {et~al.}(2013)\citenamefont
  {Refaely-Abramson}, \citenamefont {Sharifzadeh}, \citenamefont {Jain},
  \citenamefont {Baer}, \citenamefont {Neaton},\ and\ \citenamefont
  {Kronik}}]{refaely-abramson_kronik_2013}%
  \BibitemOpen
  \bibfield  {author} {\bibinfo {author} {\bibfnamefont {S.}~\bibnamefont
  {Refaely-Abramson}}, \bibinfo {author} {\bibfnamefont {S.}~\bibnamefont
  {Sharifzadeh}}, \bibinfo {author} {\bibfnamefont {M.}~\bibnamefont {Jain}},
  \bibinfo {author} {\bibfnamefont {R.}~\bibnamefont {Baer}}, \bibinfo {author}
  {\bibfnamefont {J.~B.}\ \bibnamefont {Neaton}},\ and\ \bibinfo {author}
  {\bibfnamefont {L.}~\bibnamefont {Kronik}},\ }\href@noop {} {\bibfield
  {journal} {\bibinfo  {journal} {Phys. Rev. B}\ }\textbf {\bibinfo {volume}
  {88}},\ \bibinfo {pages} {081204} (\bibinfo {year} {2013})}\BibitemShut
  {NoStop}%
\bibitem [{\citenamefont {Refaely-Abramson}\ \emph {et~al.}(2015)\citenamefont
  {Refaely-Abramson}, \citenamefont {Jain}, \citenamefont {Sharifzadeh},
  \citenamefont {Neaton},\ and\ \citenamefont
  {Kronik}}]{refaely-abramson_kronik_2015}%
  \BibitemOpen
  \bibfield  {author} {\bibinfo {author} {\bibfnamefont {S.}~\bibnamefont
  {Refaely-Abramson}}, \bibinfo {author} {\bibfnamefont {M.}~\bibnamefont
  {Jain}}, \bibinfo {author} {\bibfnamefont {S.}~\bibnamefont {Sharifzadeh}},
  \bibinfo {author} {\bibfnamefont {J.~B.}\ \bibnamefont {Neaton}},\ and\
  \bibinfo {author} {\bibfnamefont {L.}~\bibnamefont {Kronik}},\ }\href@noop {}
  {\bibfield  {journal} {\bibinfo  {journal} {Phys. Rev. B}\ }\textbf {\bibinfo
  {volume} {92}},\ \bibinfo {pages} {081204} (\bibinfo {year}
  {2015})}\BibitemShut {NoStop}%
\bibitem [{\citenamefont {Zheng}\ \emph {et~al.}(2017)\citenamefont {Zheng},
  \citenamefont {Egger}, \citenamefont {Brédas}, \citenamefont {Kronik},\ and\
  \citenamefont {Coropceanu}}]{zheng_cororpceanu_2017}%
  \BibitemOpen
  \bibfield  {author} {\bibinfo {author} {\bibfnamefont {Z.}~\bibnamefont
  {Zheng}}, \bibinfo {author} {\bibfnamefont {D.~A.}\ \bibnamefont {Egger}},
  \bibinfo {author} {\bibfnamefont {J.-L.}\ \bibnamefont {Brédas}}, \bibinfo
  {author} {\bibfnamefont {L.}~\bibnamefont {Kronik}},\ and\ \bibinfo {author}
  {\bibfnamefont {V.}~\bibnamefont {Coropceanu}},\ }\href@noop {} {\bibfield
  {journal} {\bibinfo  {journal} {J. Phys. Chem. Lett.}\ }\textbf {\bibinfo
  {volume} {8}},\ \bibinfo {pages} {3277} (\bibinfo {year} {2017})}\BibitemShut
  {NoStop}%
\bibitem [{\citenamefont {Kronik}\ and\ \citenamefont
  {K{\"u}mmel}(2018)}]{kronik_kuemmel_2018}%
  \BibitemOpen
  \bibfield  {author} {\bibinfo {author} {\bibfnamefont {L.}~\bibnamefont
  {Kronik}}\ and\ \bibinfo {author} {\bibfnamefont {S.}~\bibnamefont
  {K{\"u}mmel}},\ }\href@noop {} {\bibfield  {journal} {\bibinfo  {journal}
  {Adv. Mater.}\ }\textbf {\bibinfo {volume} {30}},\ \bibinfo {pages} {1706560}
  (\bibinfo {year} {2018})}\BibitemShut {NoStop}%
\bibitem [{\citenamefont {Kronik}\ and\ \citenamefont
  {K{\"u}mmel}(2020)}]{kronik_kummel_2020}%
  \BibitemOpen
  \bibfield  {author} {\bibinfo {author} {\bibfnamefont {L.}~\bibnamefont
  {Kronik}}\ and\ \bibinfo {author} {\bibfnamefont {S.}~\bibnamefont
  {K{\"u}mmel}},\ }\href@noop {} {\bibfield  {journal} {\bibinfo  {journal}
  {Phys. Chem. Chem. Phys.}\ }\textbf {\bibinfo {volume} {22}},\ \bibinfo
  {pages} {16467} (\bibinfo {year} {2020})}\BibitemShut {NoStop}%
\bibitem [{\citenamefont {Yang}\ \emph {et~al.}(2015)\citenamefont {Yang},
  \citenamefont {Sottile},\ and\ \citenamefont {Ullrich}}]{yang2015simple}%
  \BibitemOpen
  \bibfield  {author} {\bibinfo {author} {\bibfnamefont {Z.-h.}\ \bibnamefont
  {Yang}}, \bibinfo {author} {\bibfnamefont {F.}~\bibnamefont {Sottile}},\ and\
  \bibinfo {author} {\bibfnamefont {C.~A.}\ \bibnamefont {Ullrich}},\
  }\href@noop {} {\bibfield  {journal} {\bibinfo  {journal} {Phys. Rev. B}\
  }\textbf {\bibinfo {volume} {92}},\ \bibinfo {pages} {035202} (\bibinfo
  {year} {2015})}\BibitemShut {NoStop}%
\bibitem [{\citenamefont {Sun}\ \emph {et~al.}(2020)\citenamefont {Sun},
  \citenamefont {Yang},\ and\ \citenamefont {Ullrich}}]{sun_ullrich_2020_prr}%
  \BibitemOpen
  \bibfield  {author} {\bibinfo {author} {\bibfnamefont {J.}~\bibnamefont
  {Sun}}, \bibinfo {author} {\bibfnamefont {J.}~\bibnamefont {Yang}},\ and\
  \bibinfo {author} {\bibfnamefont {C.~A.}\ \bibnamefont {Ullrich}},\
  }\href@noop {} {\bibfield  {journal} {\bibinfo  {journal} {Phys. Rev. Res.}\
  }\textbf {\bibinfo {volume} {2}},\ \bibinfo {pages} {013091} (\bibinfo {year}
  {2020})}\BibitemShut {NoStop}%
\bibitem [{\citenamefont {Sun}\ and\ \citenamefont
  {Ullrich}(2020)}]{sun_ullrich_2020}%
  \BibitemOpen
  \bibfield  {author} {\bibinfo {author} {\bibfnamefont {J.}~\bibnamefont
  {Sun}}\ and\ \bibinfo {author} {\bibfnamefont {C.~A.}\ \bibnamefont
  {Ullrich}},\ }\href@noop {} {\bibfield  {journal} {\bibinfo  {journal} {Phys.
  Rev. Mater.}\ }\textbf {\bibinfo {volume} {4}},\ \bibinfo {pages} {095402}
  (\bibinfo {year} {2020})}\BibitemShut {NoStop}%
\bibitem [{\citenamefont {Tal}\ \emph {et~al.}(2020)\citenamefont {Tal},
  \citenamefont {Liu}, \citenamefont {Kresse},\ and\ \citenamefont
  {Pasquarello}}]{tal_pasquarello_2020}%
  \BibitemOpen
  \bibfield  {author} {\bibinfo {author} {\bibfnamefont {A.}~\bibnamefont
  {Tal}}, \bibinfo {author} {\bibfnamefont {P.}~\bibnamefont {Liu}}, \bibinfo
  {author} {\bibfnamefont {G.}~\bibnamefont {Kresse}},\ and\ \bibinfo {author}
  {\bibfnamefont {A.}~\bibnamefont {Pasquarello}},\ }\href@noop {} {\bibfield
  {journal} {\bibinfo  {journal} {Phys. Rev. Res.}\ }\textbf {\bibinfo {volume}
  {2}},\ \bibinfo {pages} {032019} (\bibinfo {year} {2020})}\BibitemShut
  {NoStop}%
\bibitem [{\citenamefont {Chen}\ \emph {et~al.}(2018)\citenamefont {Chen},
  \citenamefont {Miceli}, \citenamefont {Rignanese},\ and\ \citenamefont
  {Pasquarello}}]{chen_pasquarello_2018}%
  \BibitemOpen
  \bibfield  {author} {\bibinfo {author} {\bibfnamefont {W.}~\bibnamefont
  {Chen}}, \bibinfo {author} {\bibfnamefont {G.}~\bibnamefont {Miceli}},
  \bibinfo {author} {\bibfnamefont {G.-M.}\ \bibnamefont {Rignanese}},\ and\
  \bibinfo {author} {\bibfnamefont {A.}~\bibnamefont {Pasquarello}},\
  }\href@noop {} {\bibfield  {journal} {\bibinfo  {journal} {Phys. Rev.
  Mater.}\ }\textbf {\bibinfo {volume} {2}},\ \bibinfo {pages} {073803}
  (\bibinfo {year} {2018})}\BibitemShut {NoStop}%
\bibitem [{\citenamefont {Wing}\ \emph {et~al.}(2019)\citenamefont {Wing},
  \citenamefont {Haber}, \citenamefont {Noff}, \citenamefont {Barker},
  \citenamefont {Egger}, \citenamefont {Ramasubramaniam}, \citenamefont
  {Louie}, \citenamefont {Neaton},\ and\ \citenamefont
  {Kronik}}]{wing_kronik_2019}%
  \BibitemOpen
  \bibfield  {author} {\bibinfo {author} {\bibfnamefont {D.}~\bibnamefont
  {Wing}}, \bibinfo {author} {\bibfnamefont {J.~B.}\ \bibnamefont {Haber}},
  \bibinfo {author} {\bibfnamefont {R.}~\bibnamefont {Noff}}, \bibinfo {author}
  {\bibfnamefont {B.}~\bibnamefont {Barker}}, \bibinfo {author} {\bibfnamefont
  {D.~A.}\ \bibnamefont {Egger}}, \bibinfo {author} {\bibfnamefont
  {A.}~\bibnamefont {Ramasubramaniam}}, \bibinfo {author} {\bibfnamefont
  {S.~G.}\ \bibnamefont {Louie}}, \bibinfo {author} {\bibfnamefont {J.~B.}\
  \bibnamefont {Neaton}},\ and\ \bibinfo {author} {\bibfnamefont
  {L.}~\bibnamefont {Kronik}},\ }\href@noop {} {\bibfield  {journal} {\bibinfo
  {journal} {Phys. Rev. Mater.}\ }\textbf {\bibinfo {volume} {3}},\ \bibinfo
  {pages} {064603} (\bibinfo {year} {2019})}\BibitemShut {NoStop}%
\bibitem [{\citenamefont {Ramasubramaniam}\ \emph {et~al.}(2019)\citenamefont
  {Ramasubramaniam}, \citenamefont {Wing},\ and\ \citenamefont
  {Kronik}}]{ramasubramaniam_2019}%
  \BibitemOpen
  \bibfield  {author} {\bibinfo {author} {\bibfnamefont {A.}~\bibnamefont
  {Ramasubramaniam}}, \bibinfo {author} {\bibfnamefont {D.}~\bibnamefont
  {Wing}},\ and\ \bibinfo {author} {\bibfnamefont {L.}~\bibnamefont {Kronik}},\
  }\href@noop {} {\bibfield  {journal} {\bibinfo  {journal} {Phys. Rev.
  Mater.}\ }\textbf {\bibinfo {volume} {3}},\ \bibinfo {pages} {084007}
  (\bibinfo {year} {2019})}\BibitemShut {NoStop}%
\bibitem [{\citenamefont {Wing}\ \emph {et~al.}(2020)\citenamefont {Wing},
  \citenamefont {Neaton},\ and\ \citenamefont {Kronik}}]{wing2020narrowgap}%
  \BibitemOpen
  \bibfield  {author} {\bibinfo {author} {\bibfnamefont {D.}~\bibnamefont
  {Wing}}, \bibinfo {author} {\bibfnamefont {J.~B.}\ \bibnamefont {Neaton}},\
  and\ \bibinfo {author} {\bibfnamefont {L.}~\bibnamefont {Kronik}},\
  }\href@noop {} {\bibfield  {journal} {\bibinfo  {journal} {Adv. Theory
  Simul.}\ }\textbf {\bibinfo {volume} {3}},\ \bibinfo {pages} {2000220}
  (\bibinfo {year} {2020})}\BibitemShut {NoStop}%
\bibitem [{\citenamefont {Lewis}\ \emph {et~al.}(2020)\citenamefont {Lewis},
  \citenamefont {Ramasubramaniam},\ and\ \citenamefont
  {Sharifzadeh}}]{lewis2020tuned}%
  \BibitemOpen
  \bibfield  {author} {\bibinfo {author} {\bibfnamefont {D.~K.}\ \bibnamefont
  {Lewis}}, \bibinfo {author} {\bibfnamefont {A.}~\bibnamefont
  {Ramasubramaniam}},\ and\ \bibinfo {author} {\bibfnamefont {S.}~\bibnamefont
  {Sharifzadeh}},\ }\href@noop {} {\bibfield  {journal} {\bibinfo  {journal}
  {Physical Review Materials}\ }\textbf {\bibinfo {volume} {4}},\ \bibinfo
  {pages} {063803} (\bibinfo {year} {2020})}\BibitemShut {NoStop}%
\bibitem [{\citenamefont {Camarasa-G{\'o}mez}\ \emph
  {et~al.}(2023)\citenamefont {Camarasa-G{\'o}mez}, \citenamefont
  {Ramasubramaniam}, \citenamefont {Neaton},\ and\ \citenamefont
  {Kronik}}]{camarasa2023transferable}%
  \BibitemOpen
  \bibfield  {author} {\bibinfo {author} {\bibfnamefont {M.}~\bibnamefont
  {Camarasa-G{\'o}mez}}, \bibinfo {author} {\bibfnamefont {A.}~\bibnamefont
  {Ramasubramaniam}}, \bibinfo {author} {\bibfnamefont {J.~B.}\ \bibnamefont
  {Neaton}},\ and\ \bibinfo {author} {\bibfnamefont {L.}~\bibnamefont
  {Kronik}},\ }\href@noop {} {\bibfield  {journal} {\bibinfo  {journal} {arXiv
  preprint arXiv:2305.13049}\ } (\bibinfo {year} {2023})}\BibitemShut {NoStop}%
\bibitem [{\citenamefont {Wing}\ \emph {et~al.}(2021)\citenamefont {Wing},
  \citenamefont {Ohad}, \citenamefont {Haber}, \citenamefont {Filip},
  \citenamefont {Gant}, \citenamefont {Neaton},\ and\ \citenamefont
  {Kronik}}]{wing_2021}%
  \BibitemOpen
  \bibfield  {author} {\bibinfo {author} {\bibfnamefont {D.}~\bibnamefont
  {Wing}}, \bibinfo {author} {\bibfnamefont {G.}~\bibnamefont {Ohad}}, \bibinfo
  {author} {\bibfnamefont {J.~B.}\ \bibnamefont {Haber}}, \bibinfo {author}
  {\bibfnamefont {M.~R.}\ \bibnamefont {Filip}}, \bibinfo {author}
  {\bibfnamefont {S.~E.}\ \bibnamefont {Gant}}, \bibinfo {author}
  {\bibfnamefont {J.~B.}\ \bibnamefont {Neaton}},\ and\ \bibinfo {author}
  {\bibfnamefont {L.}~\bibnamefont {Kronik}},\ }\href@noop {} {\bibfield
  {journal} {\bibinfo  {journal} {PNAS}\ }\textbf {\bibinfo {volume} {118}}
  (\bibinfo {year} {2021})}\BibitemShut {NoStop}%
\bibitem [{\citenamefont {Ma}\ and\ \citenamefont {Wang}(2016)}]{ma_wang_2016}%
  \BibitemOpen
  \bibfield  {author} {\bibinfo {author} {\bibfnamefont {J.}~\bibnamefont
  {Ma}}\ and\ \bibinfo {author} {\bibfnamefont {L.-W.}\ \bibnamefont {Wang}},\
  }\href@noop {} {\bibfield  {journal} {\bibinfo  {journal} {Sci. Rep.}\
  }\textbf {\bibinfo {volume} {6}},\ \bibinfo {pages} {1} (\bibinfo {year}
  {2016})}\BibitemShut {NoStop}%
\bibitem [{\citenamefont {Ohad}\ \emph {et~al.}(2022)\citenamefont {Ohad},
  \citenamefont {Wing}, \citenamefont {Gant}, \citenamefont {Cohen},
  \citenamefont {Haber}, \citenamefont {Sagredo}, \citenamefont {Filip},
  \citenamefont {Neaton},\ and\ \citenamefont
  {Kronik}}]{ohad_2022_wotsrsh_haps}%
  \BibitemOpen
  \bibfield  {author} {\bibinfo {author} {\bibfnamefont {G.}~\bibnamefont
  {Ohad}}, \bibinfo {author} {\bibfnamefont {D.}~\bibnamefont {Wing}}, \bibinfo
  {author} {\bibfnamefont {S.~E.}\ \bibnamefont {Gant}}, \bibinfo {author}
  {\bibfnamefont {A.~V.}\ \bibnamefont {Cohen}}, \bibinfo {author}
  {\bibfnamefont {J.~B.}\ \bibnamefont {Haber}}, \bibinfo {author}
  {\bibfnamefont {F.}~\bibnamefont {Sagredo}}, \bibinfo {author} {\bibfnamefont
  {M.~R.}\ \bibnamefont {Filip}}, \bibinfo {author} {\bibfnamefont {J.~B.}\
  \bibnamefont {Neaton}},\ and\ \bibinfo {author} {\bibfnamefont
  {L.}~\bibnamefont {Kronik}},\ }\href@noop {} {\bibfield  {journal} {\bibinfo
  {journal} {Phys. Rev. Mater.}\ }\textbf {\bibinfo {volume} {6}},\ \bibinfo
  {pages} {104606} (\bibinfo {year} {2022})}\BibitemShut {NoStop}%
\bibitem [{\citenamefont {Gant}\ \emph {et~al.}(2022)\citenamefont {Gant},
  \citenamefont {Haber}, \citenamefont {Filip}, \citenamefont {Sagredo},
  \citenamefont {Wing}, \citenamefont {Ohad}, \citenamefont {Kronik},\ and\
  \citenamefont {Neaton}}]{gant2022optimally}%
  \BibitemOpen
  \bibfield  {author} {\bibinfo {author} {\bibfnamefont {S.~E.}\ \bibnamefont
  {Gant}}, \bibinfo {author} {\bibfnamefont {J.~B.}\ \bibnamefont {Haber}},
  \bibinfo {author} {\bibfnamefont {M.~R.}\ \bibnamefont {Filip}}, \bibinfo
  {author} {\bibfnamefont {F.}~\bibnamefont {Sagredo}}, \bibinfo {author}
  {\bibfnamefont {D.}~\bibnamefont {Wing}}, \bibinfo {author} {\bibfnamefont
  {G.}~\bibnamefont {Ohad}}, \bibinfo {author} {\bibfnamefont {L.}~\bibnamefont
  {Kronik}},\ and\ \bibinfo {author} {\bibfnamefont {J.~B.}\ \bibnamefont
  {Neaton}},\ }\href@noop {} {\bibfield  {journal} {\bibinfo  {journal}
  {Physical Review Materials}\ }\textbf {\bibinfo {volume} {6}},\ \bibinfo
  {pages} {053802} (\bibinfo {year} {2022})}\BibitemShut {NoStop}%
\bibitem [{\citenamefont {Fierro}(2005)}]{fierro2005metal}%
  \BibitemOpen
  \bibfield  {author} {\bibinfo {author} {\bibfnamefont {J.~L.~G.}\
  \bibnamefont {Fierro}},\ }\href@noop {} {\emph {\bibinfo {title} {Metal
  oxides: chemistry and applications}}}\ (\bibinfo  {publisher} {CRC press},\
  \bibinfo {year} {2005})\BibitemShut {NoStop}%
\bibitem [{\citenamefont {Yu}\ \emph {et~al.}(2016)\citenamefont {Yu},
  \citenamefont {Marks},\ and\ \citenamefont {Facchetti}}]{yu2016metal}%
  \BibitemOpen
  \bibfield  {author} {\bibinfo {author} {\bibfnamefont {X.}~\bibnamefont
  {Yu}}, \bibinfo {author} {\bibfnamefont {T.~J.}\ \bibnamefont {Marks}},\ and\
  \bibinfo {author} {\bibfnamefont {A.}~\bibnamefont {Facchetti}},\ }\href@noop
  {} {\bibfield  {journal} {\bibinfo  {journal} {Nat. Mater.}\ }\textbf
  {\bibinfo {volume} {15}},\ \bibinfo {pages} {383} (\bibinfo {year}
  {2016})}\BibitemShut {NoStop}%
\bibitem [{\citenamefont {Das}\ \emph {et~al.}(2019)\citenamefont {Das},
  \citenamefont {Di~Liberto}, \citenamefont {Tosoni},\ and\ \citenamefont
  {Pacchioni}}]{das2019band}%
  \BibitemOpen
  \bibfield  {author} {\bibinfo {author} {\bibfnamefont {T.}~\bibnamefont
  {Das}}, \bibinfo {author} {\bibfnamefont {G.}~\bibnamefont {Di~Liberto}},
  \bibinfo {author} {\bibfnamefont {S.}~\bibnamefont {Tosoni}},\ and\ \bibinfo
  {author} {\bibfnamefont {G.}~\bibnamefont {Pacchioni}},\ }\href@noop {}
  {\bibfield  {journal} {\bibinfo  {journal} {J. Chem. Theory Comput.}\
  }\textbf {\bibinfo {volume} {15}},\ \bibinfo {pages} {6294} (\bibinfo {year}
  {2019})}\BibitemShut {NoStop}%
\bibitem [{\citenamefont {Gerosa}\ \emph {et~al.}(2017)\citenamefont {Gerosa},
  \citenamefont {Bottani}, \citenamefont {Di~Valentin}, \citenamefont {Onida},\
  and\ \citenamefont {Pacchioni}}]{gerosa_2017}%
  \BibitemOpen
  \bibfield  {author} {\bibinfo {author} {\bibfnamefont {M.}~\bibnamefont
  {Gerosa}}, \bibinfo {author} {\bibfnamefont {C.}~\bibnamefont {Bottani}},
  \bibinfo {author} {\bibfnamefont {C.}~\bibnamefont {Di~Valentin}}, \bibinfo
  {author} {\bibfnamefont {G.}~\bibnamefont {Onida}},\ and\ \bibinfo {author}
  {\bibfnamefont {G.}~\bibnamefont {Pacchioni}},\ }\href@noop {} {\bibfield
  {journal} {\bibinfo  {journal} {J. Phys.: Condens. Matter}\ }\textbf
  {\bibinfo {volume} {30}},\ \bibinfo {pages} {044003} (\bibinfo {year}
  {2017})}\BibitemShut {NoStop}%
\bibitem [{\citenamefont {Chevrier}\ \emph {et~al.}(2010)\citenamefont
  {Chevrier}, \citenamefont {Ong}, \citenamefont {Armiento}, \citenamefont
  {Chan},\ and\ \citenamefont {Ceder}}]{chevrier2010hybrid}%
  \BibitemOpen
  \bibfield  {author} {\bibinfo {author} {\bibfnamefont {V.~L.}\ \bibnamefont
  {Chevrier}}, \bibinfo {author} {\bibfnamefont {S.~P.}\ \bibnamefont {Ong}},
  \bibinfo {author} {\bibfnamefont {R.}~\bibnamefont {Armiento}}, \bibinfo
  {author} {\bibfnamefont {M.~K.}\ \bibnamefont {Chan}},\ and\ \bibinfo
  {author} {\bibfnamefont {G.}~\bibnamefont {Ceder}},\ }\href@noop {}
  {\bibfield  {journal} {\bibinfo  {journal} {Phys. Rev. B}\ }\textbf {\bibinfo
  {volume} {82}},\ \bibinfo {pages} {075122} (\bibinfo {year}
  {2010})}\BibitemShut {NoStop}%
\bibitem [{\citenamefont {Li}\ \emph {et~al.}(2013)\citenamefont {Li},
  \citenamefont {Walther}, \citenamefont {Kuc},\ and\ \citenamefont
  {Heine}}]{li_2013}%
  \BibitemOpen
  \bibfield  {author} {\bibinfo {author} {\bibfnamefont {W.}~\bibnamefont
  {Li}}, \bibinfo {author} {\bibfnamefont {C.~F.}\ \bibnamefont {Walther}},
  \bibinfo {author} {\bibfnamefont {A.}~\bibnamefont {Kuc}},\ and\ \bibinfo
  {author} {\bibfnamefont {T.}~\bibnamefont {Heine}},\ }\href@noop {}
  {\bibfield  {journal} {\bibinfo  {journal} {J. Chem. Theory Comput.}\
  }\textbf {\bibinfo {volume} {9}},\ \bibinfo {pages} {2950} (\bibinfo {year}
  {2013})}\BibitemShut {NoStop}%
\bibitem [{\citenamefont {Liu}\ \emph {et~al.}(2019)\citenamefont {Liu},
  \citenamefont {Franchini}, \citenamefont {Marsman},\ and\ \citenamefont
  {Kresse}}]{liu_2019}%
  \BibitemOpen
  \bibfield  {author} {\bibinfo {author} {\bibfnamefont {P.}~\bibnamefont
  {Liu}}, \bibinfo {author} {\bibfnamefont {C.}~\bibnamefont {Franchini}},
  \bibinfo {author} {\bibfnamefont {M.}~\bibnamefont {Marsman}},\ and\ \bibinfo
  {author} {\bibfnamefont {G.}~\bibnamefont {Kresse}},\ }\href@noop {}
  {\bibfield  {journal} {\bibinfo  {journal} {J. Phys.: Condens. Matter}\
  }\textbf {\bibinfo {volume} {32}},\ \bibinfo {pages} {015502} (\bibinfo
  {year} {2019})}\BibitemShut {NoStop}%
\bibitem [{\citenamefont {Mandal}\ \emph {et~al.}(2019)\citenamefont {Mandal},
  \citenamefont {Haule}, \citenamefont {Rabe},\ and\ \citenamefont
  {Vanderbilt}}]{mandal2019systematic}%
  \BibitemOpen
  \bibfield  {author} {\bibinfo {author} {\bibfnamefont {S.}~\bibnamefont
  {Mandal}}, \bibinfo {author} {\bibfnamefont {K.}~\bibnamefont {Haule}},
  \bibinfo {author} {\bibfnamefont {K.~M.}\ \bibnamefont {Rabe}},\ and\
  \bibinfo {author} {\bibfnamefont {D.}~\bibnamefont {Vanderbilt}},\
  }\href@noop {} {\bibfield  {journal} {\bibinfo  {journal} {{NPJ} Comput.
  Mater.}\ }\textbf {\bibinfo {volume} {5}},\ \bibinfo {pages} {1} (\bibinfo
  {year} {2019})}\BibitemShut {NoStop}%
\bibitem [{\citenamefont {Massidda}\ \emph {et~al.}(1997)\citenamefont
  {Massidda}, \citenamefont {Continenza}, \citenamefont {Posternak},\ and\
  \citenamefont {Baldereschi}}]{massidda1997quasiparticle}%
  \BibitemOpen
  \bibfield  {author} {\bibinfo {author} {\bibfnamefont {S.}~\bibnamefont
  {Massidda}}, \bibinfo {author} {\bibfnamefont {A.}~\bibnamefont
  {Continenza}}, \bibinfo {author} {\bibfnamefont {M.}~\bibnamefont
  {Posternak}},\ and\ \bibinfo {author} {\bibfnamefont {A.}~\bibnamefont
  {Baldereschi}},\ }\href@noop {} {\bibfield  {journal} {\bibinfo  {journal}
  {Phys. Rev. B}\ }\textbf {\bibinfo {volume} {55}},\ \bibinfo {pages} {13494}
  (\bibinfo {year} {1997})}\BibitemShut {NoStop}%
\bibitem [{\citenamefont {{\O}str{\o}m}\ \emph {et~al.}(2022)\citenamefont
  {{\O}str{\o}m}, \citenamefont {Hossain}, \citenamefont {Burr}, \citenamefont
  {Hart},\ and\ \citenamefont {Hoex}}]{ostrom2022designing}%
  \BibitemOpen
  \bibfield  {author} {\bibinfo {author} {\bibfnamefont {I.}~\bibnamefont
  {{\O}str{\o}m}}, \bibinfo {author} {\bibfnamefont {M.~A.}\ \bibnamefont
  {Hossain}}, \bibinfo {author} {\bibfnamefont {P.~A.}\ \bibnamefont {Burr}},
  \bibinfo {author} {\bibfnamefont {J.~N.}\ \bibnamefont {Hart}},\ and\
  \bibinfo {author} {\bibfnamefont {B.}~\bibnamefont {Hoex}},\ }\href@noop {}
  {\bibfield  {journal} {\bibinfo  {journal} {Phys. Chem. Chem. Phys.}\ }
  (\bibinfo {year} {2022})}\BibitemShut {NoStop}%
\bibitem [{\citenamefont {Samsonidze}\ \emph {et~al.}(2014)\citenamefont
  {Samsonidze}, \citenamefont {Park},\ and\ \citenamefont
  {Kozinsky}}]{samsonidze2014insights}%
  \BibitemOpen
  \bibfield  {author} {\bibinfo {author} {\bibfnamefont {G.}~\bibnamefont
  {Samsonidze}}, \bibinfo {author} {\bibfnamefont {C.-H.}\ \bibnamefont
  {Park}},\ and\ \bibinfo {author} {\bibfnamefont {B.}~\bibnamefont
  {Kozinsky}},\ }\href@noop {} {\bibfield  {journal} {\bibinfo  {journal} {J.
  Phys.: Condens. Matter}\ }\textbf {\bibinfo {volume} {26}},\ \bibinfo {pages}
  {475501} (\bibinfo {year} {2014})}\BibitemShut {NoStop}%
\bibitem [{\citenamefont {Weng}\ \emph {et~al.}(2020)\citenamefont {Weng},
  \citenamefont {Pan},\ and\ \citenamefont {Wang}}]{weng_wang_2020}%
  \BibitemOpen
  \bibfield  {author} {\bibinfo {author} {\bibfnamefont {M.}~\bibnamefont
  {Weng}}, \bibinfo {author} {\bibfnamefont {F.}~\bibnamefont {Pan}},\ and\
  \bibinfo {author} {\bibfnamefont {L.-W.}\ \bibnamefont {Wang}},\ }\href@noop
  {} {\bibfield  {journal} {\bibinfo  {journal} {{NPJ} Comput. Mater.}\
  }\textbf {\bibinfo {volume} {6}},\ \bibinfo {pages} {1} (\bibinfo {year}
  {2020})}\BibitemShut {NoStop}%
\bibitem [{\citenamefont {Coulter}\ \emph {et~al.}(2013)\citenamefont
  {Coulter}, \citenamefont {Manousakis},\ and\ \citenamefont
  {Gali}}]{coulter2013limitations}%
  \BibitemOpen
  \bibfield  {author} {\bibinfo {author} {\bibfnamefont {J.~E.}\ \bibnamefont
  {Coulter}}, \bibinfo {author} {\bibfnamefont {E.}~\bibnamefont
  {Manousakis}},\ and\ \bibinfo {author} {\bibfnamefont {A.}~\bibnamefont
  {Gali}},\ }\href@noop {} {\bibfield  {journal} {\bibinfo  {journal} {Phys.
  Rev. B}\ }\textbf {\bibinfo {volume} {88}},\ \bibinfo {pages} {041107}
  (\bibinfo {year} {2013})}\BibitemShut {NoStop}%
\bibitem [{\citenamefont {Shih}\ \emph {et~al.}(2010)\citenamefont {Shih},
  \citenamefont {Xue}, \citenamefont {Zhang}, \citenamefont {Cohen},\ and\
  \citenamefont {Louie}}]{shih2010quasiparticle}%
  \BibitemOpen
  \bibfield  {author} {\bibinfo {author} {\bibfnamefont {B.-C.}\ \bibnamefont
  {Shih}}, \bibinfo {author} {\bibfnamefont {Y.}~\bibnamefont {Xue}}, \bibinfo
  {author} {\bibfnamefont {P.}~\bibnamefont {Zhang}}, \bibinfo {author}
  {\bibfnamefont {M.~L.}\ \bibnamefont {Cohen}},\ and\ \bibinfo {author}
  {\bibfnamefont {S.~G.}\ \bibnamefont {Louie}},\ }\href@noop {} {\bibfield
  {journal} {\bibinfo  {journal} {Phys. Rev. Lett.}\ }\textbf {\bibinfo
  {volume} {105}},\ \bibinfo {pages} {146401} (\bibinfo {year}
  {2010})}\BibitemShut {NoStop}%
\bibitem [{\citenamefont {Bruneval}\ \emph {et~al.}(2006)\citenamefont
  {Bruneval}, \citenamefont {Vast}, \citenamefont {Reining}, \citenamefont
  {Izquierdo}, \citenamefont {Sirotti},\ and\ \citenamefont
  {Barrett}}]{brunevalExchangeCorrelationEffects2006}%
  \BibitemOpen
  \bibfield  {author} {\bibinfo {author} {\bibfnamefont {F.}~\bibnamefont
  {Bruneval}}, \bibinfo {author} {\bibfnamefont {N.}~\bibnamefont {Vast}},
  \bibinfo {author} {\bibfnamefont {L.}~\bibnamefont {Reining}}, \bibinfo
  {author} {\bibfnamefont {M.}~\bibnamefont {Izquierdo}}, \bibinfo {author}
  {\bibfnamefont {F.}~\bibnamefont {Sirotti}},\ and\ \bibinfo {author}
  {\bibfnamefont {N.}~\bibnamefont {Barrett}},\ }\href
  {https://doi.org/10.1103/PhysRevLett.97.267601} {\bibfield  {journal}
  {\bibinfo  {journal} {Phys. Rev. Lett.}\ }\textbf {\bibinfo {volume} {97}},\
  \bibinfo {pages} {267601} (\bibinfo {year} {2006})}\BibitemShut {NoStop}%
\bibitem [{\citenamefont {Kang}\ and\ \citenamefont
  {Hybertsen}(2010)}]{kangQuasiparticleOpticalProperties2010}%
  \BibitemOpen
  \bibfield  {author} {\bibinfo {author} {\bibfnamefont {W.}~\bibnamefont
  {Kang}}\ and\ \bibinfo {author} {\bibfnamefont {M.~S.}\ \bibnamefont
  {Hybertsen}},\ }\href {https://doi.org/10.1103/PhysRevB.82.085203} {\bibfield
   {journal} {\bibinfo  {journal} {Phys. Rev. B}\ }\textbf {\bibinfo {volume}
  {82}},\ \bibinfo {pages} {085203} (\bibinfo {year} {2010})}\BibitemShut
  {NoStop}%
\bibitem [{\citenamefont {Shishkin}\ and\ \citenamefont
  {Kresse}(2007)}]{shishkinSelfconsistentGWCalculations2007}%
  \BibitemOpen
  \bibfield  {author} {\bibinfo {author} {\bibfnamefont {M.}~\bibnamefont
  {Shishkin}}\ and\ \bibinfo {author} {\bibfnamefont {G.}~\bibnamefont
  {Kresse}},\ }\href {https://doi.org/10.1103/PhysRevB.75.235102} {\bibfield
  {journal} {\bibinfo  {journal} {Phys. Rev. B}\ }\textbf {\bibinfo {volume}
  {75}},\ \bibinfo {pages} {235102} (\bibinfo {year} {2007})}\BibitemShut
  {NoStop}%
\bibitem [{\citenamefont {{van Schilfgaarde}}\ \emph
  {et~al.}(2006{\natexlab{a}})\citenamefont {{van Schilfgaarde}}, \citenamefont
  {Kotani},\ and\ \citenamefont
  {Faleev}}]{vanschilfgaardeQuasiparticleSelfConsistentGW2006b}%
  \BibitemOpen
  \bibfield  {author} {\bibinfo {author} {\bibfnamefont {M.}~\bibnamefont {{van
  Schilfgaarde}}}, \bibinfo {author} {\bibfnamefont {T.}~\bibnamefont
  {Kotani}},\ and\ \bibinfo {author} {\bibfnamefont {S.}~\bibnamefont
  {Faleev}},\ }\href {https://doi.org/10.1103/PhysRevLett.96.226402} {\bibfield
   {journal} {\bibinfo  {journal} {Phys. Rev. Lett.}\ }\textbf {\bibinfo
  {volume} {96}},\ \bibinfo {pages} {226402} (\bibinfo {year}
  {2006}{\natexlab{a}})}\BibitemShut {NoStop}%
\bibitem [{\citenamefont {Rangel}\ \emph {et~al.}(2020)\citenamefont {Rangel},
  \citenamefont {Del~Ben}, \citenamefont {Varsano}, \citenamefont {Antonius},
  \citenamefont {Bruneval}, \citenamefont {{da Jornada}}, \citenamefont {{van
  Setten}}, \citenamefont {Orhan}, \citenamefont {O'Regan}, \citenamefont
  {Canning}, \citenamefont {Ferretti}, \citenamefont {Marini}, \citenamefont
  {Rignanese}, \citenamefont {Deslippe}, \citenamefont {Louie},\ and\
  \citenamefont {Neaton}}]{rangelReproducibilityCalculationsSolids2020}%
  \BibitemOpen
  \bibfield  {author} {\bibinfo {author} {\bibfnamefont {T.}~\bibnamefont
  {Rangel}}, \bibinfo {author} {\bibfnamefont {M.}~\bibnamefont {Del~Ben}},
  \bibinfo {author} {\bibfnamefont {D.}~\bibnamefont {Varsano}}, \bibinfo
  {author} {\bibfnamefont {G.}~\bibnamefont {Antonius}}, \bibinfo {author}
  {\bibfnamefont {F.}~\bibnamefont {Bruneval}}, \bibinfo {author}
  {\bibfnamefont {F.~H.}\ \bibnamefont {{da Jornada}}}, \bibinfo {author}
  {\bibfnamefont {M.~J.}\ \bibnamefont {{van Setten}}}, \bibinfo {author}
  {\bibfnamefont {O.~K.}\ \bibnamefont {Orhan}}, \bibinfo {author}
  {\bibfnamefont {D.~D.}\ \bibnamefont {O'Regan}}, \bibinfo {author}
  {\bibfnamefont {A.}~\bibnamefont {Canning}}, \bibinfo {author} {\bibfnamefont
  {A.}~\bibnamefont {Ferretti}}, \bibinfo {author} {\bibfnamefont
  {A.}~\bibnamefont {Marini}}, \bibinfo {author} {\bibfnamefont {G.-M.}\
  \bibnamefont {Rignanese}}, \bibinfo {author} {\bibfnamefont {J.}~\bibnamefont
  {Deslippe}}, \bibinfo {author} {\bibfnamefont {S.~G.}\ \bibnamefont
  {Louie}},\ and\ \bibinfo {author} {\bibfnamefont {J.~B.}\ \bibnamefont
  {Neaton}},\ }\href {https://doi.org/10.1016/j.cpc.2020.107242} {\bibfield
  {journal} {\bibinfo  {journal} {Comput. Phys. Commun.}\ }\textbf {\bibinfo
  {volume} {255}},\ \bibinfo {pages} {107242} (\bibinfo {year}
  {2020})}\BibitemShut {NoStop}%
\bibitem [{\citenamefont {Schleife}\ \emph {et~al.}(2009)\citenamefont
  {Schleife}, \citenamefont {Fuchs}, \citenamefont {R{\"o}dl}, \citenamefont
  {Furthm{\"u}ller},\ and\ \citenamefont
  {Bechstedt}}]{schleifeBandstructureOpticaltransitionParameters2009}%
  \BibitemOpen
  \bibfield  {author} {\bibinfo {author} {\bibfnamefont {A.}~\bibnamefont
  {Schleife}}, \bibinfo {author} {\bibfnamefont {F.}~\bibnamefont {Fuchs}},
  \bibinfo {author} {\bibfnamefont {C.}~\bibnamefont {R{\"o}dl}}, \bibinfo
  {author} {\bibfnamefont {J.}~\bibnamefont {Furthm{\"u}ller}},\ and\ \bibinfo
  {author} {\bibfnamefont {F.}~\bibnamefont {Bechstedt}},\ }\href
  {https://doi.org/10.1002/pssb.200945204} {\bibfield  {journal} {\bibinfo
  {journal} {physica status solidi (b)}\ }\textbf {\bibinfo {volume} {246}},\
  \bibinfo {pages} {2150} (\bibinfo {year} {2009})}\BibitemShut {NoStop}%
\bibitem [{\citenamefont {Golze}\ \emph {et~al.}(2019)\citenamefont {Golze},
  \citenamefont {Dvorak},\ and\ \citenamefont
  {Rinke}}]{golzeGWCompendiumPractical2019}%
  \BibitemOpen
  \bibfield  {author} {\bibinfo {author} {\bibfnamefont {D.}~\bibnamefont
  {Golze}}, \bibinfo {author} {\bibfnamefont {M.}~\bibnamefont {Dvorak}},\ and\
  \bibinfo {author} {\bibfnamefont {P.}~\bibnamefont {Rinke}},\ }\href
  {https://doi.org/10.3389/fchem.2019.00377} {\bibfield  {journal} {\bibinfo
  {journal} {Front. Chem.}\ }\textbf {\bibinfo {volume} {7}},\ \bibinfo {pages}
  {377} (\bibinfo {year} {2019})}\BibitemShut {NoStop}%
\bibitem [{\citenamefont {Wu}\ \emph {et~al.}(2018)\citenamefont {Wu},
  \citenamefont {Saidi}, \citenamefont {Ohodnicki}, \citenamefont
  {Chorpening},\ and\ \citenamefont {Duan}}]{wu2018first}%
  \BibitemOpen
  \bibfield  {author} {\bibinfo {author} {\bibfnamefont {Y.-N.}\ \bibnamefont
  {Wu}}, \bibinfo {author} {\bibfnamefont {W.~A.}\ \bibnamefont {Saidi}},
  \bibinfo {author} {\bibfnamefont {P.}~\bibnamefont {Ohodnicki}}, \bibinfo
  {author} {\bibfnamefont {B.}~\bibnamefont {Chorpening}},\ and\ \bibinfo
  {author} {\bibfnamefont {Y.}~\bibnamefont {Duan}},\ }\href@noop {} {\bibfield
   {journal} {\bibinfo  {journal} {J. Phys. Chem. C}\ }\textbf {\bibinfo
  {volume} {122}},\ \bibinfo {pages} {22642} (\bibinfo {year}
  {2018})}\BibitemShut {NoStop}%
\bibitem [{\citenamefont {Wu}\ \emph {et~al.}(2020)\citenamefont {Wu},
  \citenamefont {Wuenschell}, \citenamefont {Fryer}, \citenamefont {Saidi},
  \citenamefont {Ohodnicki}, \citenamefont {Chorpening},\ and\ \citenamefont
  {Duan}}]{wu2020theoretical}%
  \BibitemOpen
  \bibfield  {author} {\bibinfo {author} {\bibfnamefont {Y.-N.}\ \bibnamefont
  {Wu}}, \bibinfo {author} {\bibfnamefont {J.~K.}\ \bibnamefont {Wuenschell}},
  \bibinfo {author} {\bibfnamefont {R.}~\bibnamefont {Fryer}}, \bibinfo
  {author} {\bibfnamefont {W.~A.}\ \bibnamefont {Saidi}}, \bibinfo {author}
  {\bibfnamefont {P.}~\bibnamefont {Ohodnicki}}, \bibinfo {author}
  {\bibfnamefont {B.}~\bibnamefont {Chorpening}},\ and\ \bibinfo {author}
  {\bibfnamefont {Y.}~\bibnamefont {Duan}},\ }\href@noop {} {\bibfield
  {journal} {\bibinfo  {journal} {J. Phys.: Condens. Matter}\ }\textbf
  {\bibinfo {volume} {32}},\ \bibinfo {pages} {405705} (\bibinfo {year}
  {2020})}\BibitemShut {NoStop}%
\bibitem [{\citenamefont {Park}\ \emph {et~al.}(2022)\citenamefont {Park},
  \citenamefont {Saidi}, \citenamefont {Chorpening},\ and\ \citenamefont
  {Duan}}]{park2022applicability}%
  \BibitemOpen
  \bibfield  {author} {\bibinfo {author} {\bibfnamefont {J.}~\bibnamefont
  {Park}}, \bibinfo {author} {\bibfnamefont {W.~A.}\ \bibnamefont {Saidi}},
  \bibinfo {author} {\bibfnamefont {B.}~\bibnamefont {Chorpening}},\ and\
  \bibinfo {author} {\bibfnamefont {Y.}~\bibnamefont {Duan}},\ }\href@noop {}
  {\bibfield  {journal} {\bibinfo  {journal} {Chem. Mater.}\ }\textbf {\bibinfo
  {volume} {34}},\ \bibinfo {pages} {6108} (\bibinfo {year}
  {2022})}\BibitemShut {NoStop}%
\bibitem [{\citenamefont {Perdew}\ and\ \citenamefont
  {Zunger}(1981)}]{perdew1981self}%
  \BibitemOpen
  \bibfield  {author} {\bibinfo {author} {\bibfnamefont {J.~P.}\ \bibnamefont
  {Perdew}}\ and\ \bibinfo {author} {\bibfnamefont {A.}~\bibnamefont
  {Zunger}},\ }\href@noop {} {\bibfield  {journal} {\bibinfo  {journal}
  {Physical Review B}\ }\textbf {\bibinfo {volume} {23}},\ \bibinfo {pages}
  {5048} (\bibinfo {year} {1981})}\BibitemShut {NoStop}%
\bibitem [{\citenamefont {Mori-S\'anchez}\ \emph {et~al.}(2008)\citenamefont
  {Mori-S\'anchez}, \citenamefont {Cohen},\ and\ \citenamefont
  {Yang}}]{mori-sanchez_yang_2008}%
  \BibitemOpen
  \bibfield  {author} {\bibinfo {author} {\bibfnamefont {P.}~\bibnamefont
  {Mori-S\'anchez}}, \bibinfo {author} {\bibfnamefont {A.~J.}\ \bibnamefont
  {Cohen}},\ and\ \bibinfo {author} {\bibfnamefont {W.}~\bibnamefont {Yang}},\
  }\href@noop {} {\bibfield  {journal} {\bibinfo  {journal} {Phys. Rev. Lett.}\
  }\textbf {\bibinfo {volume} {100}},\ \bibinfo {pages} {146401} (\bibinfo
  {year} {2008})}\BibitemShut {NoStop}%
\bibitem [{\citenamefont {Aggoune}\ \emph {et~al.}(2022)\citenamefont
  {Aggoune}, \citenamefont {Eljarrat}, \citenamefont {Nabok}, \citenamefont
  {Irmscher}, \citenamefont {Zupancic}, \citenamefont {Galazka}, \citenamefont
  {Albrecht}, \citenamefont {Koch},\ and\ \citenamefont
  {Draxl}}]{aggoune2022consistent}%
  \BibitemOpen
  \bibfield  {author} {\bibinfo {author} {\bibfnamefont {W.}~\bibnamefont
  {Aggoune}}, \bibinfo {author} {\bibfnamefont {A.}~\bibnamefont {Eljarrat}},
  \bibinfo {author} {\bibfnamefont {D.}~\bibnamefont {Nabok}}, \bibinfo
  {author} {\bibfnamefont {K.}~\bibnamefont {Irmscher}}, \bibinfo {author}
  {\bibfnamefont {M.}~\bibnamefont {Zupancic}}, \bibinfo {author}
  {\bibfnamefont {Z.}~\bibnamefont {Galazka}}, \bibinfo {author} {\bibfnamefont
  {M.}~\bibnamefont {Albrecht}}, \bibinfo {author} {\bibfnamefont
  {C.}~\bibnamefont {Koch}},\ and\ \bibinfo {author} {\bibfnamefont
  {C.}~\bibnamefont {Draxl}},\ }\href@noop {} {\bibfield  {journal} {\bibinfo
  {journal} {Commun. Mater.}\ }\textbf {\bibinfo {volume} {3}},\ \bibinfo
  {pages} {1} (\bibinfo {year} {2022})}\BibitemShut {NoStop}%
\bibitem [{\citenamefont {Wiktor}\ \emph
  {et~al.}(2017{\natexlab{a}})\citenamefont {Wiktor}, \citenamefont
  {Reshetnyak}, \citenamefont {Ambrosio},\ and\ \citenamefont
  {Pasquarello}}]{wiktor2017BiVO4}%
  \BibitemOpen
  \bibfield  {author} {\bibinfo {author} {\bibfnamefont {J.}~\bibnamefont
  {Wiktor}}, \bibinfo {author} {\bibfnamefont {I.}~\bibnamefont {Reshetnyak}},
  \bibinfo {author} {\bibfnamefont {F.}~\bibnamefont {Ambrosio}},\ and\
  \bibinfo {author} {\bibfnamefont {A.}~\bibnamefont {Pasquarello}},\
  }\href@noop {} {\bibfield  {journal} {\bibinfo  {journal} {Physical Review
  Materials}\ }\textbf {\bibinfo {volume} {1}},\ \bibinfo {pages} {022401}
  (\bibinfo {year} {2017}{\natexlab{a}})}\BibitemShut {NoStop}%
\bibitem [{\citenamefont {Madelung}(2004)}]{Madelung_2004}%
  \BibitemOpen
  \bibfield  {author} {\bibinfo {author} {\bibfnamefont {O.}~\bibnamefont
  {Madelung}},\ }\href@noop {} {\emph {\bibinfo {title} {Semiconductors Data
  Handbook}}},\ \bibinfo {edition} {3rd}\ ed.\ (\bibinfo  {publisher}
  {Springer-Verlag Berlin Heidelberg},\ \bibinfo {year} {2004})\BibitemShut
  {NoStop}%
\bibitem [{\citenamefont {Kondo}\ \emph {et~al.}(2008)\citenamefont {Kondo},
  \citenamefont {Tateishi},\ and\ \citenamefont
  {Ishizawa}}]{kondo2008structural}%
  \BibitemOpen
  \bibfield  {author} {\bibinfo {author} {\bibfnamefont {S.}~\bibnamefont
  {Kondo}}, \bibinfo {author} {\bibfnamefont {K.}~\bibnamefont {Tateishi}},\
  and\ \bibinfo {author} {\bibfnamefont {N.}~\bibnamefont {Ishizawa}},\
  }\href@noop {} {\bibfield  {journal} {\bibinfo  {journal} {Jpn. J. Appl.
  Phys.}\ }\textbf {\bibinfo {volume} {47}},\ \bibinfo {pages} {616} (\bibinfo
  {year} {2008})}\BibitemShut {NoStop}%
\bibitem [{\citenamefont {Sugiyama}\ and\ \citenamefont
  {Takeuchi}(1991)}]{sugiyama1991crystal}%
  \BibitemOpen
  \bibfield  {author} {\bibinfo {author} {\bibfnamefont {K.}~\bibnamefont
  {Sugiyama}}\ and\ \bibinfo {author} {\bibfnamefont {Y.}~\bibnamefont
  {Takeuchi}},\ }\href@noop {} {\bibfield  {journal} {\bibinfo  {journal} {Z.
  Kristallogr. - Cryst. Mater.}\ }\textbf {\bibinfo {volume} {194}},\ \bibinfo
  {pages} {305} (\bibinfo {year} {1991})}\BibitemShut {NoStop}%
\bibitem [{\citenamefont {Foo}\ \emph {et~al.}(2006)\citenamefont {Foo},
  \citenamefont {Huang}, \citenamefont {Lynn}, \citenamefont {Lee},
  \citenamefont {Klimczuk}, \citenamefont {Hagemann}, \citenamefont {Ong},\
  and\ \citenamefont {Cava}}]{foo2006synthesis}%
  \BibitemOpen
  \bibfield  {author} {\bibinfo {author} {\bibfnamefont {M.~L.}\ \bibnamefont
  {Foo}}, \bibinfo {author} {\bibfnamefont {Q.}~\bibnamefont {Huang}}, \bibinfo
  {author} {\bibfnamefont {J.}~\bibnamefont {Lynn}}, \bibinfo {author}
  {\bibfnamefont {W.-L.}\ \bibnamefont {Lee}}, \bibinfo {author} {\bibfnamefont
  {T.}~\bibnamefont {Klimczuk}}, \bibinfo {author} {\bibfnamefont
  {I.}~\bibnamefont {Hagemann}}, \bibinfo {author} {\bibfnamefont
  {N.}~\bibnamefont {Ong}},\ and\ \bibinfo {author} {\bibfnamefont {R.~J.}\
  \bibnamefont {Cava}},\ }\href@noop {} {\bibfield  {journal} {\bibinfo
  {journal} {J. Solid State Chem.}\ }\textbf {\bibinfo {volume} {179}},\
  \bibinfo {pages} {563} (\bibinfo {year} {2006})}\BibitemShut {NoStop}%
\bibitem [{\citenamefont {Garcia-Martinez}\ \emph {et~al.}(1993)\citenamefont
  {Garcia-Martinez}, \citenamefont {Rojas}, \citenamefont {Vila},\ and\
  \citenamefont {De~Vidales}}]{garcia1993microstructural}%
  \BibitemOpen
  \bibfield  {author} {\bibinfo {author} {\bibfnamefont {O.}~\bibnamefont
  {Garcia-Martinez}}, \bibinfo {author} {\bibfnamefont {R.}~\bibnamefont
  {Rojas}}, \bibinfo {author} {\bibfnamefont {E.}~\bibnamefont {Vila}},\ and\
  \bibinfo {author} {\bibfnamefont {J.~M.}\ \bibnamefont {De~Vidales}},\
  }\href@noop {} {\bibfield  {journal} {\bibinfo  {journal} {Solid State
  Ionics}\ }\textbf {\bibinfo {volume} {63}},\ \bibinfo {pages} {442} (\bibinfo
  {year} {1993})}\BibitemShut {NoStop}%
\bibitem [{\citenamefont {Mizoguchi}\ \emph {et~al.}(2004)\citenamefont
  {Mizoguchi}, \citenamefont {Woodward}, \citenamefont {Park},\ and\
  \citenamefont {Keszler}}]{mizoguchi2004strong}%
  \BibitemOpen
  \bibfield  {author} {\bibinfo {author} {\bibfnamefont {H.}~\bibnamefont
  {Mizoguchi}}, \bibinfo {author} {\bibfnamefont {P.~M.}\ \bibnamefont
  {Woodward}}, \bibinfo {author} {\bibfnamefont {C.-H.}\ \bibnamefont {Park}},\
  and\ \bibinfo {author} {\bibfnamefont {D.~A.}\ \bibnamefont {Keszler}},\
  }\href@noop {} {\bibfield  {journal} {\bibinfo  {journal} {J. Am. Chem.
  Soc.}\ }\textbf {\bibinfo {volume} {126}},\ \bibinfo {pages} {9796} (\bibinfo
  {year} {2004})}\BibitemShut {NoStop}%
\bibitem [{\citenamefont {Sleight}\ \emph {et~al.}(1979)\citenamefont
  {Sleight}, \citenamefont {Chen}, \citenamefont {Ferretti},\ and\
  \citenamefont {Cox}}]{sleight1979crystal}%
  \BibitemOpen
  \bibfield  {author} {\bibinfo {author} {\bibfnamefont {A.}~\bibnamefont
  {Sleight}}, \bibinfo {author} {\bibfnamefont {H.-Y.}\ \bibnamefont {Chen}},
  \bibinfo {author} {\bibfnamefont {A.}~\bibnamefont {Ferretti}},\ and\
  \bibinfo {author} {\bibfnamefont {D.}~\bibnamefont {Cox}},\ }\href@noop {}
  {\bibfield  {journal} {\bibinfo  {journal} {Mater. Res. Bull.}\ }\textbf
  {\bibinfo {volume} {14}},\ \bibinfo {pages} {1571} (\bibinfo {year}
  {1979})}\BibitemShut {NoStop}%
\bibitem [{\citenamefont {Perdew}\ \emph
  {et~al.}(1996{\natexlab{a}})\citenamefont {Perdew}, \citenamefont {Burke},\
  and\ \citenamefont {Ernzerhof}}]{perdew_burke_ernzerhof_1996}%
  \BibitemOpen
  \bibfield  {author} {\bibinfo {author} {\bibfnamefont {J.~P.}\ \bibnamefont
  {Perdew}}, \bibinfo {author} {\bibfnamefont {K.}~\bibnamefont {Burke}},\ and\
  \bibinfo {author} {\bibfnamefont {M.}~\bibnamefont {Ernzerhof}},\ }\href@noop
  {} {\bibfield  {journal} {\bibinfo  {journal} {Phys. Rev. Lett.}\ }\textbf
  {\bibinfo {volume} {77}},\ \bibinfo {pages} {3865} (\bibinfo {year}
  {1996}{\natexlab{a}})}\BibitemShut {NoStop}%
\bibitem [{\citenamefont {Perdew}\ \emph
  {et~al.}(1996{\natexlab{b}})\citenamefont {Perdew}, \citenamefont
  {Ernzerhof},\ and\ \citenamefont {Burke}}]{perdew_burke_hybrid_1996}%
  \BibitemOpen
  \bibfield  {author} {\bibinfo {author} {\bibfnamefont {J.~P.}\ \bibnamefont
  {Perdew}}, \bibinfo {author} {\bibfnamefont {M.}~\bibnamefont {Ernzerhof}},\
  and\ \bibinfo {author} {\bibfnamefont {K.}~\bibnamefont {Burke}},\
  }\href@noop {} {\bibfield  {journal} {\bibinfo  {journal} {J. Chem. Phys.}\
  }\textbf {\bibinfo {volume} {105}},\ \bibinfo {pages} {9982} (\bibinfo {year}
  {1996}{\natexlab{b}})}\BibitemShut {NoStop}%
\bibitem [{\citenamefont {Adamo}\ and\ \citenamefont
  {Barone}(1999)}]{adamo_barone_1999}%
  \BibitemOpen
  \bibfield  {author} {\bibinfo {author} {\bibfnamefont {C.}~\bibnamefont
  {Adamo}}\ and\ \bibinfo {author} {\bibfnamefont {V.}~\bibnamefont {Barone}},\
  }\href@noop {} {\bibfield  {journal} {\bibinfo  {journal} {J. Chem. Phys.}\
  }\textbf {\bibinfo {volume} {110}},\ \bibinfo {pages} {6158} (\bibinfo {year}
  {1999})}\BibitemShut {NoStop}%
\bibitem [{\citenamefont {Heyd}\ \emph {et~al.}(2006)\citenamefont {Heyd},
  \citenamefont {Scuseria},\ and\ \citenamefont
  {Ernzerhof}}]{heyd_ernzerhof_2006}%
  \BibitemOpen
  \bibfield  {author} {\bibinfo {author} {\bibfnamefont {J.}~\bibnamefont
  {Heyd}}, \bibinfo {author} {\bibfnamefont {G.~E.}\ \bibnamefont {Scuseria}},\
  and\ \bibinfo {author} {\bibfnamefont {M.}~\bibnamefont {Ernzerhof}},\
  }\href@noop {} {\bibfield  {journal} {\bibinfo  {journal} {J. Chem. Phys.}\
  }\textbf {\bibinfo {volume} {124}},\ \bibinfo {pages} {219906} (\bibinfo
  {year} {2006})}\BibitemShut {NoStop}%
\bibitem [{\citenamefont {Perdew}\ \emph {et~al.}(1982)\citenamefont {Perdew},
  \citenamefont {Parr}, \citenamefont {Levy},\ and\ \citenamefont
  {Balduz}}]{perdew_balduz_1982}%
  \BibitemOpen
  \bibfield  {author} {\bibinfo {author} {\bibfnamefont {J.~P.}\ \bibnamefont
  {Perdew}}, \bibinfo {author} {\bibfnamefont {R.~G.}\ \bibnamefont {Parr}},
  \bibinfo {author} {\bibfnamefont {M.}~\bibnamefont {Levy}},\ and\ \bibinfo
  {author} {\bibfnamefont {J.~L.}\ \bibnamefont {Balduz}},\ }\href@noop {}
  {\bibfield  {journal} {\bibinfo  {journal} {Phys. Rev. Lett.}\ }\textbf
  {\bibinfo {volume} {49}},\ \bibinfo {pages} {1691} (\bibinfo {year}
  {1982})}\BibitemShut {NoStop}%
\bibitem [{\citenamefont {Almbladh}\ and\ \citenamefont {von
  Barth}(1985)}]{Almbladh_von_Barth_1985}%
  \BibitemOpen
  \bibfield  {author} {\bibinfo {author} {\bibfnamefont {C.-O.}\ \bibnamefont
  {Almbladh}}\ and\ \bibinfo {author} {\bibfnamefont {U.}~\bibnamefont {von
  Barth}},\ }\href@noop {} {\bibfield  {journal} {\bibinfo  {journal} {Phys.
  Rev. B}\ }\textbf {\bibinfo {volume} {31}},\ \bibinfo {pages} {3231}
  (\bibinfo {year} {1985})}\BibitemShut {NoStop}%
\bibitem [{\citenamefont {Perdew}\ and\ \citenamefont
  {Levy}(1997)}]{perdew_levy_1997}%
  \BibitemOpen
  \bibfield  {author} {\bibinfo {author} {\bibfnamefont {J.~P.}\ \bibnamefont
  {Perdew}}\ and\ \bibinfo {author} {\bibfnamefont {M.}~\bibnamefont {Levy}},\
  }\href@noop {} {\bibfield  {journal} {\bibinfo  {journal} {Phys. Rev. B}\
  }\textbf {\bibinfo {volume} {56}},\ \bibinfo {pages} {16021} (\bibinfo {year}
  {1997})}\BibitemShut {NoStop}%
\bibitem [{\citenamefont {Levy}\ \emph {et~al.}(1984)\citenamefont {Levy},
  \citenamefont {Perdew},\ and\ \citenamefont {Sahni}}]{levy_sahni_1984}%
  \BibitemOpen
  \bibfield  {author} {\bibinfo {author} {\bibfnamefont {M.}~\bibnamefont
  {Levy}}, \bibinfo {author} {\bibfnamefont {J.~P.}\ \bibnamefont {Perdew}},\
  and\ \bibinfo {author} {\bibfnamefont {V.}~\bibnamefont {Sahni}},\
  }\href@noop {} {\bibfield  {journal} {\bibinfo  {journal} {Phys. Rev. A}\
  }\textbf {\bibinfo {volume} {30}},\ \bibinfo {pages} {2745} (\bibinfo {year}
  {1984})}\BibitemShut {NoStop}%
\bibitem [{\citenamefont {Stein}\ \emph {et~al.}(2010)\citenamefont {Stein},
  \citenamefont {Eisenberg}, \citenamefont {Kronik},\ and\ \citenamefont
  {Baer}}]{stein_baer_2010}%
  \BibitemOpen
  \bibfield  {author} {\bibinfo {author} {\bibfnamefont {T.}~\bibnamefont
  {Stein}}, \bibinfo {author} {\bibfnamefont {H.}~\bibnamefont {Eisenberg}},
  \bibinfo {author} {\bibfnamefont {L.}~\bibnamefont {Kronik}},\ and\ \bibinfo
  {author} {\bibfnamefont {R.}~\bibnamefont {Baer}},\ }\href@noop {} {\bibfield
   {journal} {\bibinfo  {journal} {Phys. Rev. Lett.}\ }\textbf {\bibinfo
  {volume} {105}},\ \bibinfo {pages} {266802} (\bibinfo {year}
  {2010})}\BibitemShut {NoStop}%
\bibitem [{\citenamefont {Kronik}\ \emph {et~al.}(2012)\citenamefont {Kronik},
  \citenamefont {Stein}, \citenamefont {Refaely-Abramson},\ and\ \citenamefont
  {Baer}}]{kronik_stein_refaely-abramson_baer_2012}%
  \BibitemOpen
  \bibfield  {author} {\bibinfo {author} {\bibfnamefont {L.}~\bibnamefont
  {Kronik}}, \bibinfo {author} {\bibfnamefont {T.}~\bibnamefont {Stein}},
  \bibinfo {author} {\bibfnamefont {S.}~\bibnamefont {Refaely-Abramson}},\ and\
  \bibinfo {author} {\bibfnamefont {R.}~\bibnamefont {Baer}},\ }\href@noop {}
  {\bibfield  {journal} {\bibinfo  {journal} {J. Chem. Theory Comp.}\ }\textbf
  {\bibinfo {volume} {8}},\ \bibinfo {pages} {1515} (\bibinfo {year}
  {2012})}\BibitemShut {NoStop}%
\bibitem [{\citenamefont {Refaely-Abramson}\ \emph {et~al.}(2011)\citenamefont
  {Refaely-Abramson}, \citenamefont {Baer},\ and\ \citenamefont
  {Kronik}}]{refaely_kronik_2011}%
  \BibitemOpen
  \bibfield  {author} {\bibinfo {author} {\bibfnamefont {S.}~\bibnamefont
  {Refaely-Abramson}}, \bibinfo {author} {\bibfnamefont {R.}~\bibnamefont
  {Baer}},\ and\ \bibinfo {author} {\bibfnamefont {L.}~\bibnamefont {Kronik}},\
  }\href@noop {} {\bibfield  {journal} {\bibinfo  {journal} {Phys. Rev. B}\
  }\textbf {\bibinfo {volume} {84}},\ \bibinfo {pages} {075144} (\bibinfo
  {year} {2011})}\BibitemShut {NoStop}%
\bibitem [{\citenamefont {Autschbach}\ and\ \citenamefont
  {Srebro}(2014)}]{autschbach_srebro_2014}%
  \BibitemOpen
  \bibfield  {author} {\bibinfo {author} {\bibfnamefont {J.}~\bibnamefont
  {Autschbach}}\ and\ \bibinfo {author} {\bibfnamefont {M.}~\bibnamefont
  {Srebro}},\ }\href@noop {} {\bibfield  {journal} {\bibinfo  {journal} {Acc.
  Chem. Res.}\ }\textbf {\bibinfo {volume} {47}},\ \bibinfo {pages} {2592}
  (\bibinfo {year} {2014})}\BibitemShut {NoStop}%
\bibitem [{\citenamefont {Phillips}\ \emph {et~al.}(2014)\citenamefont
  {Phillips}, \citenamefont {Zheng}, \citenamefont {Geva},\ and\ \citenamefont
  {Dunietz}}]{phillips_dunietz_2014}%
  \BibitemOpen
  \bibfield  {author} {\bibinfo {author} {\bibfnamefont {H.}~\bibnamefont
  {Phillips}}, \bibinfo {author} {\bibfnamefont {Z.}~\bibnamefont {Zheng}},
  \bibinfo {author} {\bibfnamefont {E.}~\bibnamefont {Geva}},\ and\ \bibinfo
  {author} {\bibfnamefont {B.~D.}\ \bibnamefont {Dunietz}},\ }\href@noop {}
  {\bibfield  {journal} {\bibinfo  {journal} {Org. Electron.}\ }\textbf
  {\bibinfo {volume} {15}},\ \bibinfo {pages} {1509 } (\bibinfo {year}
  {2014})}\BibitemShut {NoStop}%
\bibitem [{\citenamefont {Foster}\ \emph {et~al.}(2014)\citenamefont {Foster},
  \citenamefont {Azoulay}, \citenamefont {Wong},\ and\ \citenamefont
  {Allendorf}}]{foster_allendorf_2014}%
  \BibitemOpen
  \bibfield  {author} {\bibinfo {author} {\bibfnamefont {M.~E.}\ \bibnamefont
  {Foster}}, \bibinfo {author} {\bibfnamefont {J.~D.}\ \bibnamefont {Azoulay}},
  \bibinfo {author} {\bibfnamefont {B.~M.}\ \bibnamefont {Wong}},\ and\
  \bibinfo {author} {\bibfnamefont {M.~D.}\ \bibnamefont {Allendorf}},\
  }\href@noop {} {\bibfield  {journal} {\bibinfo  {journal} {Chem. Sci.}\
  }\textbf {\bibinfo {volume} {5}},\ \bibinfo {pages} {2081} (\bibinfo {year}
  {2014})}\BibitemShut {NoStop}%
\bibitem [{\citenamefont {K\"orzd\"orfer}\ and\ \citenamefont
  {Br\'{e}das}(2014)}]{korzdorfer_bredas_2014}%
  \BibitemOpen
  \bibfield  {author} {\bibinfo {author} {\bibfnamefont {T.}~\bibnamefont
  {K\"orzd\"orfer}}\ and\ \bibinfo {author} {\bibfnamefont {J.-L.}\
  \bibnamefont {Br\'{e}das}},\ }\href@noop {} {\bibfield  {journal} {\bibinfo
  {journal} {Acc. Chem. Res.}\ }\textbf {\bibinfo {volume} {47}},\ \bibinfo
  {pages} {3284} (\bibinfo {year} {2014})}\BibitemShut {NoStop}%
\bibitem [{\citenamefont {Faber}\ \emph {et~al.}(2014)\citenamefont {Faber},
  \citenamefont {Boulanger}, \citenamefont {Attaccalite}, \citenamefont
  {Duchemin},\ and\ \citenamefont {Blase}}]{faber2014excited}%
  \BibitemOpen
  \bibfield  {author} {\bibinfo {author} {\bibfnamefont {C.}~\bibnamefont
  {Faber}}, \bibinfo {author} {\bibfnamefont {P.}~\bibnamefont {Boulanger}},
  \bibinfo {author} {\bibfnamefont {C.}~\bibnamefont {Attaccalite}}, \bibinfo
  {author} {\bibfnamefont {I.}~\bibnamefont {Duchemin}},\ and\ \bibinfo
  {author} {\bibfnamefont {X.}~\bibnamefont {Blase}},\ }\href@noop {}
  {\bibfield  {journal} {\bibinfo  {journal} {Philos. Trans. R. Soc. A}\
  }\textbf {\bibinfo {volume} {372}},\ \bibinfo {pages} {20130271} (\bibinfo
  {year} {2014})}\BibitemShut {NoStop}%
\bibitem [{\citenamefont {Kraisler}\ and\ \citenamefont
  {Kronik}(2014)}]{kraisler_kronik_2014}%
  \BibitemOpen
  \bibfield  {author} {\bibinfo {author} {\bibfnamefont {E.}~\bibnamefont
  {Kraisler}}\ and\ \bibinfo {author} {\bibfnamefont {L.}~\bibnamefont
  {Kronik}},\ }\href@noop {} {\bibfield  {journal} {\bibinfo  {journal} {J.
  Chem. Phys.}\ }\textbf {\bibinfo {volume} {140}},\ \bibinfo {pages} {18A540}
  (\bibinfo {year} {2014})}\BibitemShut {NoStop}%
\bibitem [{\citenamefont {Vl{\v{c}}ek}\ \emph {et~al.}(2015)\citenamefont
  {Vl{\v{c}}ek}, \citenamefont {Eisenberg}, \citenamefont {Steinle-Neumann},
  \citenamefont {Kronik},\ and\ \citenamefont
  {Baer}}]{vlcek_eisenberg_steinle-neumann_baer_2015}%
  \BibitemOpen
  \bibfield  {author} {\bibinfo {author} {\bibfnamefont {V.}~\bibnamefont
  {Vl{\v{c}}ek}}, \bibinfo {author} {\bibfnamefont {H.~R.}\ \bibnamefont
  {Eisenberg}}, \bibinfo {author} {\bibfnamefont {G.}~\bibnamefont
  {Steinle-Neumann}}, \bibinfo {author} {\bibfnamefont {L.}~\bibnamefont
  {Kronik}},\ and\ \bibinfo {author} {\bibfnamefont {R.}~\bibnamefont {Baer}},\
  }\href@noop {} {\bibfield  {journal} {\bibinfo  {journal} {J. Chem. Phys.}\
  }\textbf {\bibinfo {volume} {142}},\ \bibinfo {eid} {034107} (\bibinfo {year}
  {2015})}\BibitemShut {NoStop}%
\bibitem [{\citenamefont {G\"orling}(2015)}]{gorling_2015}%
  \BibitemOpen
  \bibfield  {author} {\bibinfo {author} {\bibfnamefont {A.}~\bibnamefont
  {G\"orling}},\ }\href@noop {} {\bibfield  {journal} {\bibinfo  {journal}
  {Phys. Rev. B}\ }\textbf {\bibinfo {volume} {91}},\ \bibinfo {pages} {245120}
  (\bibinfo {year} {2015})}\BibitemShut {NoStop}%
\bibitem [{\citenamefont {Anisimov}\ and\ \citenamefont
  {Kozhevnikov}(2005)}]{anisimov_kozhevnikov_2005}%
  \BibitemOpen
  \bibfield  {author} {\bibinfo {author} {\bibfnamefont {V.~I.}\ \bibnamefont
  {Anisimov}}\ and\ \bibinfo {author} {\bibfnamefont {A.~V.}\ \bibnamefont
  {Kozhevnikov}},\ }\href@noop {} {\bibfield  {journal} {\bibinfo  {journal}
  {Phys. Rev. B}\ }\textbf {\bibinfo {volume} {72}},\ \bibinfo {pages} {075125}
  (\bibinfo {year} {2005})}\BibitemShut {NoStop}%
\bibitem [{\citenamefont {Weng}\ \emph {et~al.}(2017)\citenamefont {Weng},
  \citenamefont {Li}, \citenamefont {Ma}, \citenamefont {Zheng}, \citenamefont
  {Pan},\ and\ \citenamefont {Wang}}]{weng_wang_2017}%
  \BibitemOpen
  \bibfield  {author} {\bibinfo {author} {\bibfnamefont {M.}~\bibnamefont
  {Weng}}, \bibinfo {author} {\bibfnamefont {S.}~\bibnamefont {Li}}, \bibinfo
  {author} {\bibfnamefont {J.}~\bibnamefont {Ma}}, \bibinfo {author}
  {\bibfnamefont {J.}~\bibnamefont {Zheng}}, \bibinfo {author} {\bibfnamefont
  {F.}~\bibnamefont {Pan}},\ and\ \bibinfo {author} {\bibfnamefont {L.-W.}\
  \bibnamefont {Wang}},\ }\href@noop {} {\bibfield  {journal} {\bibinfo
  {journal} {Appl. Phys. Lett.}\ }\textbf {\bibinfo {volume} {111}},\ \bibinfo
  {pages} {054101} (\bibinfo {year} {2017})}\BibitemShut {NoStop}%
\bibitem [{\citenamefont {Li}\ \emph {et~al.}(2017)\citenamefont {Li},
  \citenamefont {Zheng}, \citenamefont {Su},\ and\ \citenamefont
  {Yang}}]{li_yang_2017}%
  \BibitemOpen
  \bibfield  {author} {\bibinfo {author} {\bibfnamefont {C.}~\bibnamefont
  {Li}}, \bibinfo {author} {\bibfnamefont {X.}~\bibnamefont {Zheng}}, \bibinfo
  {author} {\bibfnamefont {N.~Q.}\ \bibnamefont {Su}},\ and\ \bibinfo {author}
  {\bibfnamefont {W.}~\bibnamefont {Yang}},\ }\href@noop {} {\bibfield
  {journal} {\bibinfo  {journal} {Natl. Sci. Rev.}\ }\textbf {\bibinfo {volume}
  {5}},\ \bibinfo {pages} {203} (\bibinfo {year} {2017})}\BibitemShut {NoStop}%
\bibitem [{\citenamefont {Miceli}\ \emph {et~al.}(2018)\citenamefont {Miceli},
  \citenamefont {Chen}, \citenamefont {Reshetnyak},\ and\ \citenamefont
  {Pasquarello}}]{miceli_pasquarello_2018}%
  \BibitemOpen
  \bibfield  {author} {\bibinfo {author} {\bibfnamefont {G.}~\bibnamefont
  {Miceli}}, \bibinfo {author} {\bibfnamefont {W.}~\bibnamefont {Chen}},
  \bibinfo {author} {\bibfnamefont {I.}~\bibnamefont {Reshetnyak}},\ and\
  \bibinfo {author} {\bibfnamefont {A.}~\bibnamefont {Pasquarello}},\
  }\href@noop {} {\bibfield  {journal} {\bibinfo  {journal} {Phys. Rev. B}\
  }\textbf {\bibinfo {volume} {97}},\ \bibinfo {pages} {121112} (\bibinfo
  {year} {2018})}\BibitemShut {NoStop}%
\bibitem [{\citenamefont {Nguyen}\ \emph {et~al.}(2018)\citenamefont {Nguyen},
  \citenamefont {Colonna}, \citenamefont {Ferretti},\ and\ \citenamefont
  {Marzari}}]{nguyen_marzari_2018}%
  \BibitemOpen
  \bibfield  {author} {\bibinfo {author} {\bibfnamefont {N.~L.}\ \bibnamefont
  {Nguyen}}, \bibinfo {author} {\bibfnamefont {N.}~\bibnamefont {Colonna}},
  \bibinfo {author} {\bibfnamefont {A.}~\bibnamefont {Ferretti}},\ and\
  \bibinfo {author} {\bibfnamefont {N.}~\bibnamefont {Marzari}},\ }\href@noop
  {} {\bibfield  {journal} {\bibinfo  {journal} {Phys. Rev. X}\ }\textbf
  {\bibinfo {volume} {8}},\ \bibinfo {pages} {021051} (\bibinfo {year}
  {2018})}\BibitemShut {NoStop}%
\bibitem [{\citenamefont {Bischoff}\ \emph
  {et~al.}(2019{\natexlab{a}})\citenamefont {Bischoff}, \citenamefont
  {Reshetnyak},\ and\ \citenamefont {Pasquarello}}]{bischoff_pasquarello_2019}%
  \BibitemOpen
  \bibfield  {author} {\bibinfo {author} {\bibfnamefont {T.}~\bibnamefont
  {Bischoff}}, \bibinfo {author} {\bibfnamefont {I.}~\bibnamefont
  {Reshetnyak}},\ and\ \bibinfo {author} {\bibfnamefont {A.}~\bibnamefont
  {Pasquarello}},\ }\href@noop {} {\bibfield  {journal} {\bibinfo  {journal}
  {Phys. Rev. B}\ }\textbf {\bibinfo {volume} {99}},\ \bibinfo {pages} {201114}
  (\bibinfo {year} {2019}{\natexlab{a}})}\BibitemShut {NoStop}%
\bibitem [{\citenamefont {Bischoff}\ \emph
  {et~al.}(2019{\natexlab{b}})\citenamefont {Bischoff}, \citenamefont {Wiktor},
  \citenamefont {Chen},\ and\ \citenamefont
  {Pasquarello}}]{bischoff_pasquarello_2019_perovskites}%
  \BibitemOpen
  \bibfield  {author} {\bibinfo {author} {\bibfnamefont {T.}~\bibnamefont
  {Bischoff}}, \bibinfo {author} {\bibfnamefont {J.}~\bibnamefont {Wiktor}},
  \bibinfo {author} {\bibfnamefont {W.}~\bibnamefont {Chen}},\ and\ \bibinfo
  {author} {\bibfnamefont {A.}~\bibnamefont {Pasquarello}},\ }\href@noop {}
  {\bibfield  {journal} {\bibinfo  {journal} {Phys. Rev. Mater.}\ }\textbf
  {\bibinfo {volume} {3}},\ \bibinfo {pages} {123802} (\bibinfo {year}
  {2019}{\natexlab{b}})}\BibitemShut {NoStop}%
\bibitem [{\citenamefont {Elliott}\ \emph {et~al.}(2019)\citenamefont
  {Elliott}, \citenamefont {Colonna}, \citenamefont {Marsili}, \citenamefont
  {Marzari},\ and\ \citenamefont {Umari}}]{elliott2019koopmans}%
  \BibitemOpen
  \bibfield  {author} {\bibinfo {author} {\bibfnamefont {J.~D.}\ \bibnamefont
  {Elliott}}, \bibinfo {author} {\bibfnamefont {N.}~\bibnamefont {Colonna}},
  \bibinfo {author} {\bibfnamefont {M.}~\bibnamefont {Marsili}}, \bibinfo
  {author} {\bibfnamefont {N.}~\bibnamefont {Marzari}},\ and\ \bibinfo {author}
  {\bibfnamefont {P.}~\bibnamefont {Umari}},\ }\href@noop {} {\bibfield
  {journal} {\bibinfo  {journal} {J. Chem. Theory Comput.}\ }\textbf {\bibinfo
  {volume} {15}},\ \bibinfo {pages} {3710} (\bibinfo {year}
  {2019})}\BibitemShut {NoStop}%
\bibitem [{\citenamefont {Su}\ \emph {et~al.}(2020)\citenamefont {Su},
  \citenamefont {Mahler},\ and\ \citenamefont {Yang}}]{su2020preserving}%
  \BibitemOpen
  \bibfield  {author} {\bibinfo {author} {\bibfnamefont {N.~Q.}\ \bibnamefont
  {Su}}, \bibinfo {author} {\bibfnamefont {A.}~\bibnamefont {Mahler}},\ and\
  \bibinfo {author} {\bibfnamefont {W.}~\bibnamefont {Yang}},\ }\href@noop {}
  {\bibfield  {journal} {\bibinfo  {journal} {J. Phys. Chem. Lett.}\ }\textbf
  {\bibinfo {volume} {11}},\ \bibinfo {pages} {1528} (\bibinfo {year}
  {2020})}\BibitemShut {NoStop}%
\bibitem [{\citenamefont {Bischoff}\ \emph {et~al.}(2021)\citenamefont
  {Bischoff}, \citenamefont {Reshetnyak},\ and\ \citenamefont
  {Pasquarello}}]{bischoff2021band}%
  \BibitemOpen
  \bibfield  {author} {\bibinfo {author} {\bibfnamefont {T.}~\bibnamefont
  {Bischoff}}, \bibinfo {author} {\bibfnamefont {I.}~\bibnamefont
  {Reshetnyak}},\ and\ \bibinfo {author} {\bibfnamefont {A.}~\bibnamefont
  {Pasquarello}},\ }\href@noop {} {\bibfield  {journal} {\bibinfo  {journal}
  {Phys. Rev. Res.}\ }\textbf {\bibinfo {volume} {3}},\ \bibinfo {pages}
  {023182} (\bibinfo {year} {2021})}\BibitemShut {NoStop}%
\bibitem [{\citenamefont {Colonna}\ \emph {et~al.}(2022)\citenamefont
  {Colonna}, \citenamefont {De~Gennaro}, \citenamefont {Linscott},\ and\
  \citenamefont {Marzari}}]{colonna2022koopmans}%
  \BibitemOpen
  \bibfield  {author} {\bibinfo {author} {\bibfnamefont {N.}~\bibnamefont
  {Colonna}}, \bibinfo {author} {\bibfnamefont {R.}~\bibnamefont {De~Gennaro}},
  \bibinfo {author} {\bibfnamefont {E.}~\bibnamefont {Linscott}},\ and\
  \bibinfo {author} {\bibfnamefont {N.}~\bibnamefont {Marzari}},\ }\href@noop
  {} {\bibfield  {journal} {\bibinfo  {journal} {J. Chem. Theory Comput.}\
  }\textbf {\bibinfo {volume} {18}},\ \bibinfo {pages} {5435} (\bibinfo {year}
  {2022})}\BibitemShut {NoStop}%
\bibitem [{\citenamefont {Mahler}\ \emph {et~al.}(2022)\citenamefont {Mahler},
  \citenamefont {Williams}, \citenamefont {Su},\ and\ \citenamefont
  {Yang}}]{mahler2022localized}%
  \BibitemOpen
  \bibfield  {author} {\bibinfo {author} {\bibfnamefont {A.}~\bibnamefont
  {Mahler}}, \bibinfo {author} {\bibfnamefont {J.}~\bibnamefont {Williams}},
  \bibinfo {author} {\bibfnamefont {N.~Q.}\ \bibnamefont {Su}},\ and\ \bibinfo
  {author} {\bibfnamefont {W.}~\bibnamefont {Yang}},\ }\href@noop {} {\bibfield
   {journal} {\bibinfo  {journal} {Phys. Rev. B}\ }\textbf {\bibinfo {volume}
  {106}},\ \bibinfo {pages} {035147} (\bibinfo {year} {2022})}\BibitemShut
  {NoStop}%
\bibitem [{\citenamefont {Yang}\ \emph {et~al.}(2022)\citenamefont {Yang},
  \citenamefont {Falletta},\ and\ \citenamefont {Pasquarello}}]{yang2022one}%
  \BibitemOpen
  \bibfield  {author} {\bibinfo {author} {\bibfnamefont {J.}~\bibnamefont
  {Yang}}, \bibinfo {author} {\bibfnamefont {S.}~\bibnamefont {Falletta}},\
  and\ \bibinfo {author} {\bibfnamefont {A.}~\bibnamefont {Pasquarello}},\
  }\href@noop {} {\bibfield  {journal} {\bibinfo  {journal} {J. Phys. Chem.
  Lett.}\ }\textbf {\bibinfo {volume} {13}},\ \bibinfo {pages} {3066} (\bibinfo
  {year} {2022})}\BibitemShut {NoStop}%
\bibitem [{\citenamefont {De~Gennaro}\ \emph {et~al.}(2022)\citenamefont
  {De~Gennaro}, \citenamefont {Colonna}, \citenamefont {Linscott},\ and\
  \citenamefont {Marzari}}]{degennaro2022bloch}%
  \BibitemOpen
  \bibfield  {author} {\bibinfo {author} {\bibfnamefont {R.}~\bibnamefont
  {De~Gennaro}}, \bibinfo {author} {\bibfnamefont {N.}~\bibnamefont {Colonna}},
  \bibinfo {author} {\bibfnamefont {E.}~\bibnamefont {Linscott}},\ and\
  \bibinfo {author} {\bibfnamefont {N.}~\bibnamefont {Marzari}},\ }\href@noop
  {} {\bibfield  {journal} {\bibinfo  {journal} {Phys. Rev. B}\ }\textbf
  {\bibinfo {volume} {106}},\ \bibinfo {pages} {035106} (\bibinfo {year}
  {2022})}\BibitemShut {NoStop}%
\bibitem [{\citenamefont {Linscott}\ \emph {et~al.}(2023)\citenamefont
  {Linscott}, \citenamefont {Colonna}, \citenamefont {De~Gennaro},
  \citenamefont {Nguyen}, \citenamefont {Borghi}, \citenamefont {Ferretti},
  \citenamefont {Dabo},\ and\ \citenamefont {Marzari}}]{linscott2023koopmans}%
  \BibitemOpen
  \bibfield  {author} {\bibinfo {author} {\bibfnamefont {E.~B.}\ \bibnamefont
  {Linscott}}, \bibinfo {author} {\bibfnamefont {N.}~\bibnamefont {Colonna}},
  \bibinfo {author} {\bibfnamefont {R.}~\bibnamefont {De~Gennaro}}, \bibinfo
  {author} {\bibfnamefont {N.~L.}\ \bibnamefont {Nguyen}}, \bibinfo {author}
  {\bibfnamefont {G.}~\bibnamefont {Borghi}}, \bibinfo {author} {\bibfnamefont
  {A.}~\bibnamefont {Ferretti}}, \bibinfo {author} {\bibfnamefont
  {I.}~\bibnamefont {Dabo}},\ and\ \bibinfo {author} {\bibfnamefont
  {N.}~\bibnamefont {Marzari}},\ }\href@noop {} {\bibfield  {journal} {\bibinfo
   {journal} {J. Chem. Theory Comput.}\ }\textbf {\bibinfo {volume} {xx}},\
  \bibinfo {pages} {xx} (\bibinfo {year} {2023})}\BibitemShut {NoStop}%
\bibitem [{SM()}]{SM}%
  \BibitemOpen
  \href@noop {} {\ }\bibinfo {note} {See Supplemental Material at [URL will be
  inserted by publisher] for more computational details.}\BibitemShut {Stop}%
\bibitem [{\citenamefont {Hirata}\ and\ \citenamefont
  {Head-Gordon}(1999)}]{hirata1999time}%
  \BibitemOpen
  \bibfield  {author} {\bibinfo {author} {\bibfnamefont {S.}~\bibnamefont
  {Hirata}}\ and\ \bibinfo {author} {\bibfnamefont {M.}~\bibnamefont
  {Head-Gordon}},\ }\href@noop {} {\bibfield  {journal} {\bibinfo  {journal}
  {Chem. Phys. Lett.}\ }\textbf {\bibinfo {volume} {314}},\ \bibinfo {pages}
  {291} (\bibinfo {year} {1999})}\BibitemShut {NoStop}%
\bibitem [{\citenamefont {Sun}\ \emph {et~al.}(2021)\citenamefont {Sun},
  \citenamefont {Li},\ and\ \citenamefont {Jiang}}]{sun2021pros}%
  \BibitemOpen
  \bibfield  {author} {\bibinfo {author} {\bibfnamefont {H.-Y.}\ \bibnamefont
  {Sun}}, \bibinfo {author} {\bibfnamefont {S.-X.}\ \bibnamefont {Li}},\ and\
  \bibinfo {author} {\bibfnamefont {H.}~\bibnamefont {Jiang}},\ }\href@noop {}
  {\bibfield  {journal} {\bibinfo  {journal} {Phys. Chem. Chem. Phys.}\
  }\textbf {\bibinfo {volume} {23}},\ \bibinfo {pages} {16296} (\bibinfo {year}
  {2021})}\BibitemShut {NoStop}%
\bibitem [{\citenamefont {Luo}\ \emph {et~al.}(2002)\citenamefont {Luo},
  \citenamefont {{Ismail-Beigi}}, \citenamefont {Cohen},\ and\ \citenamefont
  {Louie}}]{luoQuasiparticleBandStructure2002a}%
  \BibitemOpen
  \bibfield  {author} {\bibinfo {author} {\bibfnamefont {W.}~\bibnamefont
  {Luo}}, \bibinfo {author} {\bibfnamefont {S.}~\bibnamefont {{Ismail-Beigi}}},
  \bibinfo {author} {\bibfnamefont {M.~L.}\ \bibnamefont {Cohen}},\ and\
  \bibinfo {author} {\bibfnamefont {S.~G.}\ \bibnamefont {Louie}},\ }\href
  {https://doi.org/10.1103/PhysRevB.66.195215} {\bibfield  {journal} {\bibinfo
  {journal} {Phys. Rev. B}\ }\textbf {\bibinfo {volume} {66}},\ \bibinfo
  {pages} {195215} (\bibinfo {year} {2002})}\BibitemShut {NoStop}%
\bibitem [{\citenamefont {Faleev}\ \emph {et~al.}(2004)\citenamefont {Faleev},
  \citenamefont {{van Schilfgaarde}},\ and\ \citenamefont
  {Kotani}}]{faleevAllElectronSelfConsistentGW2004}%
  \BibitemOpen
  \bibfield  {author} {\bibinfo {author} {\bibfnamefont {S.~V.}\ \bibnamefont
  {Faleev}}, \bibinfo {author} {\bibfnamefont {M.}~\bibnamefont {{van
  Schilfgaarde}}},\ and\ \bibinfo {author} {\bibfnamefont {T.}~\bibnamefont
  {Kotani}},\ }\href {https://doi.org/10.1103/PhysRevLett.93.126406} {\bibfield
   {journal} {\bibinfo  {journal} {Phys. Rev. Lett.}\ }\textbf {\bibinfo
  {volume} {93}},\ \bibinfo {pages} {126406} (\bibinfo {year}
  {2004})}\BibitemShut {NoStop}%
\bibitem [{\citenamefont {Kotani}\ \emph {et~al.}(2007)\citenamefont {Kotani},
  \citenamefont {{van Schilfgaarde}},\ and\ \citenamefont
  {Faleev}}]{kotaniQuasiparticleSelfconsistentGW2007}%
  \BibitemOpen
  \bibfield  {author} {\bibinfo {author} {\bibfnamefont {T.}~\bibnamefont
  {Kotani}}, \bibinfo {author} {\bibfnamefont {M.}~\bibnamefont {{van
  Schilfgaarde}}},\ and\ \bibinfo {author} {\bibfnamefont {S.~V.}\ \bibnamefont
  {Faleev}},\ }\href {https://doi.org/10.1103/PhysRevB.76.165106} {\bibfield
  {journal} {\bibinfo  {journal} {Phys. Rev. B}\ }\textbf {\bibinfo {volume}
  {76}},\ \bibinfo {pages} {165106} (\bibinfo {year} {2007})}\BibitemShut
  {NoStop}%
\bibitem [{\citenamefont {Hybertsen}\ and\ \citenamefont
  {Louie}(1985)}]{hybertsenFirstPrinciplesTheoryQuasiparticles1985}%
  \BibitemOpen
  \bibfield  {author} {\bibinfo {author} {\bibfnamefont {M.~S.}\ \bibnamefont
  {Hybertsen}}\ and\ \bibinfo {author} {\bibfnamefont {S.~G.}\ \bibnamefont
  {Louie}},\ }\href {https://doi.org/10.1103/PhysRevLett.55.1418} {\bibfield
  {journal} {\bibinfo  {journal} {Phys. Rev. Lett.}\ }\textbf {\bibinfo
  {volume} {55}},\ \bibinfo {pages} {1418} (\bibinfo {year}
  {1985})}\BibitemShut {NoStop}%
\bibitem [{\citenamefont {Adler}(1962)}]{adlerQuantumTheoryDielectric1962}%
  \BibitemOpen
  \bibfield  {author} {\bibinfo {author} {\bibfnamefont {S.~L.}\ \bibnamefont
  {Adler}},\ }\href {https://doi.org/10.1103/PhysRev.126.413} {\bibfield
  {journal} {\bibinfo  {journal} {Phys. Rev.}\ }\textbf {\bibinfo {volume}
  {126}},\ \bibinfo {pages} {413} (\bibinfo {year} {1962})}\BibitemShut
  {NoStop}%
\bibitem [{\citenamefont {Wiser}(1963)}]{wiserDielectricConstantLocal1963}%
  \BibitemOpen
  \bibfield  {author} {\bibinfo {author} {\bibfnamefont {N.}~\bibnamefont
  {Wiser}},\ }\href {https://doi.org/10.1103/PhysRev.129.62} {\bibfield
  {journal} {\bibinfo  {journal} {Phys. Rev.}\ }\textbf {\bibinfo {volume}
  {129}},\ \bibinfo {pages} {62} (\bibinfo {year} {1963})}\BibitemShut
  {NoStop}%
\bibitem [{\citenamefont {Godby}\ \emph {et~al.}(1988)\citenamefont {Godby},
  \citenamefont {Schl{\"u}ter},\ and\ \citenamefont
  {Sham}}]{godbySelfenergyOperatorsExchangecorrelation1988}%
  \BibitemOpen
  \bibfield  {author} {\bibinfo {author} {\bibfnamefont {R.~W.}\ \bibnamefont
  {Godby}}, \bibinfo {author} {\bibfnamefont {M.}~\bibnamefont
  {Schl{\"u}ter}},\ and\ \bibinfo {author} {\bibfnamefont {L.~J.}\ \bibnamefont
  {Sham}},\ }\href {https://doi.org/10.1103/PhysRevB.37.10159} {\bibfield
  {journal} {\bibinfo  {journal} {Phys. Rev. B}\ }\textbf {\bibinfo {volume}
  {37}},\ \bibinfo {pages} {10159} (\bibinfo {year} {1988})}\BibitemShut
  {NoStop}%
\bibitem [{\citenamefont {Govoni}\ and\ \citenamefont
  {Galli}(2015)}]{govoniLargeScaleGW2015a}%
  \BibitemOpen
  \bibfield  {author} {\bibinfo {author} {\bibfnamefont {M.}~\bibnamefont
  {Govoni}}\ and\ \bibinfo {author} {\bibfnamefont {G.}~\bibnamefont {Galli}},\
  }\href {https://doi.org/10.1021/ct500958p} {\bibfield  {journal} {\bibinfo
  {journal} {J. Chem. Theory Comput.}\ }\textbf {\bibinfo {volume} {11}},\
  \bibinfo {pages} {2680} (\bibinfo {year} {2015})}\BibitemShut {NoStop}%
\bibitem [{\citenamefont {Nguyen}\ \emph {et~al.}(2012)\citenamefont {Nguyen},
  \citenamefont {Pham}, \citenamefont {Rocca},\ and\ \citenamefont
  {Galli}}]{nguyenImprovingAccuracyEfficiency2012a}%
  \BibitemOpen
  \bibfield  {author} {\bibinfo {author} {\bibfnamefont {H.-V.}\ \bibnamefont
  {Nguyen}}, \bibinfo {author} {\bibfnamefont {T.~A.}\ \bibnamefont {Pham}},
  \bibinfo {author} {\bibfnamefont {D.}~\bibnamefont {Rocca}},\ and\ \bibinfo
  {author} {\bibfnamefont {G.}~\bibnamefont {Galli}},\ }\href
  {https://doi.org/10.1103/PhysRevB.85.081101} {\bibfield  {journal} {\bibinfo
  {journal} {Phys. Rev. B}\ }\textbf {\bibinfo {volume} {85}},\ \bibinfo
  {pages} {081101} (\bibinfo {year} {2012})}\BibitemShut {NoStop}%
\bibitem [{\citenamefont {Pham}\ \emph {et~al.}(2013)\citenamefont {Pham},
  \citenamefont {Nguyen}, \citenamefont {Rocca},\ and\ \citenamefont
  {Galli}}]{phamGWCalculationsUsing2013}%
  \BibitemOpen
  \bibfield  {author} {\bibinfo {author} {\bibfnamefont {T.~A.}\ \bibnamefont
  {Pham}}, \bibinfo {author} {\bibfnamefont {H.-V.}\ \bibnamefont {Nguyen}},
  \bibinfo {author} {\bibfnamefont {D.}~\bibnamefont {Rocca}},\ and\ \bibinfo
  {author} {\bibfnamefont {G.}~\bibnamefont {Galli}},\ }\href
  {https://doi.org/10.1103/PhysRevB.87.155148} {\bibfield  {journal} {\bibinfo
  {journal} {Phys. Rev. B}\ }\textbf {\bibinfo {volume} {87}},\ \bibinfo
  {pages} {155148} (\bibinfo {year} {2013})}\BibitemShut {NoStop}%
\bibitem [{\citenamefont {Wilson}\ \emph {et~al.}(2008)\citenamefont {Wilson},
  \citenamefont {Gygi},\ and\ \citenamefont
  {Galli}}]{wilsonEfficientIterativeMethod2008}%
  \BibitemOpen
  \bibfield  {author} {\bibinfo {author} {\bibfnamefont {H.~F.}\ \bibnamefont
  {Wilson}}, \bibinfo {author} {\bibfnamefont {F.}~\bibnamefont {Gygi}},\ and\
  \bibinfo {author} {\bibfnamefont {G.}~\bibnamefont {Galli}},\ }\href
  {https://doi.org/10.1103/PhysRevB.78.113303} {\bibfield  {journal} {\bibinfo
  {journal} {Phys. Rev. B}\ }\textbf {\bibinfo {volume} {78}},\ \bibinfo
  {pages} {113303} (\bibinfo {year} {2008})}\BibitemShut {NoStop}%
\bibitem [{\citenamefont {Wilson}\ \emph {et~al.}(2009)\citenamefont {Wilson},
  \citenamefont {Lu}, \citenamefont {Gygi},\ and\ \citenamefont
  {Galli}}]{wilsonIterativeCalculationsDielectric2009}%
  \BibitemOpen
  \bibfield  {author} {\bibinfo {author} {\bibfnamefont {H.~F.}\ \bibnamefont
  {Wilson}}, \bibinfo {author} {\bibfnamefont {D.}~\bibnamefont {Lu}}, \bibinfo
  {author} {\bibfnamefont {F.}~\bibnamefont {Gygi}},\ and\ \bibinfo {author}
  {\bibfnamefont {G.}~\bibnamefont {Galli}},\ }\href
  {https://doi.org/10.1103/PhysRevB.79.245106} {\bibfield  {journal} {\bibinfo
  {journal} {Phys. Rev. B}\ }\textbf {\bibinfo {volume} {79}},\ \bibinfo
  {pages} {245106} (\bibinfo {year} {2009})}\BibitemShut {NoStop}%
\bibitem [{\citenamefont {Del~Ben}\ \emph {et~al.}(2019)\citenamefont
  {Del~Ben}, \citenamefont {{da Jornada}}, \citenamefont {Antonius},
  \citenamefont {Rangel}, \citenamefont {Louie}, \citenamefont {Deslippe},\
  and\ \citenamefont {Canning}}]{delbenStaticSubspaceApproximation2019b}%
  \BibitemOpen
  \bibfield  {author} {\bibinfo {author} {\bibfnamefont {M.}~\bibnamefont
  {Del~Ben}}, \bibinfo {author} {\bibfnamefont {F.~H.}\ \bibnamefont {{da
  Jornada}}}, \bibinfo {author} {\bibfnamefont {G.}~\bibnamefont {Antonius}},
  \bibinfo {author} {\bibfnamefont {T.}~\bibnamefont {Rangel}}, \bibinfo
  {author} {\bibfnamefont {S.~G.}\ \bibnamefont {Louie}}, \bibinfo {author}
  {\bibfnamefont {J.}~\bibnamefont {Deslippe}},\ and\ \bibinfo {author}
  {\bibfnamefont {A.}~\bibnamefont {Canning}},\ }\href
  {https://doi.org/10.1103/PhysRevB.99.125128} {\bibfield  {journal} {\bibinfo
  {journal} {Phys. Rev. B}\ }\textbf {\bibinfo {volume} {99}},\ \bibinfo
  {pages} {125128} (\bibinfo {year} {2019})}\BibitemShut {NoStop}%
\bibitem [{\citenamefont {Godby}\ and\ \citenamefont
  {Needs}(1989)}]{godbyMetalinsulatorTransitionKohnSham1989a}%
  \BibitemOpen
  \bibfield  {author} {\bibinfo {author} {\bibfnamefont {R.~W.}\ \bibnamefont
  {Godby}}\ and\ \bibinfo {author} {\bibfnamefont {R.~J.}\ \bibnamefont
  {Needs}},\ }\href {https://doi.org/10.1103/PhysRevLett.62.1169} {\bibfield
  {journal} {\bibinfo  {journal} {Phys. Rev. Lett.}\ }\textbf {\bibinfo
  {volume} {62}},\ \bibinfo {pages} {1169} (\bibinfo {year}
  {1989})}\BibitemShut {NoStop}%
\bibitem [{\citenamefont {Oschlies}\ \emph {et~al.}(1995)\citenamefont
  {Oschlies}, \citenamefont {Godby},\ and\ \citenamefont
  {Needs}}]{oschliesGWSelfenergyCalculations1995}%
  \BibitemOpen
  \bibfield  {author} {\bibinfo {author} {\bibfnamefont {A.}~\bibnamefont
  {Oschlies}}, \bibinfo {author} {\bibfnamefont {R.~W.}\ \bibnamefont
  {Godby}},\ and\ \bibinfo {author} {\bibfnamefont {R.~J.}\ \bibnamefont
  {Needs}},\ }\href {https://doi.org/10.1103/PhysRevB.51.1527} {\bibfield
  {journal} {\bibinfo  {journal} {Phys. Rev. B}\ }\textbf {\bibinfo {volume}
  {51}},\ \bibinfo {pages} {1527} (\bibinfo {year} {1995})}\BibitemShut
  {NoStop}%
\bibitem [{\citenamefont {Giantomassi}\ \emph {et~al.}(2011)\citenamefont
  {Giantomassi}, \citenamefont {Stankovski}, \citenamefont {Shaltaf},
  \citenamefont {Gr{\"u}ning}, \citenamefont {Bruneval}, \citenamefont
  {Rinke},\ and\ \citenamefont
  {Rignanese}}]{giantomassiElectronicPropertiesInterfaces2011}%
  \BibitemOpen
  \bibfield  {author} {\bibinfo {author} {\bibfnamefont {M.}~\bibnamefont
  {Giantomassi}}, \bibinfo {author} {\bibfnamefont {M.}~\bibnamefont
  {Stankovski}}, \bibinfo {author} {\bibfnamefont {R.}~\bibnamefont {Shaltaf}},
  \bibinfo {author} {\bibfnamefont {M.}~\bibnamefont {Gr{\"u}ning}}, \bibinfo
  {author} {\bibfnamefont {F.}~\bibnamefont {Bruneval}}, \bibinfo {author}
  {\bibfnamefont {P.}~\bibnamefont {Rinke}},\ and\ \bibinfo {author}
  {\bibfnamefont {G.-M.}\ \bibnamefont {Rignanese}},\ }\href
  {https://doi.org/10.1002/pssb.201046094} {\bibfield  {journal} {\bibinfo
  {journal} {Phys. Status Solidi B}\ }\textbf {\bibinfo {volume} {248}},\
  \bibinfo {pages} {275} (\bibinfo {year} {2011})}\BibitemShut {NoStop}%
\bibitem [{\citenamefont {Liu}\ \emph {et~al.}(2016)\citenamefont {Liu},
  \citenamefont {Kaltak}, \citenamefont {Klime{\v s}},\ and\ \citenamefont
  {Kresse}}]{liuCubicScalingGW2016}%
  \BibitemOpen
  \bibfield  {author} {\bibinfo {author} {\bibfnamefont {P.}~\bibnamefont
  {Liu}}, \bibinfo {author} {\bibfnamefont {M.}~\bibnamefont {Kaltak}},
  \bibinfo {author} {\bibfnamefont {J.}~\bibnamefont {Klime{\v s}}},\ and\
  \bibinfo {author} {\bibfnamefont {G.}~\bibnamefont {Kresse}},\ }\href
  {https://doi.org/10.1103/PhysRevB.94.165109} {\bibfield  {journal} {\bibinfo
  {journal} {Phys. Rev. B}\ }\textbf {\bibinfo {volume} {94}},\ \bibinfo
  {pages} {165109} (\bibinfo {year} {2016})}\BibitemShut {NoStop}%
\bibitem [{\citenamefont {Wilhelm}\ \emph {et~al.}(2016)\citenamefont
  {Wilhelm}, \citenamefont {Del~Ben},\ and\ \citenamefont
  {Hutter}}]{wilhelmGWGaussianPlane2016}%
  \BibitemOpen
  \bibfield  {author} {\bibinfo {author} {\bibfnamefont {J.}~\bibnamefont
  {Wilhelm}}, \bibinfo {author} {\bibfnamefont {M.}~\bibnamefont {Del~Ben}},\
  and\ \bibinfo {author} {\bibfnamefont {J.}~\bibnamefont {Hutter}},\ }\href
  {https://doi.org/10.1021/acs.jctc.6b00380} {\bibfield  {journal} {\bibinfo
  {journal} {J. Chem. Theory Comput.}\ }\textbf {\bibinfo {volume} {12}},\
  \bibinfo {pages} {3623} (\bibinfo {year} {2016})}\BibitemShut {NoStop}%
\bibitem [{\citenamefont {Aulbur}\ \emph {et~al.}(2000)\citenamefont {Aulbur},
  \citenamefont {J{\"o}nsson},\ and\ \citenamefont
  {Wilkins}}]{aulburQuasiparticleCalculationsSolids2000a}%
  \BibitemOpen
  \bibfield  {author} {\bibinfo {author} {\bibfnamefont {W.~G.}\ \bibnamefont
  {Aulbur}}, \bibinfo {author} {\bibfnamefont {L.}~\bibnamefont
  {J{\"o}nsson}},\ and\ \bibinfo {author} {\bibfnamefont {J.~W.}\ \bibnamefont
  {Wilkins}},\ }in\ \href {https://doi.org/10.1016/S0081-1947(08)60248-9}
  {\emph {\bibinfo {booktitle} {Solid {{State Physics}}}}},\ Vol.~\bibinfo
  {volume} {54},\ \bibinfo {editor} {edited by\ \bibinfo {editor}
  {\bibfnamefont {H.}~\bibnamefont {Ehrenreich}}\ and\ \bibinfo {editor}
  {\bibfnamefont {F.}~\bibnamefont {Spaepen}}}\ (\bibinfo  {publisher}
  {{Academic Press}},\ \bibinfo {year} {2000})\ pp.\ \bibinfo {pages}
  {1--218}\BibitemShut {NoStop}%
\bibitem [{\citenamefont {Louie}\ and\ \citenamefont
  {Rubio}(2005)}]{louie2005quasiparticle}%
  \BibitemOpen
  \bibfield  {author} {\bibinfo {author} {\bibfnamefont {S.~G.}\ \bibnamefont
  {Louie}}\ and\ \bibinfo {author} {\bibfnamefont {A.}~\bibnamefont {Rubio}},\
  }in\ \href@noop {} {\emph {\bibinfo {booktitle} {Handbook of materials
  modeling}}}\ (\bibinfo  {publisher} {Springer},\ \bibinfo {year} {2005})\
  pp.\ \bibinfo {pages} {215--240}\BibitemShut {NoStop}%
\bibitem [{\citenamefont {Fuchs}\ \emph {et~al.}(2007)\citenamefont {Fuchs},
  \citenamefont {Furthm{\"u}ller}, \citenamefont {Bechstedt}, \citenamefont
  {Shishkin},\ and\ \citenamefont
  {Kresse}}]{fuchsQuasiparticleBandStructure2007b}%
  \BibitemOpen
  \bibfield  {author} {\bibinfo {author} {\bibfnamefont {F.}~\bibnamefont
  {Fuchs}}, \bibinfo {author} {\bibfnamefont {J.}~\bibnamefont
  {Furthm{\"u}ller}}, \bibinfo {author} {\bibfnamefont {F.}~\bibnamefont
  {Bechstedt}}, \bibinfo {author} {\bibfnamefont {M.}~\bibnamefont
  {Shishkin}},\ and\ \bibinfo {author} {\bibfnamefont {G.}~\bibnamefont
  {Kresse}},\ }\href {https://doi.org/10.1103/PhysRevB.76.115109} {\bibfield
  {journal} {\bibinfo  {journal} {Phys. Rev. B}\ }\textbf {\bibinfo {volume}
  {76}},\ \bibinfo {pages} {115109} (\bibinfo {year} {2007})}\BibitemShut
  {NoStop}%
\bibitem [{\citenamefont {Chen}\ and\ \citenamefont
  {Pasquarello}(2015)}]{chenAccurateBandGaps2015a}%
  \BibitemOpen
  \bibfield  {author} {\bibinfo {author} {\bibfnamefont {W.}~\bibnamefont
  {Chen}}\ and\ \bibinfo {author} {\bibfnamefont {A.}~\bibnamefont
  {Pasquarello}},\ }\href {https://doi.org/10.1103/PhysRevB.92.041115}
  {\bibfield  {journal} {\bibinfo  {journal} {Phys. Rev. B}\ }\textbf {\bibinfo
  {volume} {92}},\ \bibinfo {pages} {041115(R)} (\bibinfo {year}
  {2015})}\BibitemShut {NoStop}%
\bibitem [{\citenamefont {Jiang}\ and\ \citenamefont
  {Blaha}(2016)}]{jiangGWLinearizedAugmented2016}%
  \BibitemOpen
  \bibfield  {author} {\bibinfo {author} {\bibfnamefont {H.}~\bibnamefont
  {Jiang}}\ and\ \bibinfo {author} {\bibfnamefont {P.}~\bibnamefont {Blaha}},\
  }\href {https://doi.org/10.1103/PhysRevB.93.115203} {\bibfield  {journal}
  {\bibinfo  {journal} {Phys. Rev. B}\ }\textbf {\bibinfo {volume} {93}},\
  \bibinfo {pages} {115203} (\bibinfo {year} {2016})}\BibitemShut {NoStop}%
\bibitem [{\citenamefont {Grumet}\ \emph {et~al.}(2018)\citenamefont {Grumet},
  \citenamefont {Liu}, \citenamefont {Kaltak}, \citenamefont {Klime{\v s}},\
  and\ \citenamefont {Kresse}}]{grumetQuasiparticleApproximationFully2018}%
  \BibitemOpen
  \bibfield  {author} {\bibinfo {author} {\bibfnamefont {M.}~\bibnamefont
  {Grumet}}, \bibinfo {author} {\bibfnamefont {P.}~\bibnamefont {Liu}},
  \bibinfo {author} {\bibfnamefont {M.}~\bibnamefont {Kaltak}}, \bibinfo
  {author} {\bibfnamefont {J.}~\bibnamefont {Klime{\v s}}},\ and\ \bibinfo
  {author} {\bibfnamefont {G.}~\bibnamefont {Kresse}},\ }\href
  {https://doi.org/10.1103/PhysRevB.98.155143} {\bibfield  {journal} {\bibinfo
  {journal} {Phys. Rev. B}\ }\textbf {\bibinfo {volume} {98}},\ \bibinfo
  {pages} {155143} (\bibinfo {year} {2018})}\BibitemShut {NoStop}%
\bibitem [{\citenamefont {Rinke}\ \emph
  {et~al.}(2005{\natexlab{a}})\citenamefont {Rinke}, \citenamefont {Qteish},
  \citenamefont {Neugebauer}, \citenamefont {Freysoldt},\ and\ \citenamefont
  {Scheffler}}]{rinkeCombiningGWcalculationsExactexchangeDensityfunctional2005b}%
  \BibitemOpen
  \bibfield  {author} {\bibinfo {author} {\bibfnamefont {P.}~\bibnamefont
  {Rinke}}, \bibinfo {author} {\bibfnamefont {A.}~\bibnamefont {Qteish}},
  \bibinfo {author} {\bibfnamefont {J.}~\bibnamefont {Neugebauer}}, \bibinfo
  {author} {\bibfnamefont {C.}~\bibnamefont {Freysoldt}},\ and\ \bibinfo
  {author} {\bibfnamefont {M.}~\bibnamefont {Scheffler}},\ }\href
  {https://doi.org/10.1088/1367-2630/7/1/126} {\bibfield  {journal} {\bibinfo
  {journal} {New J. Phys.}\ }\textbf {\bibinfo {volume} {7}},\ \bibinfo {pages}
  {126} (\bibinfo {year} {2005}{\natexlab{a}})}\BibitemShut {NoStop}%
\bibitem [{\citenamefont {Bruneval}\ and\ \citenamefont
  {Marques}(2013)}]{brunevalBenchmarkingStartingPoints2013a}%
  \BibitemOpen
  \bibfield  {author} {\bibinfo {author} {\bibfnamefont {F.}~\bibnamefont
  {Bruneval}}\ and\ \bibinfo {author} {\bibfnamefont {M.~A.~L.}\ \bibnamefont
  {Marques}},\ }\href {https://doi.org/10.1021/ct300835h} {\bibfield  {journal}
  {\bibinfo  {journal} {J. Chem. Theory Comput.}\ }\textbf {\bibinfo {volume}
  {9}},\ \bibinfo {pages} {324} (\bibinfo {year} {2013})}\BibitemShut {NoStop}%
\bibitem [{\citenamefont {{van Setten}}\ \emph {et~al.}(2017)\citenamefont
  {{van Setten}}, \citenamefont {Giantomassi}, \citenamefont {Gonze},
  \citenamefont {Rignanese},\ and\ \citenamefont
  {Hautier}}]{vansettenAutomationMethodologiesLargescale2017}%
  \BibitemOpen
  \bibfield  {author} {\bibinfo {author} {\bibfnamefont {M.~J.}\ \bibnamefont
  {{van Setten}}}, \bibinfo {author} {\bibfnamefont {M.}~\bibnamefont
  {Giantomassi}}, \bibinfo {author} {\bibfnamefont {X.}~\bibnamefont {Gonze}},
  \bibinfo {author} {\bibfnamefont {G.-M.}\ \bibnamefont {Rignanese}},\ and\
  \bibinfo {author} {\bibfnamefont {G.}~\bibnamefont {Hautier}},\ }\href
  {https://doi.org/10.1103/PhysRevB.96.155207} {\bibfield  {journal} {\bibinfo
  {journal} {Phys. Rev. B}\ }\textbf {\bibinfo {volume} {96}},\ \bibinfo
  {pages} {155207} (\bibinfo {year} {2017})}\BibitemShut {NoStop}%
\bibitem [{\citenamefont {Leppert}\ \emph {et~al.}(2019)\citenamefont
  {Leppert}, \citenamefont {Rangel},\ and\ \citenamefont
  {Neaton}}]{leppertPredictiveBandGaps2019a}%
  \BibitemOpen
  \bibfield  {author} {\bibinfo {author} {\bibfnamefont {L.}~\bibnamefont
  {Leppert}}, \bibinfo {author} {\bibfnamefont {T.}~\bibnamefont {Rangel}},\
  and\ \bibinfo {author} {\bibfnamefont {J.~B.}\ \bibnamefont {Neaton}},\
  }\href {https://doi.org/10.1103/PhysRevMaterials.3.103803} {\bibfield
  {journal} {\bibinfo  {journal} {Phys. Rev. Materials}\ }\textbf {\bibinfo
  {volume} {3}},\ \bibinfo {pages} {103803} (\bibinfo {year}
  {2019})}\BibitemShut {NoStop}%
\bibitem [{\citenamefont {Marom}\ \emph {et~al.}(2012)\citenamefont {Marom},
  \citenamefont {Caruso}, \citenamefont {Ren}, \citenamefont {Hofmann},
  \citenamefont {K{\"o}rzd{\"o}rfer}, \citenamefont {Chelikowsky},
  \citenamefont {Rubio}, \citenamefont {Scheffler},\ and\ \citenamefont
  {Rinke}}]{maromBenchmarkGWMethods2012a}%
  \BibitemOpen
  \bibfield  {author} {\bibinfo {author} {\bibfnamefont {N.}~\bibnamefont
  {Marom}}, \bibinfo {author} {\bibfnamefont {F.}~\bibnamefont {Caruso}},
  \bibinfo {author} {\bibfnamefont {X.}~\bibnamefont {Ren}}, \bibinfo {author}
  {\bibfnamefont {O.~T.}\ \bibnamefont {Hofmann}}, \bibinfo {author}
  {\bibfnamefont {T.}~\bibnamefont {K{\"o}rzd{\"o}rfer}}, \bibinfo {author}
  {\bibfnamefont {J.~R.}\ \bibnamefont {Chelikowsky}}, \bibinfo {author}
  {\bibfnamefont {A.}~\bibnamefont {Rubio}}, \bibinfo {author} {\bibfnamefont
  {M.}~\bibnamefont {Scheffler}},\ and\ \bibinfo {author} {\bibfnamefont
  {P.}~\bibnamefont {Rinke}},\ }\href
  {https://doi.org/10.1103/PhysRevB.86.245127} {\bibfield  {journal} {\bibinfo
  {journal} {Phys. Rev. B}\ }\textbf {\bibinfo {volume} {86}},\ \bibinfo
  {pages} {245127} (\bibinfo {year} {2012})}\BibitemShut {NoStop}%
\bibitem [{\citenamefont
  {Sharifzadeh}(2018)}]{sharifzadehManybodyPerturbationTheory2018}%
  \BibitemOpen
  \bibfield  {author} {\bibinfo {author} {\bibfnamefont {S.}~\bibnamefont
  {Sharifzadeh}},\ }\href {https://doi.org/10.1088/1361-648X/aab0d1} {\bibfield
   {journal} {\bibinfo  {journal} {J. Phys.: Condens. Matter}\ }\textbf
  {\bibinfo {volume} {30}},\ \bibinfo {pages} {153002} (\bibinfo {year}
  {2018})}\BibitemShut {NoStop}%
\bibitem [{\citenamefont {Rohlfing}\ and\ \citenamefont
  {Louie}(1998)}]{rohlfingElectronHoleExcitationsSemiconductors1998b}%
  \BibitemOpen
  \bibfield  {author} {\bibinfo {author} {\bibfnamefont {M.}~\bibnamefont
  {Rohlfing}}\ and\ \bibinfo {author} {\bibfnamefont {S.~G.}\ \bibnamefont
  {Louie}},\ }\href {https://doi.org/10.1103/PhysRevLett.81.2312} {\bibfield
  {journal} {\bibinfo  {journal} {Phys. Rev. Lett.}\ }\textbf {\bibinfo
  {volume} {81}},\ \bibinfo {pages} {2312} (\bibinfo {year}
  {1998})}\BibitemShut {NoStop}%
\bibitem [{\citenamefont {Deslippe}\ \emph {et~al.}(2012)\citenamefont
  {Deslippe}, \citenamefont {Samsonidze}, \citenamefont {Strubbe},
  \citenamefont {Jain}, \citenamefont {Cohen},\ and\ \citenamefont
  {Louie}}]{deslippeBerkeleyGWMassivelyParallel2012b}%
  \BibitemOpen
  \bibfield  {author} {\bibinfo {author} {\bibfnamefont {J.}~\bibnamefont
  {Deslippe}}, \bibinfo {author} {\bibfnamefont {G.}~\bibnamefont
  {Samsonidze}}, \bibinfo {author} {\bibfnamefont {D.~A.}\ \bibnamefont
  {Strubbe}}, \bibinfo {author} {\bibfnamefont {M.}~\bibnamefont {Jain}},
  \bibinfo {author} {\bibfnamefont {M.~L.}\ \bibnamefont {Cohen}},\ and\
  \bibinfo {author} {\bibfnamefont {S.~G.}\ \bibnamefont {Louie}},\ }\href
  {https://doi.org/10.1016/j.cpc.2011.12.006} {\bibfield  {journal} {\bibinfo
  {journal} {Comput. Phys. Commun.}\ }\textbf {\bibinfo {volume} {183}},\
  \bibinfo {pages} {1269} (\bibinfo {year} {2012})}\BibitemShut {NoStop}%
\bibitem [{\citenamefont {Tomiki}\ \emph {et~al.}(1993)\citenamefont {Tomiki},
  \citenamefont {Ganaha}, \citenamefont {Futemma}, \citenamefont {Shikenbaru},
  \citenamefont {Aiura}, \citenamefont {Yuri}, \citenamefont {Sato},
  \citenamefont {Fukutani}, \citenamefont {Kato}, \citenamefont {Miyahara}
  \emph {et~al.}}]{tomiki1993anisotropic}%
  \BibitemOpen
  \bibfield  {author} {\bibinfo {author} {\bibfnamefont {T.}~\bibnamefont
  {Tomiki}}, \bibinfo {author} {\bibfnamefont {Y.}~\bibnamefont {Ganaha}},
  \bibinfo {author} {\bibfnamefont {T.}~\bibnamefont {Futemma}}, \bibinfo
  {author} {\bibfnamefont {T.}~\bibnamefont {Shikenbaru}}, \bibinfo {author}
  {\bibfnamefont {Y.}~\bibnamefont {Aiura}}, \bibinfo {author} {\bibfnamefont
  {M.}~\bibnamefont {Yuri}}, \bibinfo {author} {\bibfnamefont {S.}~\bibnamefont
  {Sato}}, \bibinfo {author} {\bibfnamefont {H.}~\bibnamefont {Fukutani}},
  \bibinfo {author} {\bibfnamefont {H.}~\bibnamefont {Kato}}, \bibinfo {author}
  {\bibfnamefont {T.}~\bibnamefont {Miyahara}}, \emph {et~al.},\ }\href@noop {}
  {\bibfield  {journal} {\bibinfo  {journal} {J. Phys. Soc. Jpn.}\ }\textbf
  {\bibinfo {volume} {62}},\ \bibinfo {pages} {1372} (\bibinfo {year}
  {1993})}\BibitemShut {NoStop}%
\bibitem [{\citenamefont {Bortz}\ \emph {et~al.}(1990)\citenamefont {Bortz},
  \citenamefont {French}, \citenamefont {Jones}, \citenamefont {Kasowski},\
  and\ \citenamefont {Ohuchi}}]{bortz1990temperature}%
  \BibitemOpen
  \bibfield  {author} {\bibinfo {author} {\bibfnamefont {M.}~\bibnamefont
  {Bortz}}, \bibinfo {author} {\bibfnamefont {R.}~\bibnamefont {French}},
  \bibinfo {author} {\bibfnamefont {D.}~\bibnamefont {Jones}}, \bibinfo
  {author} {\bibfnamefont {R.}~\bibnamefont {Kasowski}},\ and\ \bibinfo
  {author} {\bibfnamefont {F.}~\bibnamefont {Ohuchi}},\ }\href@noop {}
  {\bibfield  {journal} {\bibinfo  {journal} {Phys. Scr.}\ }\textbf {\bibinfo
  {volume} {41}},\ \bibinfo {pages} {537} (\bibinfo {year} {1990})}\BibitemShut
  {NoStop}%
\bibitem [{\citenamefont {Tiwald}\ and\ \citenamefont
  {Schubert}(2000)}]{tiwald2000measurement}%
  \BibitemOpen
  \bibfield  {author} {\bibinfo {author} {\bibfnamefont {T.~E.}\ \bibnamefont
  {Tiwald}}\ and\ \bibinfo {author} {\bibfnamefont {M.}~\bibnamefont
  {Schubert}},\ }in\ \href@noop {} {\emph {\bibinfo {booktitle} {Optical
  Diagnostic Methods For Inorganic Materials II}}},\ Vol.\ \bibinfo {volume}
  {4103}\ (\bibinfo {organization} {SPIE},\ \bibinfo {year} {2000})\ pp.\
  \bibinfo {pages} {19--29}\BibitemShut {NoStop}%
\bibitem [{\citenamefont {Whited}\ and\ \citenamefont
  {Walker}(1969)}]{whited_walker_1969_cao}%
  \BibitemOpen
  \bibfield  {author} {\bibinfo {author} {\bibfnamefont {R.}~\bibnamefont
  {Whited}}\ and\ \bibinfo {author} {\bibfnamefont {W.}~\bibnamefont
  {Walker}},\ }\href@noop {} {\bibfield  {journal} {\bibinfo  {journal} {Phys.
  Rev.}\ }\textbf {\bibinfo {volume} {188}},\ \bibinfo {pages} {1380} (\bibinfo
  {year} {1969})}\BibitemShut {NoStop}%
\bibitem [{\citenamefont {Gori}\ \emph {et~al.}(2010)\citenamefont {Gori},
  \citenamefont {Rakel}, \citenamefont {Cobet}, \citenamefont {Richter},
  \citenamefont {Esser}, \citenamefont {Hoffmann}, \citenamefont {Del~Sole},
  \citenamefont {Cricenti},\ and\ \citenamefont {Pulci}}]{gori2010optical}%
  \BibitemOpen
  \bibfield  {author} {\bibinfo {author} {\bibfnamefont {P.}~\bibnamefont
  {Gori}}, \bibinfo {author} {\bibfnamefont {M.}~\bibnamefont {Rakel}},
  \bibinfo {author} {\bibfnamefont {C.}~\bibnamefont {Cobet}}, \bibinfo
  {author} {\bibfnamefont {W.}~\bibnamefont {Richter}}, \bibinfo {author}
  {\bibfnamefont {N.}~\bibnamefont {Esser}}, \bibinfo {author} {\bibfnamefont
  {A.}~\bibnamefont {Hoffmann}}, \bibinfo {author} {\bibfnamefont
  {R.}~\bibnamefont {Del~Sole}}, \bibinfo {author} {\bibfnamefont
  {A.}~\bibnamefont {Cricenti}},\ and\ \bibinfo {author} {\bibfnamefont
  {O.}~\bibnamefont {Pulci}},\ }\href@noop {} {\bibfield  {journal} {\bibinfo
  {journal} {Phys. Rev. B}\ }\textbf {\bibinfo {volume} {81}},\ \bibinfo
  {pages} {125207} (\bibinfo {year} {2010})}\BibitemShut {NoStop}%
\bibitem [{\citenamefont {Haidu}\ \emph {et~al.}(2011)\citenamefont {Haidu},
  \citenamefont {Fronk}, \citenamefont {Gordan}, \citenamefont {Scarlat},
  \citenamefont {Salvan},\ and\ \citenamefont {Zahn}}]{haidu2011dielectric}%
  \BibitemOpen
  \bibfield  {author} {\bibinfo {author} {\bibfnamefont {F.}~\bibnamefont
  {Haidu}}, \bibinfo {author} {\bibfnamefont {M.}~\bibnamefont {Fronk}},
  \bibinfo {author} {\bibfnamefont {O.~D.}\ \bibnamefont {Gordan}}, \bibinfo
  {author} {\bibfnamefont {C.}~\bibnamefont {Scarlat}}, \bibinfo {author}
  {\bibfnamefont {G.}~\bibnamefont {Salvan}},\ and\ \bibinfo {author}
  {\bibfnamefont {D.~R.}\ \bibnamefont {Zahn}},\ }\href@noop {} {\bibfield
  {journal} {\bibinfo  {journal} {Phys. Rev. B}\ }\textbf {\bibinfo {volume}
  {84}},\ \bibinfo {pages} {195203} (\bibinfo {year} {2011})}\BibitemShut
  {NoStop}%
\bibitem [{\citenamefont {Wiktor}\ \emph
  {et~al.}(2017{\natexlab{b}})\citenamefont {Wiktor}, \citenamefont
  {Rothlisberger},\ and\ \citenamefont {Pasquarello}}]{wiktor2017perovskites}%
  \BibitemOpen
  \bibfield  {author} {\bibinfo {author} {\bibfnamefont {J.}~\bibnamefont
  {Wiktor}}, \bibinfo {author} {\bibfnamefont {U.}~\bibnamefont
  {Rothlisberger}},\ and\ \bibinfo {author} {\bibfnamefont {A.}~\bibnamefont
  {Pasquarello}},\ }\href@noop {} {\bibfield  {journal} {\bibinfo  {journal}
  {J. Phys. Chem. Lett.}\ }\textbf {\bibinfo {volume} {8}},\ \bibinfo {pages}
  {5507} (\bibinfo {year} {2017}{\natexlab{b}})}\BibitemShut {NoStop}%
\bibitem [{\citenamefont {Wang}\ \emph {et~al.}(2022)\citenamefont {Wang},
  \citenamefont {Tal}, \citenamefont {Bischoff}, \citenamefont {Gono},\ and\
  \citenamefont {Pasquarello}}]{wang2022accurate}%
  \BibitemOpen
  \bibfield  {author} {\bibinfo {author} {\bibfnamefont {H.}~\bibnamefont
  {Wang}}, \bibinfo {author} {\bibfnamefont {A.}~\bibnamefont {Tal}}, \bibinfo
  {author} {\bibfnamefont {T.}~\bibnamefont {Bischoff}}, \bibinfo {author}
  {\bibfnamefont {P.}~\bibnamefont {Gono}},\ and\ \bibinfo {author}
  {\bibfnamefont {A.}~\bibnamefont {Pasquarello}},\ }\href@noop {} {\bibfield
  {journal} {\bibinfo  {journal} {{NPJ} Comput. Mater.}\ }\textbf {\bibinfo
  {volume} {8}},\ \bibinfo {pages} {237} (\bibinfo {year} {2022})}\BibitemShut
  {NoStop}%
\bibitem [{\citenamefont {Giustino}(2017)}]{giustino2017electron}%
  \BibitemOpen
  \bibfield  {author} {\bibinfo {author} {\bibfnamefont {F.}~\bibnamefont
  {Giustino}},\ }\href@noop {} {\bibfield  {journal} {\bibinfo  {journal} {Rev.
  Mod. Phys.}\ }\textbf {\bibinfo {volume} {89}},\ \bibinfo {pages} {015003}
  (\bibinfo {year} {2017})}\BibitemShut {NoStop}%
\bibitem [{\citenamefont {Karsai}\ \emph {et~al.}(2018)\citenamefont {Karsai},
  \citenamefont {Engel}, \citenamefont {Flage-Larsen},\ and\ \citenamefont
  {Kresse}}]{Karsai_kresse_2018}%
  \BibitemOpen
  \bibfield  {author} {\bibinfo {author} {\bibfnamefont {F.}~\bibnamefont
  {Karsai}}, \bibinfo {author} {\bibfnamefont {M.}~\bibnamefont {Engel}},
  \bibinfo {author} {\bibfnamefont {E.}~\bibnamefont {Flage-Larsen}},\ and\
  \bibinfo {author} {\bibfnamefont {G.}~\bibnamefont {Kresse}},\ }\href@noop {}
  {\bibfield  {journal} {\bibinfo  {journal} {New J. Phys.}\ }\textbf {\bibinfo
  {volume} {20}},\ \bibinfo {pages} {123008} (\bibinfo {year}
  {2018})}\BibitemShut {NoStop}%
\bibitem [{\citenamefont {Miglio}\ \emph {et~al.}(2020)\citenamefont {Miglio},
  \citenamefont {Brousseau-Couture}, \citenamefont {Godbout}, \citenamefont
  {Antonius}, \citenamefont {Chan}, \citenamefont {Louie}, \citenamefont
  {C{\^o}t{\'e}}, \citenamefont {Giantomassi},\ and\ \citenamefont
  {Gonze}}]{miglio2020predominance}%
  \BibitemOpen
  \bibfield  {author} {\bibinfo {author} {\bibfnamefont {A.}~\bibnamefont
  {Miglio}}, \bibinfo {author} {\bibfnamefont {V.}~\bibnamefont
  {Brousseau-Couture}}, \bibinfo {author} {\bibfnamefont {E.}~\bibnamefont
  {Godbout}}, \bibinfo {author} {\bibfnamefont {G.}~\bibnamefont {Antonius}},
  \bibinfo {author} {\bibfnamefont {Y.-H.}\ \bibnamefont {Chan}}, \bibinfo
  {author} {\bibfnamefont {S.~G.}\ \bibnamefont {Louie}}, \bibinfo {author}
  {\bibfnamefont {M.}~\bibnamefont {C{\^o}t{\'e}}}, \bibinfo {author}
  {\bibfnamefont {M.}~\bibnamefont {Giantomassi}},\ and\ \bibinfo {author}
  {\bibfnamefont {X.}~\bibnamefont {Gonze}},\ }\href@noop {} {\bibfield
  {journal} {\bibinfo  {journal} {{NPJ} Comput. Mater.}\ }\textbf {\bibinfo
  {volume} {6}},\ \bibinfo {pages} {1} (\bibinfo {year} {2020})}\BibitemShut
  {NoStop}%
\bibitem [{\citenamefont {Antonius}\ and\ \citenamefont
  {Louie}(2022)}]{antonius2022theory}%
  \BibitemOpen
  \bibfield  {author} {\bibinfo {author} {\bibfnamefont {G.}~\bibnamefont
  {Antonius}}\ and\ \bibinfo {author} {\bibfnamefont {S.~G.}\ \bibnamefont
  {Louie}},\ }\href@noop {} {\bibfield  {journal} {\bibinfo  {journal} {Phys.
  Rev. B}\ }\textbf {\bibinfo {volume} {105}},\ \bibinfo {pages} {085111}
  (\bibinfo {year} {2022})}\BibitemShut {NoStop}%
\bibitem [{\citenamefont {Engel}\ \emph {et~al.}(2022)\citenamefont {Engel},
  \citenamefont {Miranda}, \citenamefont {Chaput}, \citenamefont {Togo},
  \citenamefont {Verdi}, \citenamefont {Marsman},\ and\ \citenamefont
  {Kresse}}]{engel2022zero}%
  \BibitemOpen
  \bibfield  {author} {\bibinfo {author} {\bibfnamefont {M.}~\bibnamefont
  {Engel}}, \bibinfo {author} {\bibfnamefont {H.}~\bibnamefont {Miranda}},
  \bibinfo {author} {\bibfnamefont {L.}~\bibnamefont {Chaput}}, \bibinfo
  {author} {\bibfnamefont {A.}~\bibnamefont {Togo}}, \bibinfo {author}
  {\bibfnamefont {C.}~\bibnamefont {Verdi}}, \bibinfo {author} {\bibfnamefont
  {M.}~\bibnamefont {Marsman}},\ and\ \bibinfo {author} {\bibfnamefont
  {G.}~\bibnamefont {Kresse}},\ }\href@noop {} {\bibfield  {journal} {\bibinfo
  {journal} {Phys. Rev. B}\ }\textbf {\bibinfo {volume} {106}},\ \bibinfo
  {pages} {094316} (\bibinfo {year} {2022})}\BibitemShut {NoStop}%
\bibitem [{\citenamefont {Cardona}\ and\ \citenamefont
  {Thewalt}(2005)}]{cardona_thewalt_2005}%
  \BibitemOpen
  \bibfield  {author} {\bibinfo {author} {\bibfnamefont {M.}~\bibnamefont
  {Cardona}}\ and\ \bibinfo {author} {\bibfnamefont {M.~L.~W.}\ \bibnamefont
  {Thewalt}},\ }\href@noop {} {\bibfield  {journal} {\bibinfo  {journal} {Rev.
  Mod. Phys.}\ }\textbf {\bibinfo {volume} {77}},\ \bibinfo {pages} {1173}
  (\bibinfo {year} {2005})}\BibitemShut {NoStop}%
\bibitem [{\citenamefont {Zacharias}\ and\ \citenamefont
  {Giustino}(2016)}]{zacharias&giustino2016}%
  \BibitemOpen
  \bibfield  {author} {\bibinfo {author} {\bibfnamefont {M.}~\bibnamefont
  {Zacharias}}\ and\ \bibinfo {author} {\bibfnamefont {F.}~\bibnamefont
  {Giustino}},\ }\href@noop {} {\bibfield  {journal} {\bibinfo  {journal}
  {Phys. Rev. B}\ }\textbf {\bibinfo {volume} {94}},\ \bibinfo {pages} {075125}
  (\bibinfo {year} {2016})}\BibitemShut {NoStop}%
\bibitem [{\citenamefont {Zacharias}\ and\ \citenamefont
  {Giustino}(2020)}]{zacharias&giustino2020}%
  \BibitemOpen
  \bibfield  {author} {\bibinfo {author} {\bibfnamefont {M.}~\bibnamefont
  {Zacharias}}\ and\ \bibinfo {author} {\bibfnamefont {F.}~\bibnamefont
  {Giustino}},\ }\href@noop {} {\bibfield  {journal} {\bibinfo  {journal}
  {Phys. Rev. Res.}\ }\textbf {\bibinfo {volume} {2}},\ \bibinfo {pages}
  {013357} (\bibinfo {year} {2020})}\BibitemShut {NoStop}%
\bibitem [{\citenamefont {Filip}\ \emph {et~al.}(2021)\citenamefont {Filip},
  \citenamefont {Haber},\ and\ \citenamefont {Neaton}}]{filip2021phonon}%
  \BibitemOpen
  \bibfield  {author} {\bibinfo {author} {\bibfnamefont {M.~R.}\ \bibnamefont
  {Filip}}, \bibinfo {author} {\bibfnamefont {J.~B.}\ \bibnamefont {Haber}},\
  and\ \bibinfo {author} {\bibfnamefont {J.~B.}\ \bibnamefont {Neaton}},\
  }\href@noop {} {\bibfield  {journal} {\bibinfo  {journal} {Phys. Rev. Lett.}\
  }\textbf {\bibinfo {volume} {127}},\ \bibinfo {pages} {067401} (\bibinfo
  {year} {2021})}\BibitemShut {NoStop}%
\bibitem [{\citenamefont {Blase}\ \emph {et~al.}(2011)\citenamefont {Blase},
  \citenamefont {Attaccalite},\ and\ \citenamefont {Olevano}}]{blase2011first}%
  \BibitemOpen
  \bibfield  {author} {\bibinfo {author} {\bibfnamefont {X.}~\bibnamefont
  {Blase}}, \bibinfo {author} {\bibfnamefont {C.}~\bibnamefont {Attaccalite}},\
  and\ \bibinfo {author} {\bibfnamefont {V.}~\bibnamefont {Olevano}},\
  }\href@noop {} {\bibfield  {journal} {\bibinfo  {journal} {Physical Review
  B}\ }\textbf {\bibinfo {volume} {83}},\ \bibinfo {pages} {115103} (\bibinfo
  {year} {2011})}\BibitemShut {NoStop}%
\bibitem [{\citenamefont {Usuda}\ \emph {et~al.}(2002)\citenamefont {Usuda},
  \citenamefont {Hamada}, \citenamefont {Kotani},\ and\ \citenamefont {van
  Schilfgaarde}}]{usuda2002all}%
  \BibitemOpen
  \bibfield  {author} {\bibinfo {author} {\bibfnamefont {M.}~\bibnamefont
  {Usuda}}, \bibinfo {author} {\bibfnamefont {N.}~\bibnamefont {Hamada}},
  \bibinfo {author} {\bibfnamefont {T.}~\bibnamefont {Kotani}},\ and\ \bibinfo
  {author} {\bibfnamefont {M.}~\bibnamefont {van Schilfgaarde}},\ }\href@noop
  {} {\bibfield  {journal} {\bibinfo  {journal} {Phys. Rev. B}\ }\textbf
  {\bibinfo {volume} {66}},\ \bibinfo {pages} {125101} (\bibinfo {year}
  {2002})}\BibitemShut {NoStop}%
\bibitem [{\citenamefont {Rinke}\ \emph
  {et~al.}(2005{\natexlab{b}})\citenamefont {Rinke}, \citenamefont {Qteish},
  \citenamefont {Neugebauer}, \citenamefont {Freysoldt},\ and\ \citenamefont
  {Scheffler}}]{Rinke_2005}%
  \BibitemOpen
  \bibfield  {author} {\bibinfo {author} {\bibfnamefont {P.}~\bibnamefont
  {Rinke}}, \bibinfo {author} {\bibfnamefont {A.}~\bibnamefont {Qteish}},
  \bibinfo {author} {\bibfnamefont {J.}~\bibnamefont {Neugebauer}}, \bibinfo
  {author} {\bibfnamefont {C.}~\bibnamefont {Freysoldt}},\ and\ \bibinfo
  {author} {\bibfnamefont {M.}~\bibnamefont {Scheffler}},\ }\href@noop {}
  {\bibfield  {journal} {\bibinfo  {journal} {New J. Phys.}\ }\textbf {\bibinfo
  {volume} {7}},\ \bibinfo {pages} {126} (\bibinfo {year}
  {2005}{\natexlab{b}})}\BibitemShut {NoStop}%
\bibitem [{\citenamefont {{van Schilfgaarde}}\ \emph
  {et~al.}(2006{\natexlab{b}})\citenamefont {{van Schilfgaarde}}, \citenamefont
  {Kotani},\ and\ \citenamefont
  {Faleev}}]{vanschilfgaardeAdequacyApproximationsMathbitGW2006}%
  \BibitemOpen
  \bibfield  {author} {\bibinfo {author} {\bibfnamefont {M.}~\bibnamefont {{van
  Schilfgaarde}}}, \bibinfo {author} {\bibfnamefont {T.}~\bibnamefont
  {Kotani}},\ and\ \bibinfo {author} {\bibfnamefont {S.~V.}\ \bibnamefont
  {Faleev}},\ }\href {https://doi.org/10.1103/PhysRevB.74.245125} {\bibfield
  {journal} {\bibinfo  {journal} {Phys. Rev. B}\ }\textbf {\bibinfo {volume}
  {74}},\ \bibinfo {pages} {245125} (\bibinfo {year}
  {2006}{\natexlab{b}})}\BibitemShut {NoStop}%
\bibitem [{\citenamefont {Stankovski}\ \emph {et~al.}(2011)\citenamefont
  {Stankovski}, \citenamefont {Antonius}, \citenamefont {Waroquiers},
  \citenamefont {Miglio}, \citenamefont {Dixit}, \citenamefont {Sankaran},
  \citenamefont {Giantomassi}, \citenamefont {Gonze}, \citenamefont
  {C{\^o}t{\'e}},\ and\ \citenamefont {Rignanese}}]{stankovski0WBandGap2011}%
  \BibitemOpen
  \bibfield  {author} {\bibinfo {author} {\bibfnamefont {M.}~\bibnamefont
  {Stankovski}}, \bibinfo {author} {\bibfnamefont {G.}~\bibnamefont
  {Antonius}}, \bibinfo {author} {\bibfnamefont {D.}~\bibnamefont
  {Waroquiers}}, \bibinfo {author} {\bibfnamefont {A.}~\bibnamefont {Miglio}},
  \bibinfo {author} {\bibfnamefont {H.}~\bibnamefont {Dixit}}, \bibinfo
  {author} {\bibfnamefont {K.}~\bibnamefont {Sankaran}}, \bibinfo {author}
  {\bibfnamefont {M.}~\bibnamefont {Giantomassi}}, \bibinfo {author}
  {\bibfnamefont {X.}~\bibnamefont {Gonze}}, \bibinfo {author} {\bibfnamefont
  {M.}~\bibnamefont {C{\^o}t{\'e}}},\ and\ \bibinfo {author} {\bibfnamefont
  {G.-M.}\ \bibnamefont {Rignanese}},\ }\href
  {https://doi.org/10.1103/PhysRevB.84.241201} {\bibfield  {journal} {\bibinfo
  {journal} {Phys. Rev. B}\ }\textbf {\bibinfo {volume} {84}},\ \bibinfo
  {pages} {241201} (\bibinfo {year} {2011})}\BibitemShut {NoStop}%
\bibitem [{\citenamefont {Friedrich}\ \emph {et~al.}(2012)\citenamefont
  {Friedrich}, \citenamefont {Betzinger}, \citenamefont {Schlipf},
  \citenamefont {Bl{\"u}gel},\ and\ \citenamefont
  {Schindlmayr}}]{friedrich2012hybrid}%
  \BibitemOpen
  \bibfield  {author} {\bibinfo {author} {\bibfnamefont {C.}~\bibnamefont
  {Friedrich}}, \bibinfo {author} {\bibfnamefont {M.}~\bibnamefont
  {Betzinger}}, \bibinfo {author} {\bibfnamefont {M.}~\bibnamefont {Schlipf}},
  \bibinfo {author} {\bibfnamefont {S.}~\bibnamefont {Bl{\"u}gel}},\ and\
  \bibinfo {author} {\bibfnamefont {A.}~\bibnamefont {Schindlmayr}},\
  }\href@noop {} {\bibfield  {journal} {\bibinfo  {journal} {J. Phys.: Condens.
  Matter}\ }\textbf {\bibinfo {volume} {24}},\ \bibinfo {pages} {293201}
  (\bibinfo {year} {2012})}\BibitemShut {NoStop}%
\bibitem [{\citenamefont {Cooper}\ \emph {et~al.}(2015)\citenamefont {Cooper},
  \citenamefont {Gul}, \citenamefont {Toma}, \citenamefont {Chen},
  \citenamefont {Liu}, \citenamefont {Guo}, \citenamefont {Ager}, \citenamefont
  {Yano},\ and\ \citenamefont {Sharp}}]{cooper2015indirect}%
  \BibitemOpen
  \bibfield  {author} {\bibinfo {author} {\bibfnamefont {J.~K.}\ \bibnamefont
  {Cooper}}, \bibinfo {author} {\bibfnamefont {S.}~\bibnamefont {Gul}},
  \bibinfo {author} {\bibfnamefont {F.~M.}\ \bibnamefont {Toma}}, \bibinfo
  {author} {\bibfnamefont {L.}~\bibnamefont {Chen}}, \bibinfo {author}
  {\bibfnamefont {Y.-S.}\ \bibnamefont {Liu}}, \bibinfo {author} {\bibfnamefont
  {J.}~\bibnamefont {Guo}}, \bibinfo {author} {\bibfnamefont {J.~W.}\
  \bibnamefont {Ager}}, \bibinfo {author} {\bibfnamefont {J.}~\bibnamefont
  {Yano}},\ and\ \bibinfo {author} {\bibfnamefont {I.~D.}\ \bibnamefont
  {Sharp}},\ }\href@noop {} {\bibfield  {journal} {\bibinfo  {journal} {J.
  Phys. Chem. C}\ }\textbf {\bibinfo {volume} {119}},\ \bibinfo {pages} {2969}
  (\bibinfo {year} {2015})}\BibitemShut {NoStop}%
\bibitem [{\citenamefont {Tezuka}\ \emph {et~al.}(1994)\citenamefont {Tezuka},
  \citenamefont {Shin}, \citenamefont {Ishii}, \citenamefont {Ejima},
  \citenamefont {Suzuki},\ and\ \citenamefont
  {Sato}}]{tezuka1994photoemission}%
  \BibitemOpen
  \bibfield  {author} {\bibinfo {author} {\bibfnamefont {Y.}~\bibnamefont
  {Tezuka}}, \bibinfo {author} {\bibfnamefont {S.}~\bibnamefont {Shin}},
  \bibinfo {author} {\bibfnamefont {T.}~\bibnamefont {Ishii}}, \bibinfo
  {author} {\bibfnamefont {T.}~\bibnamefont {Ejima}}, \bibinfo {author}
  {\bibfnamefont {S.}~\bibnamefont {Suzuki}},\ and\ \bibinfo {author}
  {\bibfnamefont {S.}~\bibnamefont {Sato}},\ }\href@noop {} {\bibfield
  {journal} {\bibinfo  {journal} {J. Phys. Soc. Jpn.}\ }\textbf {\bibinfo
  {volume} {63}},\ \bibinfo {pages} {347} (\bibinfo {year} {1994})}\BibitemShut
  {NoStop}%
\bibitem [{\citenamefont {Zimmermann}\ \emph {et~al.}(1999)\citenamefont
  {Zimmermann}, \citenamefont {Steiner}, \citenamefont {Claessen},
  \citenamefont {Reinert}, \citenamefont {H{\"u}fner}, \citenamefont {Blaha},\
  and\ \citenamefont {Dufek}}]{zimmermann1999electronic}%
  \BibitemOpen
  \bibfield  {author} {\bibinfo {author} {\bibfnamefont {R.}~\bibnamefont
  {Zimmermann}}, \bibinfo {author} {\bibfnamefont {P.}~\bibnamefont {Steiner}},
  \bibinfo {author} {\bibfnamefont {R.}~\bibnamefont {Claessen}}, \bibinfo
  {author} {\bibfnamefont {F.}~\bibnamefont {Reinert}}, \bibinfo {author}
  {\bibfnamefont {S.}~\bibnamefont {H{\"u}fner}}, \bibinfo {author}
  {\bibfnamefont {P.}~\bibnamefont {Blaha}},\ and\ \bibinfo {author}
  {\bibfnamefont {P.}~\bibnamefont {Dufek}},\ }\href@noop {} {\bibfield
  {journal} {\bibinfo  {journal} {J. Phys.: Condens. Matter}\ }\textbf
  {\bibinfo {volume} {11}},\ \bibinfo {pages} {1657} (\bibinfo {year}
  {1999})}\BibitemShut {NoStop}%
\bibitem [{\citenamefont {Sharifzadeh}\ \emph {et~al.}(2012)\citenamefont
  {Sharifzadeh}, \citenamefont {Biller}, \citenamefont {Kronik},\ and\
  \citenamefont {Neaton}}]{sharifzadeh2012quasiparticle}%
  \BibitemOpen
  \bibfield  {author} {\bibinfo {author} {\bibfnamefont {S.}~\bibnamefont
  {Sharifzadeh}}, \bibinfo {author} {\bibfnamefont {A.}~\bibnamefont {Biller}},
  \bibinfo {author} {\bibfnamefont {L.}~\bibnamefont {Kronik}},\ and\ \bibinfo
  {author} {\bibfnamefont {J.~B.}\ \bibnamefont {Neaton}},\ }\href@noop {}
  {\bibfield  {journal} {\bibinfo  {journal} {Phys. Rev. B}\ }\textbf {\bibinfo
  {volume} {85}},\ \bibinfo {pages} {125307} (\bibinfo {year}
  {2012})}\BibitemShut {NoStop}%
\bibitem [{\citenamefont {Whited}\ \emph {et~al.}(1973)\citenamefont {Whited},
  \citenamefont {Flaten},\ and\ \citenamefont {Walker}}]{whited_walker_1973}%
  \BibitemOpen
  \bibfield  {author} {\bibinfo {author} {\bibfnamefont {R.}~\bibnamefont
  {Whited}}, \bibinfo {author} {\bibfnamefont {C.~J.}\ \bibnamefont {Flaten}},\
  and\ \bibinfo {author} {\bibfnamefont {W.}~\bibnamefont {Walker}},\
  }\href@noop {} {\bibfield  {journal} {\bibinfo  {journal} {Solid State
  Commun.}\ }\textbf {\bibinfo {volume} {13}},\ \bibinfo {pages} {1903}
  (\bibinfo {year} {1973})}\BibitemShut {NoStop}%
\bibitem [{\citenamefont {French}(1990)}]{french1990electronic}%
  \BibitemOpen
  \bibfield  {author} {\bibinfo {author} {\bibfnamefont {R.~H.}\ \bibnamefont
  {French}},\ }\href@noop {} {\bibfield  {journal} {\bibinfo  {journal} {J. Am.
  Ceram. Soc.}\ }\textbf {\bibinfo {volume} {73}},\ \bibinfo {pages} {477}
  (\bibinfo {year} {1990})}\BibitemShut {NoStop}%
\bibitem [{\citenamefont {Pascual}\ \emph {et~al.}(1977)\citenamefont
  {Pascual}, \citenamefont {Camassel},\ and\ \citenamefont
  {Mathieu}}]{pascual1977resolved}%
  \BibitemOpen
  \bibfield  {author} {\bibinfo {author} {\bibfnamefont {J.}~\bibnamefont
  {Pascual}}, \bibinfo {author} {\bibfnamefont {J.}~\bibnamefont {Camassel}},\
  and\ \bibinfo {author} {\bibfnamefont {H.}~\bibnamefont {Mathieu}},\
  }\href@noop {} {\bibfield  {journal} {\bibinfo  {journal} {Phys. Rev. Lett.}\
  }\textbf {\bibinfo {volume} {39}},\ \bibinfo {pages} {1490} (\bibinfo {year}
  {1977})}\BibitemShut {NoStop}%
\bibitem [{\citenamefont {Takahata}\ and\ \citenamefont
  {Naka}(2018)}]{takahata2018photoluminescence}%
  \BibitemOpen
  \bibfield  {author} {\bibinfo {author} {\bibfnamefont {M.}~\bibnamefont
  {Takahata}}\ and\ \bibinfo {author} {\bibfnamefont {N.}~\bibnamefont
  {Naka}},\ }\href@noop {} {\bibfield  {journal} {\bibinfo  {journal} {Phys.
  Rev. B}\ }\textbf {\bibinfo {volume} {98}},\ \bibinfo {pages} {195205}
  (\bibinfo {year} {2018})}\BibitemShut {NoStop}%
\bibitem [{\citenamefont {Tsoi}\ \emph {et~al.}(2006)\citenamefont {Tsoi},
  \citenamefont {Lu}, \citenamefont {Ramdas}, \citenamefont {Alawadhi},
  \citenamefont {Grimsditch}, \citenamefont {Cardona},\ and\ \citenamefont
  {Lauck}}]{tsoi2006isotopic}%
  \BibitemOpen
  \bibfield  {author} {\bibinfo {author} {\bibfnamefont {S.}~\bibnamefont
  {Tsoi}}, \bibinfo {author} {\bibfnamefont {X.}~\bibnamefont {Lu}}, \bibinfo
  {author} {\bibfnamefont {A.}~\bibnamefont {Ramdas}}, \bibinfo {author}
  {\bibfnamefont {H.}~\bibnamefont {Alawadhi}}, \bibinfo {author}
  {\bibfnamefont {M.}~\bibnamefont {Grimsditch}}, \bibinfo {author}
  {\bibfnamefont {M.}~\bibnamefont {Cardona}},\ and\ \bibinfo {author}
  {\bibfnamefont {R.}~\bibnamefont {Lauck}},\ }\href@noop {} {\bibfield
  {journal} {\bibinfo  {journal} {Phys. Rev. B}\ }\textbf {\bibinfo {volume}
  {74}},\ \bibinfo {pages} {165203} (\bibinfo {year} {2006})}\BibitemShut
  {NoStop}%
\bibitem [{\citenamefont {Kazimierczuk}\ \emph {et~al.}(2014)\citenamefont
  {Kazimierczuk}, \citenamefont {Fr{\"o}hlich}, \citenamefont {Scheel},
  \citenamefont {Stolz},\ and\ \citenamefont {Bayer}}]{kazimierczuk2014giant}%
  \BibitemOpen
  \bibfield  {author} {\bibinfo {author} {\bibfnamefont {T.}~\bibnamefont
  {Kazimierczuk}}, \bibinfo {author} {\bibfnamefont {D.}~\bibnamefont
  {Fr{\"o}hlich}}, \bibinfo {author} {\bibfnamefont {S.}~\bibnamefont
  {Scheel}}, \bibinfo {author} {\bibfnamefont {H.}~\bibnamefont {Stolz}},\ and\
  \bibinfo {author} {\bibfnamefont {M.}~\bibnamefont {Bayer}},\ }\href@noop {}
  {\bibfield  {journal} {\bibinfo  {journal} {Nature}\ }\textbf {\bibinfo
  {volume} {514}},\ \bibinfo {pages} {343} (\bibinfo {year}
  {2014})}\BibitemShut {NoStop}%
\bibitem [{\citenamefont {Uihlein}\ \emph {et~al.}(1981)\citenamefont
  {Uihlein}, \citenamefont {Fr{\"o}hlich},\ and\ \citenamefont
  {Kenklies}}]{uihlein1981investigation}%
  \BibitemOpen
  \bibfield  {author} {\bibinfo {author} {\bibfnamefont {C.}~\bibnamefont
  {Uihlein}}, \bibinfo {author} {\bibfnamefont {D.}~\bibnamefont
  {Fr{\"o}hlich}},\ and\ \bibinfo {author} {\bibfnamefont {R.}~\bibnamefont
  {Kenklies}},\ }\href@noop {} {\bibfield  {journal} {\bibinfo  {journal}
  {Phys. Rev. B}\ }\textbf {\bibinfo {volume} {23}},\ \bibinfo {pages} {2731}
  (\bibinfo {year} {1981})}\BibitemShut {NoStop}%
\bibitem [{\citenamefont {Towns}\ \emph {et~al.}(2014)\citenamefont {Towns},
  \citenamefont {Cockerill}, \citenamefont {Dahan}, \citenamefont {Foster},
  \citenamefont {Gaither}, \citenamefont {Grimshaw}, \citenamefont {Hazlewood},
  \citenamefont {Lathrop}, \citenamefont {Lifka}, \citenamefont {Peterson}
  \emph {et~al.}}]{towns2014xsede}%
  \BibitemOpen
  \bibfield  {author} {\bibinfo {author} {\bibfnamefont {J.}~\bibnamefont
  {Towns}}, \bibinfo {author} {\bibfnamefont {T.}~\bibnamefont {Cockerill}},
  \bibinfo {author} {\bibfnamefont {M.}~\bibnamefont {Dahan}}, \bibinfo
  {author} {\bibfnamefont {I.}~\bibnamefont {Foster}}, \bibinfo {author}
  {\bibfnamefont {K.}~\bibnamefont {Gaither}}, \bibinfo {author} {\bibfnamefont
  {A.}~\bibnamefont {Grimshaw}}, \bibinfo {author} {\bibfnamefont
  {V.}~\bibnamefont {Hazlewood}}, \bibinfo {author} {\bibfnamefont
  {S.}~\bibnamefont {Lathrop}}, \bibinfo {author} {\bibfnamefont
  {D.}~\bibnamefont {Lifka}}, \bibinfo {author} {\bibfnamefont {G.~D.}\
  \bibnamefont {Peterson}}, \emph {et~al.},\ }\href@noop {} {\bibfield
  {journal} {\bibinfo  {journal} {Comput. Sci. Eng.}\ }\textbf {\bibinfo
  {volume} {16}},\ \bibinfo {pages} {62} (\bibinfo {year} {2014})}\BibitemShut
  {NoStop}%
\end{thebibliography}%
\end{document}